\documentclass[aps,prx,reprint,superscriptaddress,floatfix]{revtex4-2}

\usepackage{amsmath,amssymb,amsfonts}
\usepackage{graphicx}   
\usepackage{upgreek}    
\usepackage{dcolumn}    
\usepackage{bm}
\usepackage{mathtools}
\usepackage{xcolor}
\usepackage{float} 
\usepackage[normalem]{ulem}
\usepackage{enumitem}

\usepackage{placeins}

\definecolor{linkColor}{RGB}{0,70,120}
\usepackage[colorlinks=true, allcolors=linkColor,pdfborder={0 0 0},pdfencoding = auto]{hyperref}

\definecolor{fb_color}{rgb}{0.8,0.5,0.}

\definecolor{py_color}{rgb}{0.8,0.0,0.6}

\definecolor{new_color}{rgb}{0,0.3,0.7}

\definecolor{gray}{rgb}{0.5,0.5,0.5}

\begin{document}

\title{Feedback-controlled mechanics of tissue vertex model: emergent soft elasticity and active yielding}

\author{Pengyu Yu}
\affiliation{Department of Physics, University of California Santa Barbara, Santa Barbara, CA 93106, USA}
%\affiliation{Institute of Biomechanics and Medical Engineering, Department of Engineering Mechanics, Tsinghua University, Beijing 100084, China}

\author{Fridtjof Brauns}
\email[]{fbrauns@kitp.ucsb.edu}
\affiliation{Kavli Institute for Theoretical Physics, University of California Santa Barbara, Santa Barbara, CA 93106, USA}

\author{M.\ Cristina Marchetti}
\email[]{cmarchetti@ucsb.edu}
\affiliation{Department of Physics, University of California Santa Barbara, Santa Barbara, CA 93106, USA}

\begin{abstract}

Biological tissues exhibit diverse mechanical and rheological behaviors during morphogenesis. 
While much is known about tissue phase transitions controlled by structural order and cell mechanics, key questions regarding how tissue-scale nematic order emerges from cell-scale processes and influences tissue rheology remain unclear.
Here, we develop a minimal vertex model that incorporates a coupling between active forces generated by cytoskeletal fibers and their alignment with local elastic stress in solid epithelial tissues. 
Combining theory and simulations, we show that this feedback loop induces an isotropic--nematic transition, leading to an ordered solid state that exhibits soft elasticity. Further increasing activity drives collective self-yielding, that is the ability to spontaneously transition form elastic to plastic deformations, with tissue flows that are correlated across the entire system. We argue that this remarkable state, that we dub plastic nematic solid, is uniquely suited to facilitate active tissue remodeling during morphogenesis. It fundamentally differs from the well-studied fluid regime where macroscopic elastic stresses vanish and the velocity correlations remain short-ranged.
Altogether, our theoretical results reveal a rich spectrum of tissue states jointly governed by activity and passive cell deformability, with important implications for understanding tissue mechanics and morphogenesis.

\end{abstract}

\maketitle

\section{\label{sec:intro}Introduction}

During morphogenesis, tissues undergo dramatic shape changes, associated with extensive remodeling of their cell-scale architecture and changes in their rheological properties \cite{Petridou2019,Lenne2022}. A central challenge in understanding these processes is how tissue-scale properties and dynamics emerge from the cell scale.
This challenge involves not only to the rich physics of matter at the interface between crystalline and amorphous, but also the active forces generated by cells and the mechanosensitive feedback loops that allow cells to collectively control tissue mechanics.
Thus, activity and feedback on the cell scale make living tissue fundamentally different from ``passive'' materials.

Living cells collectively organize into states with distinct rheological properties. They may sustain pre-stress to form stable solid structures \cite{Blanchard2018,Mongera2018}, or transition into a fluid state to facilitate collective tissue flows \cite{Mitchel2020,Hannezo2022}.
A solid tissue refers to a state that maintains mechanical integrity through persistent cell-cell contacts and supports finite elastic stresses.
In the last decade, experimental advances have provided access to the mechanical properties on the cell and tissue scales \cite{Villeneuve2025}. Techniques such as laser ablation \cite{Fernandez-Gonzalez2009,Shivakumar2016}, optical traps \cite{Bambardekar2015,Nishizawa2023}, magnetic droplets \cite{Mongera2018} allow measuring stresses inside tissues.
Collective tissue flows and shape changes during morphogenesis have often been interpreted through the lens of a so-called solid-to-fluid transition \cite{Bi2015,Wang2020a}, where tissue fluidization allows cells to rearrange and drive large-scale remodeling. However, experimental evidence suggests that tissue flow does not necessarily require a complete loss of solid-like mechanical properties. Recoil after laser cutting and force measurements using optical traps reveal that junctions are under tension even in flowing tissues \cite{Etournay2015,Piscitello-Gomez2023,Rigato2024}.
These studies suggest that morphogenetic flow should be understood as plastic deformation of a \emph{solid} tissue.
Further motivation for studying morphogenesis of solid tissue comes from regenerating \textit{Hydra}, where cell divisions are very rare and cells do not exchange neighbors as the nematic texture remodels \cite{Gierer1972,Bode2003}, suggesting that the tissue behaves like an elastic solid.

Experimental and theoretical work in the last decade has revealed that nematic order plays a vital role in coordinating spatiotemporal dynamics during tissue morphogenesis. 
Such nematic order arises when cells collectively align anisotropic cytoskeletal structures such as stress fibers, protrusions, or molecular motors localized along cell-cell interfaces \cite{Streichan2018}.
The stress generated along the direction of nematic order can drive collective tissue flows \cite{Marchetti2013,Streichan2018}.
Moreover, nematic order can control biological structure and function \cite{Lopez-Gay2020,Maroudas-Sacks2021}.
A striking example is the polyp-shaped organism \textit{Hydra}, whose body plan is tightly coupled to nematic order of muscle-like actomyosin fibers along its surface.
Recent work has shown that feedback loops between this nematic order, mechanics and morphogen signaling play a key role in \textit{Hydra}'s ability to regenerate from small tissue fragments and even from aggregates of dissociated cells \cite{Lengfeld2009,Aufschnaiter2017,Maroudas-Sacks2025}.

Our work is motivated by the combination of the two findings outlined above—the solid-like properties and the pervasiveness of nematic order in tissue morphogenesis.
Continuum models of active nematic solids have examined the interplay between internally generated stresses, elastic deformations and morphogen activation in driving tissue structure \cite{Wang2023a,Weevers2025,Ibrahimi2025}.
Many questions, however, remain on how order at the tissue scale originates from processes at the cell scale. In particular, we focus here on two key questions: (i) How does tissue-scale nematic order emerge from cell-scale processes? (ii) How do active stresses and nematic order influence tissue rheology?

Various theoretical works have begun to bridge the gap between cell-scale properties and tissue-scale behaviors.
Studies using agent-based models (in particular vertex models) have shown that active tensions \cite{Rozman2023,Claussen2024} or intercellular forces \cite{Mueller2019,Zhang2023,Chiang2024a}, mediated by cell-cell junctions, can produce nematic order in tissues. Motivated by the intrinsic extensile or contractile nematic activity of different cell types \cite{Balasubramaniam2021,Blanch-Mercader2021,Lv2024a}, recent studies have incorporated anisotropic shape-dependent bulk stresses into vertex models to account for self-organized tissue flows \cite{Lin2023,Rozman2025} and cell sorting \cite{Rozman2024a}. 
The role of nematic order on tissue rheology is, however, largely unexplored. Moreover, while much work has examined the transition between fluid-like and solid-like behavior in epithelial layers on substrates—where dissipation is primarily governed by propulsive forces \cite{Bi2016,F.Staddon2022,Ray2025}—far less attention has been paid to how intercellular forces regulate tissue rheology, including the yielding and plasticity of solid-like tissues. 

In this paper, we address these questions in the context of a modified vertex model of tissue. Key ingredients in the model are a nematic degree of freedom for each cell and a feedback loop coupling the nematic to local elastic stresses. In the remainder of the introduction we first briefly describe the model and then summarize our main results.

\begin{figure*}[htb]
    \centering
    \includegraphics[width=0.95 \textwidth]{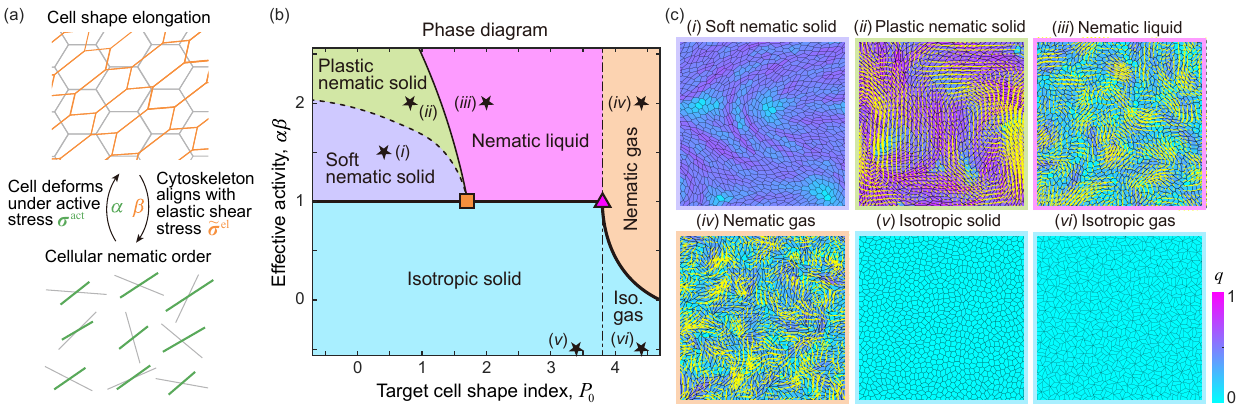}
    \caption{(a)~Schematic of the active feedback loop between the active extensile stress \({\bm{\sigma }}^{\mathrm{act}}\) generated by cellular cytoskeletal structures and their alignment with the local elastic shear stress \(\tilde{\bm{\sigma }}^{\mathrm{el}}\) [see model Eqs.~(\ref{EqEnergy}--\ref{EqEdgeTension})]. Here \(\alpha\) is the strength of active stress and \(\beta\) is the alignment strength. Their product $\alpha\beta$ determines an effective activity that controls the phase behavior.
    (b)~Phase diagram in terms of the effective activity \(\alpha\beta\) and the target cell shape index \(P_0\). The phase boundaries are obtained from numerical simulations in Figs.~\ref{Figure6}(a--e).
    (c)~Representative snapshots with velocity fields (yellow arrows) corresponding to the parameter combinations labeled by black stars in the phase diagram~[see Video~S3 in Supplemental Material].
    Cell shading (cyan to magenta) shows the magnitude of the nematic order parameter $q$. %The longest arrows correspond to a cell speed of 0.013 in panel (\textit{ii}) and 0.065 in panels (\textit{iii}) and (\textit{iv}).. 
     }
    \label{Figure1}
\end{figure*}

\setlength{\tabcolsep}{0.5em}
\renewcommand{\arraystretch}{1.5}
\begin{table*}
    \centering
    \begin{tabular}{cccccc}
    \hline
         & Isotropic solid & Soft nematic solid & Plastic nematic solid & Nematic liquid & Gas \\
         \hline
        Junctional tension & finite & finite & finite & finite & vanishing \\
        Elastic stress & finite & finite & finite & vanishing & vanishing \\
        Flow correlation length & no flow & no flow & large scale & active length & cell scale \\
        Distinctive feature & --- & soft elasticity & self-yielding & active nematic & extensive\\[-0.5em]
        & & & avalanches & turbulence & floppy modes\\
        \hline
    \end{tabular}
    \caption{Mechanical and rheological properties of diverse tissue states.}
    \label{table}
\end{table*} 

\subsection{2D vertex model of nematic tissue}
\label{subsec:model}

Vertex models are a well-established model class for epithelial tissue, where cells are described as a 2D network of (irregular) polygons tiling the plane and vertices define the fundamental geometric degrees of freedom \cite{Weliky1990,Nagai2001,Alt2017}. 
Unlike continuum descriptions, they explicitly define tensions along cell-cell junctions,
%A key advantage of the vertex model is that cell-cell junctions and interfacial tensions are explicitly defined, 
making it particularly well suited for describing junction-level mechanics and topological rearrangements \cite{Kafer2007,Fletcher2014,Osborne2017}.
In the area--perimeter formulation of the vertex model, a target cell shape index $P_0$ governs cell deformability and controls a solid-to-fluid transition; the fluid regime found above a critical target shape index is characterized by vanishing junctional tensions \cite{Bi2015,Yan2019a}. We extend this area--perimeter vertex model by incorporating mechanochemical feedback loops that couple mechanical stimuli to intracellular organization (mechanosensation and mechanotransduction), processes ubiquitous in living tissue and essential for morphogenesis \cite{Collinet2021}.
Specifically, we introduce two key ingredients:
(i) internal nematic order \cite{Dye2021,Maroudas-Sacks2025}, representing anisotropy of intracellular structures such as stress fibers, actomyosin protrusions, or junctional myosin localization \cite{Fernandez-Gonzalez2009,Kim2011,Maroudas-Sacks2021,Weng2022};
(ii) a feedback mechanism that acts to align the cells' nematic order along the local elastic stress \cite{Dye2021}, a coupling motivated by experiments on a wide range of tissues \cite{Mao2022,Maroudas-Sacks2025,Bailles2025}.
Cells generate local active stress along their nematic direction, thus closing a feedback loop as sketched in Fig.~\ref{Figure1}(a).

\subsection{Summary of results}
\label{subsec:results}

By numerically studying the dynamics of this model, we show that the feedback between activity and mechanical deformations gives rise to emergent nematic order and a range of new tissue states with a rich rheology. In particular, it provides a mechanism for fluidity and plasticity qualitatively different from the vanishing of edge tensions.
We show how tissue dynamics and rheology are jointly governed by an effective activity and the passive cell deformability as parametrized by $P_0$ [Fig.~\ref{Figure1}(b)].

At low effective activity, the system remains isotropic and solid-like for $P_0<3.81$ and it ``melts'' at the well-studied solid-to-fluid transition at $P_0 = 3.81$ \cite{Bi2015}, above which junctional tensions vanish. In this regime, an extensive number of degrees of freedom is floppy, suggesting that it should be considered a gas, rather than a liquid~\cite{Moshe2018,Merkel2019}. Accordingly, we denote this threshold as $P_0^\mathrm{G}$ to signify this transition to the gas phase.
A range of new rheological phases summarized in Table~\ref{table} emerge when the effective activity exceeds a critical threshold in the solid regime of the passive vertex model ($P_0<P_0^{\mathrm G}$). Here, alignment to mechanical stress induces emergent nematic order. 
\begin{enumerate}[label=(\roman*), leftmargin=\parindent]
    \item Tissue first transitions to a soft nematic solid. This state exhibits soft elasticity as sufficiently small but finite strains can be accommodated by    reorientation of the nematic texture—a mechanism reminiscent of soft nematic elastomers \cite{Golubovic1989,Warner1994}. For larger strains, it exhibits shear-induced rigidity \cite{Koski2012,Huang2022a,Fielding2023}.
    \item Upon further increasing the effective activity, tissue transitions to a plastic nematic solid with long-range correlated, internally-driven tissue flows~\cite{Brauns2024a}. These flows emerge because active stress locally drives the tissue beyond the yield threshold, leading to plastic rearrangements, while the tissue continues to maintain large elastic internal stresses.
    \item For higher cell deformability or higher activity, a transition to an active nematic liquid regime takes place. Here, macroscopic elastic stresses vanish, while microscopic junctional tensions are still finite, and tissue exhibits the turbulent-like dynamics ubiquitously observed in active nematic liquid crystals~\cite{Alert2020a}~[see Video~S4 in Supplemental Material].
    \item Finally for $P_0>P_0^{\mathrm G}$ and above a critical activity, we find an active nematic gas, with elongated cells that continuously exchange neighbors.
    In the nematic gas, junctional tensions vanish, distinguishing it from the liquid, where tensions remain finite. This distinction is crucial in the light of experimental evidence indicating finite tensions \cite{Piscitello-Gomez2023,Grigas2025}.
\end{enumerate}

As shown in Fig.~\ref{Figure1}, we have mapped out a phase diagram that organizes these phases and the transitions between them. We have additionally examined the response of the solid phases to quasi-static shear deformations, and have quantified the rheology of the tissue in each of these phases.
Notably, several of the phenomena found in the phase diagram, such as soft elasticity \cite{Warner1994}, active plastic flow \cite{Claussen2024}, and active turbulence \cite{Kawaguchi2017,Alert2020a}, have previously been studied in nematic materials, but had remained disconnected. Our model unifies these phenomena and allows one to study transitions between them.
Taken together, our work shows how active stresses and mechanical feedback jointly control nematic order and give rise to diverse rheological behaviors, including an active plastic solid phase. It reveals a multi-stage route from solid-like to fluid-like tissue behavior that is fundamentally different from the well-studied solid-to-fluid transition at $P_0^{\mathrm G}\!\approx\!3.81$ associated with the vanishing of junctional tensions. We show here that cell shape alone is not a good indicator of fluid vs solid-like rheology. Instead we suggest that tissues flow through active plasticity, while remaining solid.

%Our findings have important implications for tissue mechanics and morphogenesis. Previous work has suggested that during development cells may utilize the rigidity transition at $P_0^{\mathrm G}$ to switch between fluid and solid in order to facilitate morphogenetic flows \cite{Bi2015,Wang2020a}. The increased observed shape index of shape-changing tissue has been taken as evidence of this. However, the observation of tension on junctions is in conflict with the floppy regime \cite{Kim2021,Brauns2024a,Shen2025}.  We show here that cell shape alone is not a good indicator of fluid vs solid-like rheology. Instead we suggest that tissues flow through active plasticity, while remaining solid.

In the remainder of the paper we first introduce the model (Sec.~\ref{sec:model}), then present a mean-field calculation of the isotropic--nematic transition and the numerical results quantifying the various states and their transitions (Sec.~\ref{subsec:transition}). We next discuss the dynamics of topological defects in nematic solids (Sec.~\ref{subsec:defects}), and examine the response of tissues in different states to quasi-static shear deformations (Sec.~\ref{sec:rheology}). Finally, we reveal an overall phase diagram jointly governed by effective activity and target cell shape index (Secs.~\ref{sec:phases} and \ref{sec:phase_mechanics}). We conclude with an extensive discussion of our results and their implications in Sec.~\ref{sec:conclusion}. Details on the mean-field analysis, the effect of nematic alignment range, the implementation of shear deformations, and the method for defect identification are described in Appendices~\ref{Appendix_EquilAnalysis}-\ref{Appendix_Calculation}.

\section{\label{sec:model}Model}

\begin{figure}[t]
    \centering
    \includegraphics[width=0.48 \textwidth]{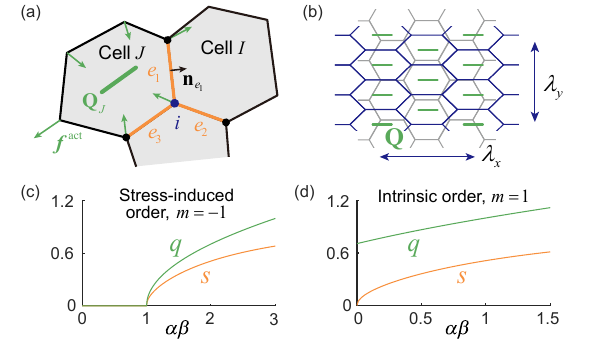}
    \caption{(a)~Diagram of active force induced by extensile nematic stress. The green stick represents the nematic order \({\bf{Q}}_J\) of cell \(J\) and the green arrows denote the extensile active forces \({\bm{f}}^{\rm{act}}\) induced by \({\bf{Q}}_J\). \({\bf n}_{e_1}\) is the unit vector normal to edge \(e_1\) pointing from cell \(J\) to cell \(I\).
    (b)~Schematic of the uniform deformation of hexagonal cells used for the mean-field analysis. 
    (c,d)~Mean-field values of the nematic order parameter \(q\) and the cell shape anisotropy \(s\) versus \(\alpha\beta\), for (c)~\(m\!=\!-1\) and (d)~\(m\!=\!1\).}
    \label{Figure2}
\end{figure}

We use the 2D vertex model \cite{Fletcher2014,Alt2017} to describe solid epithelial tissues and additionally endow each cell with a nematic degree of freedom to account for cellular active fibers~\cite{Dye2021,Maroudas-Sacks2025}. The passive mechanical energy of the 2D vertex model \cite{Farhadifar2007,Staple2010,Bi2015} reads
\begin{equation}
    E = \frac{1}{2}\,\sum_{J}\! {\left[ {{K_{\rm{a}}}{{({A_J} - {A_0})}^2} + {K_\mathrm{p}}{{({P_{\!J}} - {P_0})}^2}} \right]}\,,
\label{EqEnergy}
\end{equation}
where \(K_{\rm{a}}\) and \(K_\mathrm{p}\) represent the area and perimeter rigidity, respectively, which constrain the $J$-th cell's area \(A_J\) and perimeter \(P_J\) to their target values \(A_0\) and $P_0$. The sum is over all cells. Here and below we measure lengths in units of $\sqrt{A_0}$. The target perimeter $P_0$ then defines the target cell shape index, which characterizes the intrinsic cell deformability in the passive model. A lower shape index corresponds to a higher effective contractility along the cell perimeter, signifying a regime dominated by cortical tension \cite{Brodland2002,Lecuit2007}. Conversely, a higher index represents a reduction or even a total loss of junctional tension, leading the system toward a floppy state.
The geometric degrees of freedom are the vertex positions, in contrast to the cell centroid-based approach used in the Voronoi model \cite{Yang2017,Lawson-Keister2022a}\footnote{While Voronoi-type models offer greater simplicity due to their reduced number of geometric degrees of freedom, they do not correctly capture the (im)balance of forces at vertices. They are therefore not suitable to describe the subtle interplay of active and passive mechanical forces that governs the rich rheological behavior found in our model.}.
The vertex positions \({\bf{r}}_i\) evolve according to overdamped dynamics
\begin{equation}
     \nu \frac{d}{dt}\mathbf{r}_i = %{\bf{f}}_{\!i}^{\mathrm{el}} 
     -\frac{\partial E}{\partial \mathbf{r}_i}
     + {\bm{f}}_{i}^{{\rm{act}}},
    \label{EqOverdamped}
\end{equation}
where \(\nu\) is a friction and \({\bm{f}}_{i}^{{\rm{act}}}\) is the active force on the vertices induced by the cellular nematic stress.
Its form is based on the definition of the Cauchy stress  \cite{Tlili2019a,Lin2022}. The stress generates a force on each edge, which is equally partitioned between its two end vertices, giving
\begin{equation}
    {\bm{f}}_{i}^{{\rm{act}}} =  - \frac{1}{2}\sum\limits_e {{{\ell}_e}} ({\bm{\sigma }}_{\!J}^{{\rm{act}}} - {\bm{\sigma }}_{\!I}^{{\rm{act}}}) \cdot {{\bf{n}}_e},
    \label{EqActiveForce}
\end{equation}
where \({\bm{\sigma}}_{\!J}^{\rm{act}}\) represents the active nematic stress (defined below) exerted by cell \(J\)  and the summation is taken over the three edges connected to vertex \(i\) [Fig.~\ref{Figure2}(a)]. Here \(J\) and \(I\) denote the two neighboring cells sharing the interfacial edge \(e\), and \({\bf{n}}_e\) is the unit normal vector to edge \(e\) pointing from cell \(J\) to cell \(I\).

Cells generate active stresses through molecular motors in the cytoskeleton. The anisotropy of cytoskeletal structures such as stress fibers, adherens junctions, and actomyosin protrusions is represented by an independent nematic order parameter expressed via a traceless symmetric tensor $\mathbf{Q}_J$ \cite{Dye2021,Maroudas-Sacks2025}. The magnitude $q_J = |\mathbf{Q}_J|$ encodes the degree of intracellular order and the eigenvector of $\mathbf{Q}_J$ associated with the positive eigenvalue indicates the spatial orientation. Thus, $\mathbf{Q}_J$ vanishes when the intracellular organization is isotropic.
Force generation by the oriented cytoskeletal structures is reflected in the familiar deviatoric active stress
\begin{equation}
     {\bm{\sigma}}_{\!J}^{\rm{act}} = \alpha {\bf{Q}}_J,
    \label{EqSigma}
\end{equation}
which arises from the dipolar nature of active forces~ \cite{AditiSimha2002,Giomi2013,Lin2023,Rozman2024a}. Here, \(\alpha\) is the strength of activity. A positive (negative) sign of \(\alpha\) corresponds to contractile (extensile) activity, 
producing equal and opposite stresses along the nematic and perpendicular axes.
Note that, because of their linear relation, $\mathbf{Q}$ and ${\bm{\sigma}}^{\rm{act}}$ are mathematically interchangeable, and our model could in principle be formulated in terms of either one alone. However, these two quantities have distinct physical meanings and could have independent dynamics in more general models that we plan to explore in the future. For this reason, and for clarity, we keep them distinct.

We assume that the dynamics of $\mathbf{Q}$ is governed by three contributions: an intrinsic tendency of individual cells to acquire a nematic axis (i.e.,\ an anisotropically organized cytoskeleton), parametrized by the coefficient $m$; a saturation that limits the magnitude of cellular order; and alignment with the local elastic shear stress with alignment strength $\beta$ [Fig.~\ref{Figure1}(a)] \cite{Dye2021,Bailles2025,Brauns2026} 
\begin{equation}
    \tau_q \frac{d}{dt}\mathbf{Q}_J = 
    \left[m-2\,\mathrm{Tr}({\bf{Q}}_J^2) \right]{{\bf{Q}}_J} 
    -\beta
    \left\langle \tilde{\bm{\sigma}}^\mathrm{el}_{\!J} \right\rangle_{\!N_{\!J}}.
    \label{EqQJ}
\end{equation}
The timescale \(\tau_q\) of the nematic evolution is proportional to a rotational friction.
The elastic stress \({\bm{\sigma }}_{\!J}^{\mathrm{el}}\) for cell \(J\) \cite{Batchelor1970} is given by
\begin{equation}
    \boldsymbol{\sigma}^\mathrm{el}_{\!J} =\frac{\partial E}{\partial A_J}\,\mathbf{I} +\frac{1}{2A_J}\sum_{e \in J} T_e\, \frac{\boldsymbol{\ell}_e\otimes \boldsymbol{\ell}_e}{{\ell}_e},
    \label{EqElasticStress_E}
\end{equation}
where $\boldsymbol{\ell}_e$ is the vector along edge $e$, the summation runs over all edges of cell $J$, and 
\begin{equation}
T_e \equiv \frac{\partial E}{\partial \ell_e}
= K_{\rm p}(P_I + P_{\!J} - 2P_0)
\label{EqEdgeTension}
\end{equation}
defines the effective tension along edge \(e\). Negative values of $T_e$, while mathematically allowed, lie outside its physical interpretation as junctional tensions \footnote{Negative $T_e$ would correspond to compressive stress on junctions which would cause them to buckle/wrinkle. This is experimentally observed only in rare exceptions.}.
The shear stress tensor is the traceless part of the elastic stress tensor $\tilde{\bm{\sigma}}_{\!J}^{\mathrm{el}} = \bm{\sigma}_{\!J}^{\mathrm{el}}-\tfrac12 (\mathrm{Tr}\,{\bm{\sigma }}_{\!J}^{\mathrm{el}})\mathbf{I}$.
A negative value of $m$ corresponds to cells that intrinsically (i.e.,\ without coupling to neighbors) don't acquire cytoskeletal anisotropy, while a positive value means that cells spontaneously develop anisotropy.
A negative (positive) \(\beta\) causes \({\bf Q}_{J}\) to align parallel (perpendicular) to the stress. Given that cytoskeletal fibers often form supracellular structures and that protrusions (filopodia) can probe the local environment of a cell \cite{Lecuit2011,Maroudas-Sacks2021}, we adopt an alignment rule where \({\bf Q}_J\) orients along the local average stress of its neighborhood \(N_J\) which includes both the cell \(J\) and its nearest neighbors (defined as those sharing common edges with cell \(J\)).

Equations~(\ref{EqEnergy}--\ref{EqActiveForce}) are nondimensionalized using \(\nu\), \(K_{\rm{a}}\), and the length scale \(\sqrt{A_0}\). The model then contains two characteristic time scales: \(\nu/(A_0 K_{\rm{a}})\) controls the mechanical relaxation and \(\tau_q\) governs the remodeling of nematic order. In our simulations we set both equal to unity.
The competition between these two timescales will be examined elsewhere.
The dimensionless perimeter rigidity is fixed at \(K_\mathrm{p}\!=\!0.1\) throughout the manuscript.
In the following, we first focus on the dynamics deep in the solid regime, setting the target cell shape index \(P_0\!=\!-0.5\), which corresponds to a situation where perimeter elasticity contributes positively to junctional tension \cite{Farhadifar2007,Bi2015}.
Further below, we will also examine the effect of varying $P_0$, including values corresponding to the fluid state of the vertex model.
The values of cell activity \(\alpha\) and order alignment strength \(\beta\) will be discussed below.

We solve the coupled equations for vertex motion [Eq.~\eqref{EqOverdamped}] and cellular nematic order  \({\bf{Q}}_J\) [Eq.~\eqref{EqQJ}] using a simple Euler integration scheme with time step \(\Delta t=0.02\). T1 transitions \cite{Fletcher2014,Alt2017} are implemented to account for cell rearrangements with a length threshold \(\Delta \ell_{\rm{T1}}=0.01\). We initialize the system with \(N=1000\) cells using a random Voronoi tessellation with periodic boundary conditions in a box of length \(\sqrt{N}\). At \(t=0\), the system is isotropic under mechanical equilibrium with \({\bf Q}_J=\bf 0\) for all cells. Here, the isotropic state refers to the absence of nematic order within the tissue.
Unless stated otherwise, simulations are run for a total time \(t=4 \times 10^4\) to achieve a dynamical steady state.

\section{Results}

\subsection{Nematic transition and active yielding}
\label{subsec:transition}

We first analyze the ground states  of our model under an external deformation analytically. To do this, we consider a uniform affine deformation of a hexagonal tissue, with stretches \(\lambda_x\) and \(\lambda_y\) along the \(x\)- and \(y\)-axes, respectively. Accordingly, we choose the ansatz \(\mathbf{Q} \!=\! q/\sqrt{2} \, \text{diag}(1, -1)\) 
%\fb{Could we change the ansatz to $q/2 \text{diag}(1,-1)$, to avoid the inconvenient factor $\sqrt{2}$ below? I also don't think we should introduce $q_0$ as a separate quantity from $q$} \py{I originally made this change to make the notation consistent with Fig.2(c,d), where the plotted nematic order is $q=|\mathbf{Q}|$. In the previous version, the same symbol $q$ was used both in $\mathbf{Q}=q\text{diag}(1, -1)$ and for the nematic order parameter. But I agree that introducing a separate $q_0$ is also inconvenient. I would use $q/\sqrt{2}$ so that $q$ itself is the order parameter we have defined. Thanks!} 
for the nematic tensor of all cells [Fig.~\ref{Figure2}(b)]. Under these assumptions, the equilibrium state is controlled by \(2{q^2} = \alpha \beta  + m\) [see Appendix~\ref{Appendix_EquilAnalysis}]. Nematic order, quantified by $q\!>\!0$ %\fb{I find this confusing and suggest we simply write $q_0 > 0$} \py{Addressed}
, emerges through a pitchfork bifurcation at \(\alpha\beta \! > \!-m\) [Fig.~\ref{Figure2}(c)]. Therefore, we define \(\alpha\beta\) as effective activity and refer to it as “activity” thereafter.
Clearly, \(\alpha\) and \(\beta\) must be of the same sign to induce a nematic transition. In all simulations, we set \(\alpha=\beta<0\), unless stated otherwise.
This corresponds to the scenario where cytoskeletal fibers align parallel to the stress field, and cells actively extend along the director axis [Fig.~\ref{Figure1}(a)].
This activity can be driven by actomyosin protrusions \cite{Caswell2018,Maroudas-Sacks2021,Weng2022}
or by microtubules \cite{Picone2010,Singh2018}. 

For negative \(m\), cells do not spontaneously polarize and a critical activity is required to induce nematic order. In contrast, a positive \(m\) leads to spontaneous intrinsic order even in the absence of mechanical coupling [Fig.~\ref{Figure2}(d)].
This may describe, for instance, muscle or fibroblast cells which intrinsically take on elongated shapes.
Hereafter, we focus on the regime of emergent nematic order induced by mechanical stresses and fix \(m\!=\!-1\) unless noted otherwise.
This corresponds to tissues like epithelia that collectively develop orientational order.

The isotropic-to-nematic transition is accompanied by cell elongation. This is quantified by the cell shape tensor defined as
\begin{equation}
    {{\bf{S}}_J} = \frac{1}{P_{\!J}}\sum\nolimits_e {\boldsymbol{\ell}_e} \otimes {\boldsymbol{\ell}_e}/{\ell_e} - \frac{1}{2} \bf{I}\;.
    \label{eq:shape}
\end{equation} 
The mean shape anisotropy is then measured by the scalar \(s\!=\!{\langle \sqrt{{2{\rm{Tr(}}{\bf{S}}_J^2{\rm{)}}}} \rangle_{\!J}^{}} \!\in\! \left[ {0,1} \right)\), where the average is taken over all cells. Larger $s$ indicates tissues composed of more elongated cells. We note that $\mathbf{S}$ is computed a posteriori from cell shapes and is therefore distinct from the nematic tensor $\mathbf{Q}$, which is an intrinsic variable evolving independently of cell geometry.

\begin{figure}[t]
    \centering
    \includegraphics[width=0.48 \textwidth]{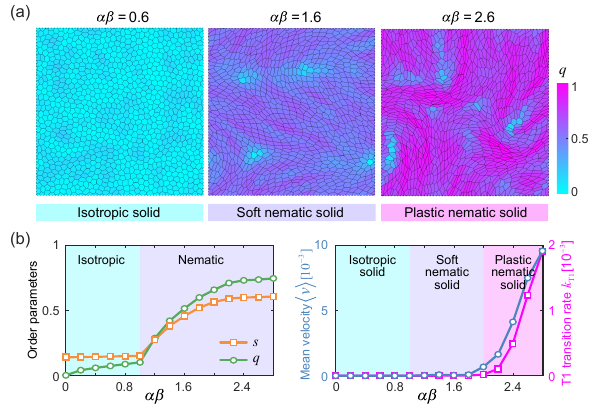}
    \caption{(a)~Tissue snapshots at different activity for \(m\!=\!-1\). The color is the magnitude of the nematic order parameter.
    (b)~Left: nematic order parameter \(q\) (green circles) and  mean shape anisotropy \(s\) (orange squares); right: mean cell velocity \(\langle v \rangle\) (blue circles) and T1 transition rate \(k_{\rm T1}\) (magenta squares) as functions of \(\alpha\beta\). All results are for $\alpha=\beta$, $P_0=-0.5$ and $N=1000$ cells.}
    \label{Figure3}
\end{figure}

Vertex model simulations confirm the isotropic-to-nematic transition mediated by the active feedback loop [Video~S1 in Supplemental Material]. When \(\alpha\beta \!<\! -m\), the cells remain static and isotropic, and the tissue nematic order parameter \(q\!=\!{\langle \sqrt{{{\rm{Tr(}}{\bf{Q}}_{\!J}^2{\rm{)}}}} \rangle_{\!J}^{}}\) is close to 0 [Fig.~\ref{Figure3}(a)]. When \(\alpha\beta \!>\! -m\), the active stress and mechanical feedback together destabilize the isotropic state and induce nematic order accompanied by steady cell elongation [Fig.~\ref{Figure3}(b)].

\begin{figure*}[htb]
    \centering
    \includegraphics[width=0.99 \textwidth]{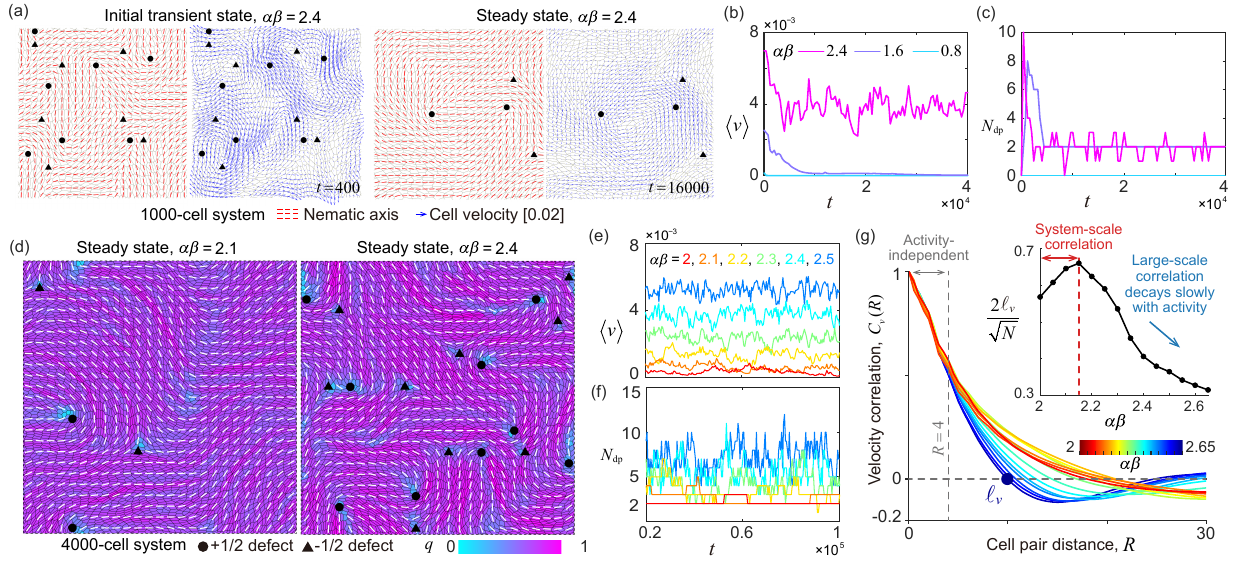}
    \caption{(a)~Snapshots of the nematic field (red lines) and the velocity field (blue arrows) in the initial transient state (\(t\!=\!400\)) and the dynamically steady flowing state (\(t\!=\!1.6\times10^4\)) in the plastic solid regime (\(\alpha\beta\!=\!2.4\)).
    (b,c)~Temporal evolution of (b) mean cell velocity \(\langle v \rangle\) and (c) number of \(\pm 1/2\) defect pairs \(N_{\rm{dp}}\) in isotropic, soft nematic, and plastic nematic solids (\(\alpha\beta=0.8, 1.6, 2.4\), respectively). 
    In the soft nematic solid ($\alpha\beta\!=\!1.6$), \(N_{\rm{dp}}\!=\!2\) in the steady state, and the curve partly overlaps with that of the plastic solid. (d)~Nematic fields of the plastic nematic solid for larger systems ($N\!=\!4000$ cells) at $\alpha\beta=2.1$ and $2.4$. (e--g) Activity dependence of plastic tissue flows ($N\!=\!4000$): (e) mean velocity \(\langle v \rangle\), (f) number of defect pairs \(N_{\rm{dp}}\), and (g) velocity correlation $C_v(R)$, versus $\alpha\beta$. The characteristic correlation length $\ell_v$ is identified as the first zero crossing of $C_v$. Inset in (g) shows a nonmonotonic dependence of $\ell_v$ on the activity. All results are for $\alpha=\beta$, $P_0=-0.5$ and $m=-1$.
     }
    \label{Figure4}
\end{figure*}

Intriguingly, further increasing \(\alpha\beta\) induces collective tissue flows [Video~S1 in Supplemental Material]. We calculate the mean cell velocity \(\langle v \rangle\) and the T1 transition rate \(k_{\rm T1}\) as indicators of the dynamical behavior [Appendix~\ref{Appendix_Calculation}]. \textit{A priori}, \(\langle v \rangle\) and \(k_{\rm T1}\) are not rigorous observables to quantify the unjamming transition—their validity is supported, however, by our rheological measurements presented in Sec.~\ref{sec:rheology}. As shown in Fig.~\ref{Figure3}(b), an intermediate state is identified at \(1\!<\!\alpha\beta\!<\!2\), where cells arrest after developing nematic order, with both \(\langle v \rangle\) and \(k_{\rm T1}\) close to 0 in the steady state. We call this the \emph{soft nematic solid} regime (we suppress the qualifier \emph{active} to keep the name short). Its distinct mechanical properties and formation mechanism will be discussed later.

Once \(\alpha\beta\) exceeds a second threshold of $\sim2$, active cell rearrangements and collective cell motion emerge, as characterized by elevated values of \(k_{\rm T1}\) and \(\langle v \rangle\) [Fig.~\ref{Figure3}(b)]. 
In this regime, high active stress destabilizes locally jammed states and facilitates neighbor exchanges, leading to persistent \emph{active plastic flow} \cite{Brauns2024a,Brauns2026}.  Passive yielding materials  transition from elastic to plastic response under the action of externally applied stresses. Our model tissue undergoes the same type of transition \emph{spontaneously}, due to internally generated active stresses. We therefore refer to this behavior as ``self-yielding''.
We refer to the tissue in this regime as a \emph{plastic nematic solid} (again, we suppress the qualifier \emph{active}). In Sec.~\ref{sec:phase_mechanics} we will show that elastic stresses are sustained in this regime while the tissue flows plastically. This distinguishes the plastic nematic solid from the nematic fluid regime where elastic stresses vanish.

We present additional simulations where we map out the phase diagrams of \(q\), \(\langle v \rangle\), and \(k_{\rm T1}\) while varying \(\alpha\) and \(\beta\) separately, to verify that the isotropic--nematic transition occurs precisely at the analytical bifurcation point \(\alpha\beta\!=\!-m\) [Fig.~\ref{FigureS1}(b)].
Moreover, simulation results for the intrinsically ordered case (\(m\!=\!1\)) are shown in Figs.~\ref{FigureS1}(a,c), where the soft and plastic nematic solid states emerge under active coupling, consistent with the analysis in Fig.~\ref{Figure2}(d).

\subsection{Emergence and dynamics of topological defects}
\label{subsec:defects}

The emergence of nematic order is accompanied by the nucleation of \(\pm 1/2\) topological defect pairs in the nematic field $\mathbf{Q}$ [Fig.~\ref{Figure4}(a)].
Topological defects are identified by first locating candidate cores from the zeros (defined as values below a threshold) of the cellular nematic order parameter, and then computing the winding number (topological charge) around each candidate on a coarse-grained grid [see Appendix~\ref{Appendix_Calculation} for details].
In the soft nematic solid, the defects eventually annihilate and cells revert to a quiescent state, leaving behind two stationary defect pairs [Figs.~\ref{Figure4}(b,c)].
In a domain with periodic boundary conditions—enforcing that the average strain must vanish—these two defect pairs are required to accommodate nematic order which is coupled to cell elongation [see Fig.~\ref{FigureS6}(a)]. 
In passing, we note that for mechanically free boundary conditions, the tissue could undergo uniform shear, thus supporting a globally ordered state with either finite or continually increasing elongation [see Fig.~\ref{FigureS6}(b)].

The plastic solid regime at higher \(\alpha\beta\) exhibits sustained flows accompanied by defect motion [Fig.~\ref{Figure4}(a)], during which both the mean velocity \(\langle v \rangle\) and the number of defect pairs $N_{\rm dp}$ fluctuate [Figs.~\ref{Figure4}(b,c)].
Interestingly, the defects are found to propagate much faster than the constituent cells [Video~S1 in Supplemental Material], which indicates a decoupling between the dynamics of the nematic texture and the material motion. 
Defect motion relative to the tissue has been experimentally observed in \emph{Hydra} \cite{Maroudas-Sacks2021} and was recently studied in a continuum active-nematic-solid model by some of us~\cite{Brauns2026}.

In a system with $N\!=\!1000$ cells, the plastic nematic solid mainly contains just two defect pairs [Fig.~\ref{Figure4}(c)]. Occasionally, an additional pair unbinds, or a pair annihilates, leaving an unstable domain wall that rapidly nucleates a new defect pair. The spontaneous unbinding of defect pairs suggests that larger systems might accommodate more defects in steady state. Indeed, in simulations with $N\!=\!4000$ cells, we find that more defects persist at sufficiently large activity [Fig.~\ref{Figure4}(d)].
Both the mean velocity and the number of defect pairs increase with the activity [Figs.~\ref{Figure4}(e,f)].

In active nematics, the defect spacing is closely linked to the velocity correlation length and both are controlled by the competition between active stresses and energetic cost of distortions of the nematic texture (Frank elasticity) \cite{Alert2020a}. We measure the correlation length $\ell_v$ as the first zero crossing of the velocity correlation function [see Appendix~\ref{Appendix_Calculation} for details] and find that it shows the expected decrease as a function of activity $\alpha\beta$ above $\alpha\beta \approx 2.2$ [Fig.~\ref{Figure4}(g)]. 
Near the self-yielding transition, $\ell_v$ exhibits a nonmonotonic dependence on $\alpha\beta$. For $\alpha\beta < 2.2$, the correlation length grows slightly with increasing $\alpha\beta$, suggesting that higher activity drives the coordinated onset of plastic flow. In this regime, $N_{\rm dp}$ remains fixed at $2$ as required by topology [Fig.~\ref{Figure4}(f)] and tissue flows are correlated across the entire system. We have verified this system-scale correlation for various system sizes with up to $10^4$ cells [see Fig.~\ref{FigureS2}]. This observation suggests that the plastic solid possesses no intrinsic length scale at the onset of the self-yielding transition.

\begin{figure*}[htb]
    \centering
    \includegraphics[width=0.98\textwidth]{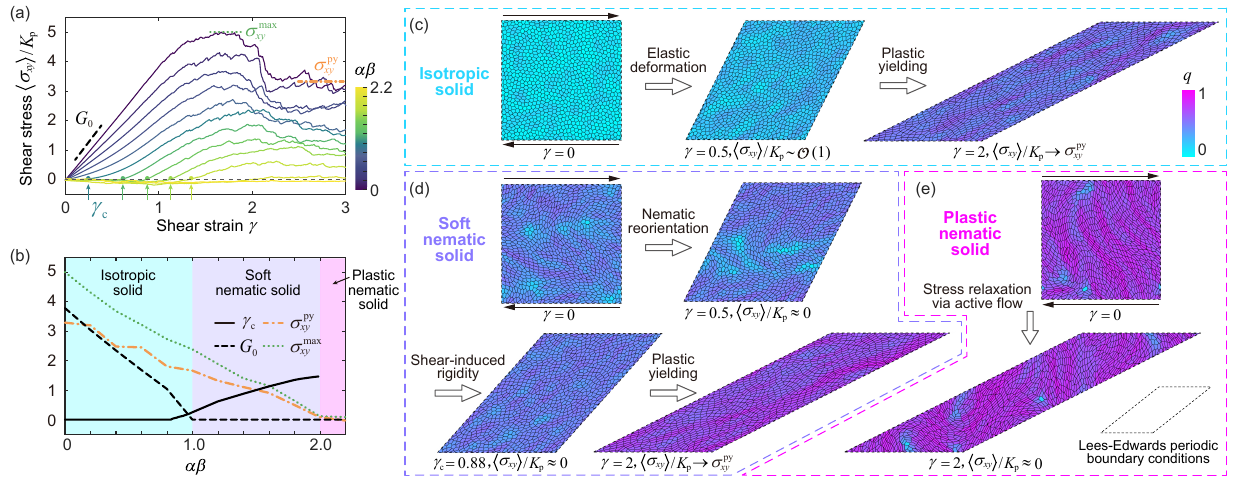}
    \caption{(a)~Shear stress–strain curves for varying \(\alpha\beta\) showing the mean shear stress $\langle \sigma_{xy} \rangle$
    as a function of strain $\gamma$. The initial shear modulus \(G_0\) is obtained from the slope under an infinitesimal strain, as highlighted by the dashed black line. \(\gamma_c\) is the critical strain at which rigidity emerges, as shown by the arrows along the strain axis. \(\sigma_{xy}^\mathrm{max}\) and \(\sigma_{xy}^\mathrm{py}\) denote the peak and post-yield shear stress, respectively. 
    (b)~Evolution of \(G_0\) (dashed black line), \(\gamma_c\) (solid black line), \(\sigma_{xy}^\mathrm{max}\) (dotted green line), and \(\sigma_{xy}^\mathrm{py}\) (dash-dotted orange line) versus \(\alpha\beta\).
    (c--e)~Representative snapshots of (c) isotropic (\(\alpha\beta\!=\!0.6\)), (d) soft nematic (\(\alpha\beta\!=\!1.4\)), and (e) plastic nematic (\(\alpha\beta\!=\!2.2\)) solid tissues under shear deformation. The corresponding snapshots of the shear stress field are shown in Fig.~\ref{FigureS4}. All results are for $\alpha=\beta$, $P_0=-0.5$, $m=-1$ and $N=1000$ cells.
    }
    \label{Figure5}
\end{figure*}

\subsection{Shear rheological response}
\label{sec:rheology}

Amorphous epithelial tissues exhibit complex mechanical responses to shear deformation, such as nonlinear elasticity and rate-dependent shear-thinning or thickening 
\cite{Sadeghipour2018,Huang2022a,Fielding2023,Hertaeg2024,Grossman2025,Nguyen2025}. Yet, how such mechanical behaviors manifest in tissues endowed with long-range nematic order remains poorly understood. 
Here, we perform rheological measurements on the tissues in steady state at various activities $\alpha\beta$ (as characterized in Fig.~\ref{Figure3}). We apply quasi-static simple shear strains to the tissues by using Lees--Edwards periodic boundary conditions [Video~S2 in Supplemental Material, see Appendix~\ref{Appendix_Shear} for details].
We measure the resulting total cell stress as ${\boldsymbol{\sigma}}_{\!J} = {\boldsymbol{\sigma}}_{\!J}^{\mathrm{el}} + {\boldsymbol{\sigma}}_{\!J}^{{\rm{act}}}$. The mean shear stress, denoted as $\langle \sigma_{xy} \rangle$, is calculated by averaging the off-diagonal component $\sigma_{xy}$ of ${\boldsymbol{\sigma}}_{\!J}$ over all cells.
The resulting stress-strain relations and representative snapshots of both nematic and stress fields for a range of \(\alpha\beta\) are shown in Fig.~\ref{Figure5} and Fig.~\ref{FigureS4}. 

In the isotropic solid state, the shear stress initially increases linearly with a finite shear modulus \(G_0\) and subsequently reaches a maximum at a critical strain, marking the onset of yielding and irreversible plastic cell rearrangements [Fig.~\ref{Figure5}(c) and Fig.~\ref{FigureS4}(a)]. After reaching the peak stress \(\sigma_{xy}^\mathrm{max}\), the tissue enters a post-yielding regime characterized by a post-yield stress plateau \(\sigma_{xy}^\mathrm{py}\), defined as the average shear stress over the strain interval \(\gamma\in[2.5,3]\). In contrast to passive materials, here tissue yielding and plastic deformations are accompanied by the emergence of nematic order and cell elongation [Fig.~\ref{Figure5}(c)], resembling shear-induced nematic order in stretched elastomers \cite{Berret1994}. In this process, elongated cells align along the principal shear direction, thereby accommodating the applied shear.

As \(\alpha\beta\) is increased, the initial shear modulus \(G_0\) decreases and vanishes at the isotropic--nematic transition [Figs.~\ref{Figure5}(a,b)]. The vanishing of $G_0$ motivates the name soft nematic solid.
Soft elasticity of the nematically ordered tissue arises because elongated cells can reorient and accommodate the applied shear, while the mean shear stress $\langle \sigma_{xy} \rangle$ remains zero [Fig.~\ref{Figure5}(d)].
This soft nematic elasticity is reminiscent of nematic elastomers \cite{Golubovic1989,Warner1994}. However, in contrast to such passive materials, here, softness arises dynamically from the interplay of active stresses and mechanical feedback that gives rise to emergent nematic order. 
When the applied shear reaches a critical value \(\gamma_c\), the shear stress becomes nonzero, indicating stiffening of the tissue [Fig.~\ref{FigureS4}(b)].
This occurs because the strain that can be accommodated by reorientation of the nematic texture is exhausted once the nematic aligns with the principal shear axis.
Further shearing builds up elastic stress and eventually leads to cell rearrangements, i.e.\ plastic yielding. 
Shear-induced rigidity and yielding have been reported in the passive vertex model in its ``floppy'' regime ($P_0\!>\!P_0^{\mathrm G}\!\approx\!3.81$)~\cite{Huang2022a,Fielding2023}. 
In Sec.~\ref{sec:phases}, we map out the full phase diagram in the $(P_0,\alpha\beta)$ plane and show that the floppy regime is fundamentally distinct from this soft nematic solid.

Finally, in the plastic nematic solid for \(\alpha\beta \gtrsim 2\), active stresses persistently cause self-yielding. The resulting sustained tissue flows rapidly relax stresses arising from externally applied shear, leading to a complete loss of mechanical rigidity to externally applied shear [Fig.~\ref{Figure5}(e) and Fig.~\ref{FigureS4}(c)]. Despite the loss of macroscopic rigidity, however, the tissue retains finite local elasticity, with elastic shear stresses that are canceled by active stresses in the steady state. This important distinction from the fluid state, where elastic shear stresses vanish on all scales, is discussed further in Sec.~\ref{sec:phase_mechanics}.

These rheological quantifications provide a complementary characterization of the distinct tissue states [Fig.~\ref{Figure5}(b)]. In the isotropic regime, \(G_0\) decreases almost linearly with increasing \(\alpha\beta\). When \(\alpha\beta\) is close to the critical value 1, the disappearance of \(G_0\) and a nonzero value of  \(\gamma_c\) indicate the transition from isotropic to nematic with soft elasticity.
When \(\alpha\beta \gtrsim 2\), the persistent plastic flow leads to the complete loss of tissue rigidity. Throughout this process, both \(\sigma_{xy}^\mathrm{max}\) and \(\sigma_{xy}^\mathrm{py}\) remain finite and decrease with higher \(\alpha\beta\), but drop to zero after the transition to the plastic nematic solid.

In Appendix~\ref{Appendix_AlignmentEffect} we examine the behavior in a setting where nematic order aligns with locally averaged stress over different coarse-graining radii. In the case of single-cell stress alignment without local averaging, no long-range nematic order emerges [Figs.~\ref{FigureS3}(a,f)] and the soft elasticity of the nematic solid state becomes less pronounced, with the critical strain $\gamma_\mathrm{c} \ll 1$ and independent of \(\alpha\beta\) [Fig.~\ref{FigureS3}(e)].
This is in line with the behavior of nematic elastomers which exhibit soft elasticity only if nematic domains are sufficiently large \cite{Finkelmann1997}. Only then can reorientation of nematic directors accommodate externally applied shear strains without incurring an energetic cost.

Together, our theoretical results suggest that nematic solid tissues possess a dual mechanical nature: actomyosin networks actively generate forces to maintain tissue integrity, while simultaneously aligning with stress fields to self-organize supracellular networks and enable active remodeling. 

\subsection{Phase diagram reveals roles of activity and target shape index}
\label{sec:phases}

\begin{figure*}[htb]
    \centering
    \includegraphics[width=0.98\textwidth]{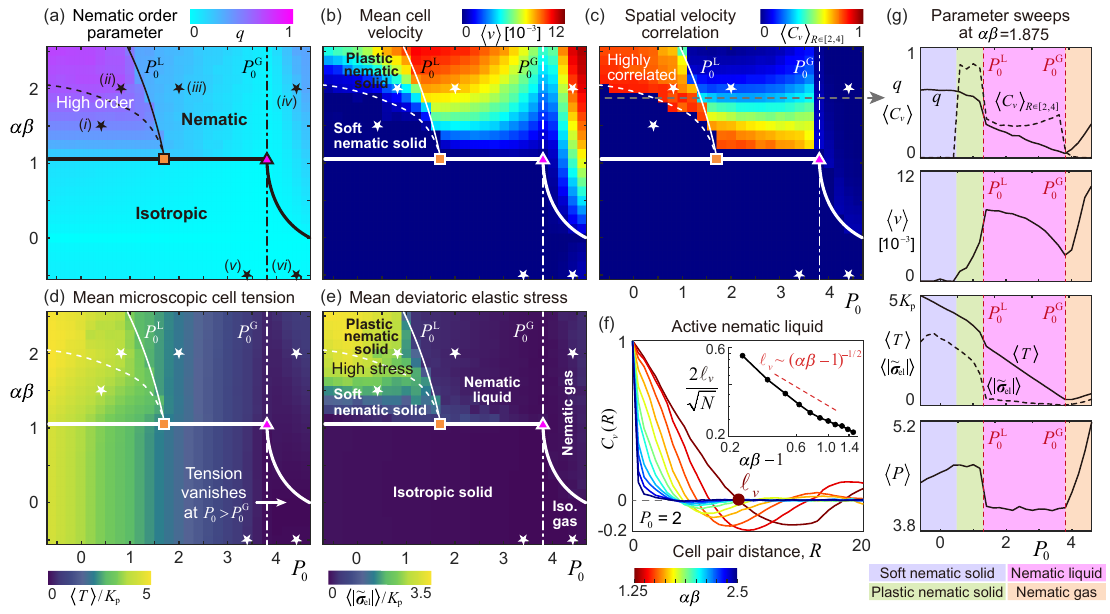}
    \caption{
    (a--e)~Diagrams depending on the effective activity \(\alpha\beta\) and the target cell shape index \(P_0\), showing (a)~nematic order parameter \(q\), (b)~mean cell velocity \(\langle v \rangle\), (c)~spatial velocity correlation \(\langle C_v \rangle_{R \in [2,4]}\), (d)~mean cell line tension \(\langle T \rangle \), and (e)~mean deviatoric elastic stress $\langle\lvert\tilde{\boldsymbol{\sigma}}_\mathrm{\!el}\rvert\rangle$. The dashed line separates the soft and plastic nematic solid phases. The thin solid line represents a critical \(P_{\!0}^{\mathrm{L}}\) dependent on \(\alpha\beta\), which marks the active melting transition from nematic solid to fluid. The dash-dotted line indicates the threshold \(P_0\!=\!P_{\!0}^{\mathrm G}\). Two triple points, at $(P_0, \alpha\beta) \approx (1.7,1)$ and $(P_0, \alpha\beta) = (3.81,1)$, are marked by orange square and magenta triangle, respectively.
    (f)~Velocity correlation function for the nematic liquid regime under varying \(\alpha\beta\) at fixed $P_0=2$. The inset shows the dependence of characteristic correlation length on the activity, with a power-law fit indicated by the red dashed line.
    (g)~Parameter sweeps of the diagrams~(a--e) and Fig.~\ref{FigureS5}(b) at $\alpha\beta=1.875$.
    The overall phase diagram is shown in Fig.~\ref{Figure1}(b).
    All results are for $\alpha=\beta$, $m=-1$ and $N=1000$ cells.
    }
    \label{Figure6}
\end{figure*}

So far, we have focused on the role of active stresses in tissues which, when passive, are deep in the solid regime of the vertex model. Previous studies have investigated the role of the target shape index $P_0$ on the ``passive'' rheology of the vertex model, in particular, the rigidity transition that happens when $P_0$ is increased beyond $P_0^{\mathrm G} \approx 3.81$~\cite{Bi2015,Huang2022a,Hertaeg2024}. 
What is the role of the target shape index on the stress-mediated nematic ordering mechanism proposed here?
What is the relation between the ``passive'' rigidity transition and the active self-yielding transition investigated above?
To address these questions, we map out the parameter space of $\alpha \beta$ and $P_0$ using numerical simulations, and quantify nematic order as well as kinematic and rheological features (Fig.~\ref{Figure6} and Fig.~\ref{FigureS5}, Video~S3 in Supplemental Material).

We find that, when \(P_0\!<\!P_{\!0}^{\mathrm G}\), the isotropic--nematic transition remains at the critical value \(\alpha\beta\!=\!1\) [Fig.~\ref{Figure6}(a)]. 
Nematic order gradually vanishes as $P_0$ approaches the well-known rigidity transition at \(P_{\!0}^{\mathrm G}\). At this transition, which occurs independently of activity \(\alpha\beta\), the model enters a passive floppy regime where microscopic junctional tensions vanish [see Fig.~\ref{Figure6}(d)] because cells have excess perimeter \cite{Farhadifar2007,Bi2015}.
Above $P_{\!0}^{\mathrm G}$, the isotropic--nematic transition shifts to lower $\alpha\beta$, as cells are spontaneously elongated due to their excess perimeter [cf.\ Figs.~\ref{FigureS5}(a,b)].
The soft and plastic nematic solids that we identified in the previous sections occur in a distinct region with high nematic order bounded by \(P_0^{\mathrm{L}}(\alpha\beta)\) [Figs.~\ref{Figure6}(a,b)].
The plastic nematic solid at \(P_0\!<\!P_0^{\mathrm{L}}\) exhibits significantly higher nematic order than the nematic fluid at \(P_0 \!>\!P_0^{\mathrm{L}}\).

Remarkably, we find that the active-stress-driven transitions from a soft nematic solid to a plastic nematic solid, and eventually to a nematic fluid, both meet at a ``triple point'' at $(P_0, \alpha\beta) \approx (1.7,1.0)$ [marked by an orange square in Figs.~\ref{Figure6}(a--e)].
Notably, this critical point lies way below the ``passive'' rigidity transition of the vertex model at $P_0^{\mathrm G} \approx 3.81$.
We conclude that the active-stress-driven melting transition at $P_0^\mathrm{L}$ is fundamentally different from the rigidity transition at $P_0^{\mathrm G}$.

In nematic solids, upon approaching the transition at $P_0^\mathrm{L}$, the average cell velocity \(\langle v \rangle\) increases gradually while $q$ remains high [Fig.~\ref{Figure6}(b)]. Upon crossing $P_0^\mathrm{L}$, nematic order \(q\) decreases significantly in an apparently discontinuous fashion while \(\langle v \rangle\) reaches a maximum. With further increase of \(P_0\), both \(q\) and \(\langle v \rangle\) decrease continuously, but rise again once \(P_0\) exceeds \(P_{\!0}^{\mathrm G}\).

We calculate the mean value of the velocity correlation $C_v(R)$ over the range $R \! \in\![2,4]$ as a measure of short-range correlation strength [Fig.~\ref{Figure6}(c), Appendix~\ref{Appendix_Calculation}].
Velocity correlations are maximal (i.e.\ most long-ranged) in the plastic nematic solid, in sharp contrast to the weaker and shorter-range correlations observed in the nematic liquid (\(P_0^{\mathrm{L}}\!<\!P_0\!<\!P_0^{\mathrm{G}}\)) and they vanish almost entirely in the nematic gas (\(P_0\!>\!P_0^{\mathrm{G}}\)).
In the liquid, velocity correlations decrease with increasing activity $\alpha\beta$, suggesting the existence of an ``active length'' that decreases with activity [Fig.~\ref{Figure6}(f), Video~S4 in Supplemental Material]. We find that the characteristic velocity-correlation length follows \(\ell_v\sim (\alpha\beta\!-\!1)^{-1/2}\) in the nematic liquid, consistent with the characteristic length scale of active nematic turbulence \cite{Alert2020a}.
Mechanistically, this connection arises because rapid cell rearrangements in the liquid phase relax elastic deformations, rendering the small local elastic stress an indirect measure of the strain rate. Thus, the stress-alignment parameter $\beta$ in Eq.~\eqref{EqQJ} effectively plays a role analogous to the flow-alignment parameter in active nematic theories.

There are several important differences in the  behavior of velocity correlations in the nematic liquid as compared to the plastic nematic solid discussed in Sec.~\ref{subsec:defects}.
In the plastic solid the velocity correlation length $\ell_v$ is a nonmonotonic function of activity near the onset of spontaneous motion [cf.\ Fig.~\ref{Figure4}(g), inset], and is much larger than in the liquid at large activity [$\alpha\beta \gtrsim 2$; see Fig.~\ref{Figure6}(c) and Fig.~\ref{FigureS5}(c)]. 
We hypothesize that these differences result from different effective Frank elastic constants in the two regimes.
In our model, the Frank constant is not an independent control parameter. Instead an effective Frank elasticity arises from the coupling of nematic order across space mediated by the alignment of cells to elastic stress ($\beta$ term in Eq.~\eqref{EqQJ}).  The plastic nematic solid exhibits strong, sustained elastic stresses (see next section below) and high nematic order, which suggests a larger Frank elastic constant compared to the nematic liquid, resulting in a larger correlation length. Further investigations of the transition from the solid to the liquid and independent measures of the effective Frank constants in the two regimes are important directions for future research.

% \del{However, as discussed in Sec.~\ref{subsec:defects}, the correlation length in the plastic nematic solid exhibits a nonmonotonic dependence on the effective activity [Fig.~\ref{Figure4}(g)]. We hypothesize that near the onset of plastic tissue flow ( $\alpha\beta \lesssim 2.15$ ), where cell rearrangements occur infrequently, elastic stresses accumulate persistently and maintain long-range coherence. In our model, the spatial coupling of nematic order, which is the origin of Frank elasticity, arises from the alignment of cells to elastic stress. Thus the elastic energy cost of nematic distortions is not an independent material elastic parameter, but is intrinsically coupled to the strength of active stresses. Because elastic forces are inherently long-ranged, this coupling cancels the activity dependence of correlation length, leaving no intrinsic length scale governing defect separation [Fig.~\ref{FigureS2}]. % Near the transition from the plastic solid to the soft nematic solid, the effective activity vanishes, causing the correlation length to diverge toward the system size. As $\alpha\beta$ increases further, the activity begins to outweigh and decouple with the Frank elasticity, which leads to a decrease in the correlation length, reminiscent of the behavior in the nematic liquid. Nevertheless, because the nematic order is much stronger in the plastic solid than in the liquid, the enhanced Frank elasticity extends the correlation length in plastic tissue flows.}

\subsection{Mechanical origins of diverse tissue phases} 
\label{sec:phase_mechanics}

To elucidate the mechanical origin of these distinct dynamical regimes, we calculate the average tension acting along cell junctions
\(\langle T \rangle := {\left\langle T_{e} \right\rangle}_{\!e}\), 
where $T_e$ is given by Eq.~\eqref{EqEdgeTension} and the average is taken over edges $e$ [Fig.~\ref{Figure6}(d)].
In addition, we calculate the average magnitude of the deviatoric elastic stress over all cells as $\langle\lvert\tilde{\boldsymbol{\sigma}}_\mathrm{\!el}\rvert\rangle := {\left\langle\lvert \tilde{\boldsymbol{\sigma}}_{\!J}^{\mathrm{el}} \rvert\right\rangle}_{\!J}$, which quantifies macroscopic stresses on the tissue scale [Fig.~\ref{Figure6}(e)]. 
Importantly, the deviatoric elastic stress can vanish while $\langle T \rangle$ remains finite. This is because pressure, due to the bulk elasticity $\sim K_\mathrm{a}(A_J - A_0)$ can compensate the isotropic component of the tensional stress.

We find that \(\langle T \rangle\) decreases with increasing \(P_0\), reflecting a reduction in the effective junctional tension that leads to tissue softening. 
Above \(P_{\!0}^{\mathrm G}\), the junctional tension \(\langle T \rangle\) vanishes [Fig.~\ref{Figure6}(d)], implying that an extensive number of degrees of freedom becomes unconstrained.
It has previously been recognized that the vanishing tensions are responsible for the loss of rigidity above \(P_{\!0}^{\mathrm G}\) \cite{Yan2019a}.
The underconstrained nature of this state, together with the lack of velocity correlations, suggests that the regime \(P_0\!>\!P_{\!0}^{\mathrm G}\) should be called a nematic \emph{gas}.

The map of $\langle\lvert\tilde{\boldsymbol{\sigma}}_\mathrm{\!el}\rvert\rangle$ shows that the nematic solid exhibits high macroscopic elastic stress [Fig.~\ref{Figure6}(e)], consistent with the high nematic order in Fig.~\ref{Figure6}(a).
These stresses facilitate long-range  nematic order and are therefore responsible for the long-range correlation of plastic tissue flows and the lack of an ``active length'' in the plastic solid. 
Indeed, collective, system-scale actuation is a common feature of active solids \cite{Berezney2026,Baconnier2022}.
In contrast to these previously studied systems, the plastic nematic solid dynamically remodels while sustaining active stresses. This remodeling appears intermittent, with rapid avalanche-like bursts of cell rearrangements, which are accompanied by reorientation of nematic order and fast motion of topological defects [see Videos~S1 and S3 in Supplemental Material]. Such avalanches are reminiscent of intermittent plastic events observed in the vertex models subjected to \emph{external shear} \cite{Amiri2023,Nguyen2025}. The statistical properties of internally driven avalanches in the plastic nematic solid and their role in facilitating long-range correlated flows will be further studied in a forthcoming manuscript.

In the nematic liquid (\(P_0^{\mathrm{L}}\!<\!P_0\!<\!P_{\!0}^{\mathrm G}\)), the elastic stress $\langle\lvert\tilde{\boldsymbol{\sigma}}_\mathrm{\!el}\rvert\rangle$ nearly vanishes despite the non-vanishing junctional tensions $\langle T \rangle$ [Figs.~\ref{Figure6}(d,e)].
Here, active stresses drive fluidization via continual self-yielding, so that rearrangements dissipate elastic stress faster than it builds up.
Since this active-stress-mediated fluidization necessitates nematic order, the transition from isotropic solid to nematic liquid coincides with the onset of nematic order along the line $\alpha\beta = 1$.
Finally, the nematic liquid is distinct from the nematic gas, where \emph{both} macroscopic elastic stress and junctional tensions vanish.
The isotropic solid, nematic liquid, and gas phases meet at a triple point at $(P_0, \alpha\beta) = (3.81, 1)$, marked by a magenta triangle in Figs.~\ref{Figure6}(a--e).

Together, our results demonstrate that when effective activity exceeds a threshold ($\alpha\beta\!>\!1$), increasing $P_0$ drives the tissue through a sequence of transitions: from a nematic solid (soft, or plastic, or both in sequence) to a nematic liquid, and finally to a nematic gas [see the representative case at $\alpha\beta\!=\!1.875$ in Fig.~\ref{Figure6}(g)]. Crucially, whether macroscopic elastic stress and microscopic junctional tension are maintained or vanish governs these emergent behaviors and distinguishes each regime [see Table~\ref{table}]. This multi-stage transition framework unifies previously disconnected observations of tissue phenomena and elucidates their distinctive rheological features as well as underlying mechanical origins.

\section{Conclusion and discussion}
\label{sec:conclusion}

%\fb{We could cite this paper https://www.nature.com/articles/s41467-022-28151-9 for mechanical phase transitions in tissues.}

In the introduction, we raised two questions central to the mechanics underlying tissue morphogenesis. First, how is force generation coordinated between cells across large tissues? Second, how do tissues change shape, i.e.\ flow, while resisting external forces and perturbations.
Our theory and simulations show that a simple and experimentally motivated mechanical feedback loop—aligning active stress generation with the axis of cell stretching—allows cells to self-organize into a state with large-scale flows while sustaining internal elastic stresses.
This remarkable state, which we call an (active) plastic nematic solid, is ideally suited to facilitate morphogenesis and provides a simple answer to the two questions above. 
Indeed, experiments have demonstrated a key role of mechanical feedback and nematic order for morphogenesis in many different organisms \cite{Maroudas-Sacks2021,Zhang2021b,Guillamat2022,Ravichandran2025,Li2025a}.
The exotic rheological features of the plastic nematic solid we theoretically predict are related to the recently proposed concepts of active plastic flow \cite{Claussen2024,Claussen2026} and ``fluid under tension'' \cite{Grigas2025}.
All three share the key property of maintained microscopic tensions in flowing tissue, and all three rely on mechanical feedback loops. Together, these theoretical developments hint at a fundamentally new understanding of flow and rearrangements in tissue morphogenesis.

The key parameters in our model are the activity and the cell deformability. In the phase diagram spanned by these parameters, the plastic nematic solid phase lies in-between a (soft) solid and an active nematic liquid. 
The soft solid exhibits key properties of nematic elastomers, including a vanishing elastic modulus up to a critical strain, followed by a strain-stiffening response \cite{Golubovic1989,Warner1994}.
The nematic liquid exhibits the hallmarks of active nematic turbulence—spatiotemporally chaotic flows with a correlation length that scales with the inverse square root of the activity \cite{Alert2020a}.
Taken together, our work provides a unifying theoretical framework for understanding tissues with nematic order, bridging solid and fluid regimes, with intermediate phases that go beyond conventional rheological categories \cite{Lenne2022}.
Importantly, our model allows the study of the transitions between these different phases, which have previously been studied individually and in separate physical systems. This is a promising direction for future research.
The transition from the active nematic liquid to the plastic nematic solid is particularly interesting due to their dramatically different velocity correlation lengths. The long-ranged correlations near the self-yielding transition suggest that the plastic nematic solid may constitute an extended critical phase, akin to flocks and swarms \cite{Gonzalez-Albaladejo2024} or the hexatic phase in 2D melting \cite{Nelson1979}.
The intermittent dynamics observed in the plastic nematic solid further suggest a possible connection to scale-free avalanches in sheared granular media and self-organized criticality. This hypothesis will be addressed in a forthcoming manuscript.

How could the phases described above be identified in experiments?
In recent years, the \emph{observed} shape index $\langle P \rangle$ [see Fig.~\ref{FigureS5}(b)] has been widely adopted to characterize tissue mechanics \cite{Wang2020a,Lin2025}. Based on previous models, a high shape index has been associated with the regime that we call gas here. In this regime junctional tensions vanish, at odds with the experimental observation of taut (rather than wrinkled) junctions that recoil after laser ablation, indicating that they are under tension \cite{Mitchel2020,Piscitello-Gomez2023,Rigato2024}.
Our model offers an alternative explanation for the high shape index: tissue deformation driven by active stresses that induce plasticity deep in the solid regime \cite{Brauns2024a,Tahaei2025,Shen2025}.
Notably, the transition from the arrested plastic nematic solid to the flowing liquid state is marked by a reduction in $\langle P \rangle$, at odds with the common conception that a higher shape index marks a more fluid tissue state.
In summary, the observed shape index alone is not sufficient to distinguish these regimes. More detailed quantifications, such as spatial correlations of cell elongation \cite{Zhang2023} and cell velocities, are needed to distinguish these mechanical regimes.
Ultimately what's needed are experimental measurements of stress and rheology both on the cell and the tissue scale~\cite{Bambardekar2015,Shivakumar2016,Petridou2019,Mongera2018,Dye2021,Michaut2025}.
Specifically, junctional tensions and tissue-scale elastic stresses can be measured via laser cutting and other methods at the respective scales. Such measurements might allow one to distinguish between the active plastic nematic solid, where both tissue-scale stresses and junctional tension are finite, from the active nematic liquid where tissue-scale shear stress vanishes.

Two distinct scenarios of anisotropic active forces in epithelial tissues have been proposed previously: \emph{bulk stress} vs \emph{junctional tension} \cite{Duclut2022a}. Their mechanisms differ in how they drive cell shape changes and rearrangements. If the total tissue strain is constrained, anisotropic junctional tensions promote active T1 transitions that generate tension cables across adjacent interfaces, driving collective cell elongation along the axis of high junctional tensions \cite{Sknepnek2023,Brauns2024a,Claussen2024}. Thus, contractile junctional forces can appear extensile from the perspective of cell shape change, as seen in closed curved tissues such as protrusions in \emph{Hydra} ectoderm \cite{Aufschnaiter2017} and convergent extension in \emph{Xenopus} mesoderm \cite{Weng2022}.
By contrast, anisotropic bulk contractile stress mainly drives cell elongation perpendicularly to the nematic axis \cite{Duclut2022a}. 
In our study we focused on extensile bulk stress aligned with local elastic deformation. If the stress is contractile along the cell long axis (\(\alpha\beta\!<\!0\)), it would suppress elongation. Since \emph{in vivo} actomyosin fibers are predominantly contractile \cite{Weng2022,Brauns2024a,Bailles2025}, one could reformulate the model by replacing bulk stress with junctional tension \cite{Sknepnek2023,Rozman2023}, or by introducing an intrinsic energy term for cell elongation \cite{Lin2023}. 
The effective rheology of these active stress modalities remains to be studied.

Throughout the present study we assumed that the timescales of mechanical relaxation and nematic ordering are comparable. In tissues, relaxation to quasi-static force balance may be much faster than reorganization of the cytoskeleton. Systematically exploring the role of these different timescales remains an important direction for future research.
A related question is the role of internal (i.e.\ viscous) dissipation vs substrate friction. 
Recent theory work has shown that dominant viscous dissipation can facilitate correlated flows in the ``gas'' phase ($P_0 > 3.81$) of a vertex model where active nematic stress is directly linked to cell shape \cite{Rozman2025}. This internal dissipation promotes local coordination between neighboring cells, playing a role analogous to the stress-driven nematic alignment in our model. Combining viscous dissipation and high active stress also induces correlated flows and a nematic liquid phase at $P_0 < 3.81$ \cite{Rozman2025,Lin2026}. Recent experiments have also observed the correlated fluidization induced by intercellular friction in epithelial tissues \cite{Bera2026}. It remains an open question whether a ``plastic nematic solid'' phase exists in such ``wet'' formulations, and how the interplay between internal dissipation and mechanical feedback would further shape the tissue mechanics and dynamics.

Another simplification in our analysis is the choice $\alpha=\beta$ adopted in most of our simulations. 
In our minimal model, the active stress is linearly slaved to the nematic order through $\alpha$, while the mechanical feedback to local stress is linear in $\beta$. Their product $\alpha\beta$ therefore acts as an effective activity controlling the isotropic-to-nematic transition. Varying $\alpha$ and $\beta$ independently in simulations [Fig.~\ref{FigureS1}(b)] confirms that this transition occurs at the mean-field bifurcation point $\alpha\beta=-m$. 
Beyond the onset of nematic order, however, the two parameters play different roles: the soft-to-plastic transition depends more strongly on $\alpha$ than on $\beta$ [Figs.~\ref{FigureS1}(b,c)]. This likely reflects the need for sufficiently strong active stresses to drive plastic rearrangements. Thus, the distinct roles of $\alpha$ and $\beta$ in driving the active plastic-solid regime remain to be further investigated.
More generally, the reduction to a single effective activity $\alpha\beta$ breaks down if the active stress depends nonlinearly on $\mathbf Q$, or if nematic reorientation responds to mechanical stress with a finite delay. How such more general feedback laws modify tissue phase behaviors remains to be explored.

Our detailed quantification of kinematics in a minimal setting provides a foundation for future studies on more complex scenarios.
First, in developing tissues, reaction--diffusion of biochemical factors (morphogens) has been found to couple with cell deformation through mechanochemical feedback \cite{Howard2011,Bailles2022,Maroudas-Sacks2025}. How to theoretically describe the self-organized mechanochemical pattern is an important direction for future exploration. Second, real tissues reside in complex 3D architectures, where curvature, topology, and boundary constraints may influence the alignment of stress fibers \cite{Sussman2020,Wang2023a,Luciano2024,Ravichandran2025}, necessitating the development of fully 3D deformable cell models \cite{Torres-Sanchez2022,Yu2024,Runser2024,Yu2025}.

Finally, in our current model active stress and passive cell deformability are treated as independent parameters. In reality, however, the cytoskeletal networks that generate active stress also determine the mechanical properties of cells \cite{Salbreux2012,Mulla2019}. Turnover of cytoskeletal networks on the timescale of minutes means that elastic stresses rapidly relax and must be maintained by the activity of molecular motors. Thus, a sharp distinction of active and passive stresses on the cell scale is not possible, calling for new approaches to tissue mechanics, such as models treating all junctional tensions as active \cite{Kim2021,Noll2017,Claussen2024,Brauns2024a,Claussen2026}.
Ultimately, experimental quantification of tissue rheology across scales and further investigation of the feedback loops controlling active stress generation will be required to quantitatively understand dynamical tissue remodeling.

\begin{acknowledgments}
We thank the anonymous referees for their constructive feedback which helped to significantly improve the manuscript.
P.Y. acknowledges support from Prof. Bo Li and the Tsinghua Scholarship for Overseas Graduate Studies, and is grateful for the hospitality of the UCSB Department of Physics. F.B.\ acknowledges support by the Gordon and Betty Moore Foundation post-doctoral fellowship (grant \#2919). M.C.M. was supported by the National Science Foundation award DMR-2528734.
\end{acknowledgments}

\appendix

\section{\label{Appendix_EquilAnalysis}Equilibrium theoretical analysis}

\begin{figure}[htb]
    \centering
    \includegraphics[width=0.48 \textwidth]{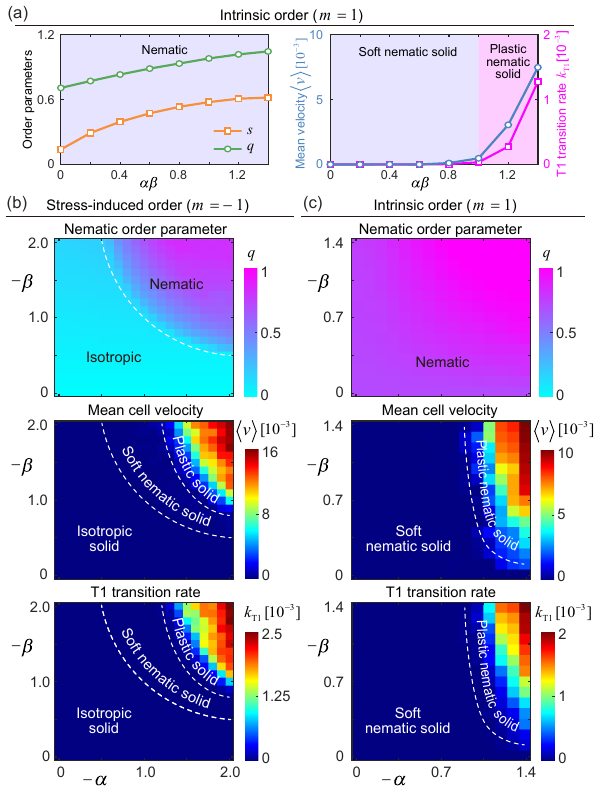}
    \caption{(a)~Nematic order parameter \(q\), mean cell shape anisotropy \(s\), mean cell velocity \(\langle v \rangle\), and T1 transition rate \(k_{\rm T1}\) versus \(\alpha\beta\), with $\alpha=\beta$, for the intrinsic order (\(m\!=\!1\)).
    (b,c)~Phase diagrams of \(q\), \(\langle v \rangle\), and \(k_{\rm T1}\) upon varying \(\alpha\) and \(\beta\), for (b) stress-induced order (\(m\!=\!-1\)) and (c) intrinsic order (\(m\!=\!1\)). All results are for $P_0=-0.5$ and $N=1000$ cells.}
    \label{FigureS1}
\end{figure}

The cell elastic stress induced by the mechanical energy defined by Eq.~(\ref{EqElasticStress_E}) is explicitly given by 
\begin{equation}
\begin{aligned}
        {\bm{\sigma }}_{\!J}^{\mathrm{el}} &= K_{\rm a}(A_J-A_0) \,\mathbf{I} \\
        &\quad + \frac{1}{2A_J}\sum\limits_e K_\mathrm{p}(P_{\!J}+P_I-2 P_0) \,\frac{{\boldsymbol{\ell}_e} \otimes {\boldsymbol{\ell}_e}}{{{{\ell}_e}}},
\end{aligned} 
\label{EqElasticStress}
\end{equation}
where \(P_{\!J}\) and \(P_I\) are the perimeters of cells \(J\) and \(I\) that share the edge \(e\). The sum is over all edges of cell \(J\). 
The total stress is
\begin{equation}
   {\bm{\sigma }}_{\!J} = {\bm{\sigma }}_{\!J}^{\mathrm{el}} + {\bm{\sigma }}_{\!J}^{{\rm{act}}} = {\bm{\sigma }}_{\!J}^{\mathrm{el}}+\alpha {\bf{Q}}_J.
\label{EqTotalStress2}
\end{equation}
We analyze the equilibrium state of the cell collective in the presence of active stress. Considering a uniform affine deformation of the hexagonal pattern [Fig.~\ref{Figure2}(b)], the active cell elongation along the \(x\)-axis and \(y\)-axis is described by the stretches of the vertices as \({\bf{r}}_i = (x_i,y_i) \xrightarrow{\lambda_x,\lambda_y} (\lambda_x x_i,\lambda_y y_i)\), where \(\lambda_x\) and \(\lambda_y\) are the stretches along the \(x\)- and \(y\)-axes, respectively. Under this uniform deformation, the nematic order parameter is also uniform and given by \(\mathbf{Q}_J = q/\sqrt{2} \, \text{diag}(1, -1)\), where positive \(q\) %\fb{see comment in the main text regarding the choice of ansatz here.}\py{Have revised the relevant text here.} 
corresponds to cell elongation along the \(x\)-axis. The total cell stress can be expressed as 
\begin{equation}
\begin{aligned}
        {\bm{\sigma }}_{\!J}  = & \left [ K_{\rm a}(A_J-A_0)+ \frac{K_\mathrm{p}P_{\!J}(P_{\!J}-P_0)}{2A_J} \right ]{\bf{I}}  \\
        & + \frac{K_\mathrm{p}P_{\!J}(P_{\!J}-P_0)}{A_J} {{\bf{S}}_J} +\alpha {\bf{Q}}_J,
\end{aligned} 
\label{EqStressUnderDeform}
\end{equation}
where \({{\bf{S}}_J} = (\frac{{{\lambda _x}}}{{\sqrt {\lambda _x^2 + 3\lambda _y^2} }} - \frac{1}{2}) \, \text{diag}(1, -1)\) is the cell shape anisotropy tensor under the affine deformation. Substituting the deviatoric elastic stress from Eq.~(\ref{EqStressUnderDeform}) into Eq.~(\ref{EqQJ}), we obtain the dynamics of \({\bf{Q}}_J\): 
\begin{equation}
    \tau_q  \frac{d}{dt}{\mathbf{Q}}_J = \left[ {m - 2{\rm{Tr}}({\bf{Q}}_J^2)} \right]{{\bf{Q}}_J} - \beta \frac{{K_\mathrm{p}{P_{\!J}}({P_{\!J}}-P_0)}}{{{A_J}}}{{\bf{S}}_J}.
\label{EqQJUnderDeform}
\end{equation}
At the uniform equilibrium state with free boundaries, stress must vanish, \({\bm{\sigma }}_{\!J}=\bf0\), and \(\tfrac{d}{dt}\mathbf{Q}_J=\mathbf{0}\), which yields the equation
\begin{equation}
    2{q^2} = \alpha \beta  + m.
    \label{EqCouplingControl}
\end{equation}
The tissue undergoes a pitchfork bifurcation as a function of the effective activity \(\alpha\beta\). The tissue is isotropic for \(\alpha\beta < -m\) and undergoes a transition to a nematic state when \(\alpha\beta > -m\). 
The mean shape anisotropy of the tissue is quantified by the scalar \(s\!=\!{\langle \sqrt{{2{\rm{Tr(}}{\bf{S}}_J^2{\rm{)}}}} \rangle _J} \! \in\! \left[ {0,1} \right)\). One can obtain explicit expressions for both the nematic order parameter $q$ and the mean shape anisotropy $s$. These are shown in Figs.~\ref{Figure2}(c,d) and Fig.~\ref{FigureS1}(a) for the parameters used in the simulations.

\begin{figure}[htb]
    \centering
    \includegraphics[width=0.48\textwidth]{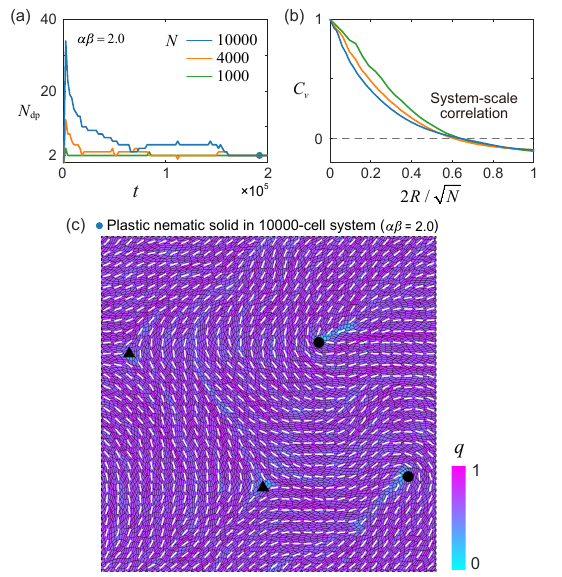}
    \caption{
    (a)~Evolution of the number of \(\pm 1/2\) defect pairs \(N_{\rm{dp}}\) over time and (b) velocity correlation function $C_v$ in the plastic nematic solid regime (\(\alpha\beta\!=\!2\)) for different system sizes (cell number \(N=1000, 4000, 10000\)). 
    (c)~Snapshots of the nematic order (white lines) for $N = 10000$ cells, where two defect pairs exist.
    All results are for $\alpha=\beta$, $P_0=-0.5$ and $m=-1$.
    }
    \label{FigureS2}
\end{figure}

\section{Range of nematic alignment to elastic stress}
\label{Appendix_AlignmentEffect}

Given that supracellular actomyosin networks are typically interwoven across neighboring cells, we have implemented a mechanism that allows  fibers to sense and respond to the local average stress. Here we investigate the effect of different coarse-graining radii for the local stress alignment. The averaged elastic shear stress in the alignment term of Eq.~(\ref{EqQJ}) is calculated over the spatial averaging disk with radius $nR_0$, where $R_0\!=\! 1.075$ corresponds to the distance between cell centroids in a hexagonal pattern. 

\begin{figure}[htb]
    \centering
    \includegraphics[width=0.48\textwidth]{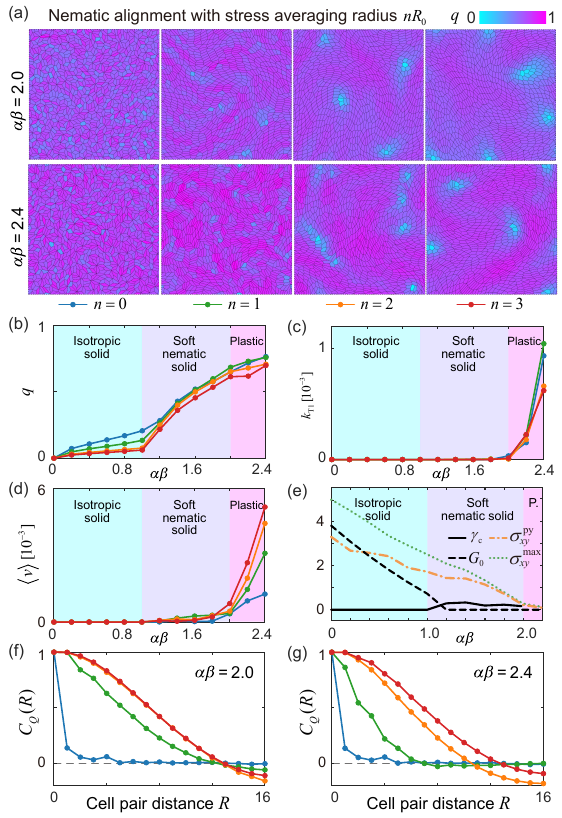}
    \caption{Influence of stress averaging radius on nematic alignment.
    (a)~Representative snapshots for different averaging radii $nR_0$ at $\alpha\beta=2$ and $\alpha\beta=2.4$.
    (b--d)~Evolution of (b)~nematic order parameter $q$, (c)~T1 transition rate $k_{\rm T1}$, and (d)~mean cell velocity $\langle v \rangle$ as a function of $\alpha\beta$ for varying $n$.
    (e)~Evolution of \(G_0\), \(\gamma_c\), \(\tilde{\sigma}_{\!\mathrm{max}}\), and \(\tilde{\sigma}_{\!\mathrm{py}}\) versus \(\alpha\beta\) for the case of single-cell stress alignment ($n=0$) under externally applied shear deformation. 
    (f,g)~Nematic correlation function $C_Q$ versus cell pair distance $R$ for different $n$ at (f) $\alpha\beta=2$ and (g) $\alpha\beta=2.4$. % \py{Velocity correlation at higher activity should be done, which shows higher averaging radius could promote the correlation at high activity.}
    All results are for $\alpha=\beta$, $P_0 = -0.5$, $m = -1$, and $N=1000$ cells.
   % \fb{Sorry to keep complaining, but could you swap orange and green. Then the colors will follow a more natural sequence. Thank you!}    
    %\py{Thanks for your reminding! Addressed.}
    }
    \label{FigureS3}
\end{figure}

Simulation results show that when alignment operates solely at the single-cell level ($n=0$), the supracellular effects vanish and no long-range nematic order emerges [Fig.~\ref{FigureS3}(a)].
Conversely, for a finite interaction range ($n>0$), long-range correlated soft and plastic nematic solids emerge with increasing activity $\alpha\beta$. In the soft nematic solid regime ($1 < \alpha\beta < 2 $) and plastic solid near the yielding transition ($ \alpha\beta \rightarrow 2$), the system-scale correlations occur, regardless of the specific averaging radius [Figs.~\ref{FigureS3}(a,f)]. While deep within the plastic solid regime at higher activities (e.g., $\alpha\beta=2.4$), where the correlation length decreases, an increased averaging radius is found to enhance the correlations [Fig.~\ref{FigureS3}(g)]. In this regime, the characteristic length is $\sim 10$ at $n=2$, matching the length scale governed by the neighboring alignment mechanism found in Fig.~\ref{Figure4}(g). This verifies the importance of local mechanical feedback in mediating large-scale correlations.

In addition, we find that varying the alignment radii $n=1$--$3$ has only a minor effect on nematic order \(q\) and T1 transition rate \(k_{\rm{T1}}\) [Figs.~\ref{FigureS3}(b,c)]. The mean cell velocity \(\langle v \rangle\) increases with higher $n$ in the plastic solid phase [Fig.~\ref{FigureS3}(d)] as larger averaging radii smooth out local stress fluctuations and drive more coherent cell motion. These results suggest that our effective model is insensitive to the microscopic details of the local alignment mechanism.

%single-cell stress alignment strongly suppresses correlations in nematic order  As a result, the topological defects disappear and plastic tissue flow significantly slows down, with cell trajectories exhibiting localized oscillations rather than sustained motion [Figs.~\ref{FigureS3}(a,c)]. 
Furthermore, we also perform rheological measurements on the steady-state tissues for the $n=0$ case. The critical strain \(\gamma_c\) for the emergence of shear-induced rigidity becomes very small and independent of \(\alpha\beta\) in the soft nematic solid [Fig.~\ref{FigureS3}(e)]. This indicates that single-cell stress alignment is not sufficient to drive correlated nematic reorientation to accommodate externally applied strain.
These results highlight the critical role of supracellular mechanical response in inducing long-range nematic order and active rheological properties in solid epithelia.

\section{\label{Appendix_Shear}Shear deformation scheme}

\begin{figure*}[htb]
    \centering
    \includegraphics[width=0.75\textwidth]{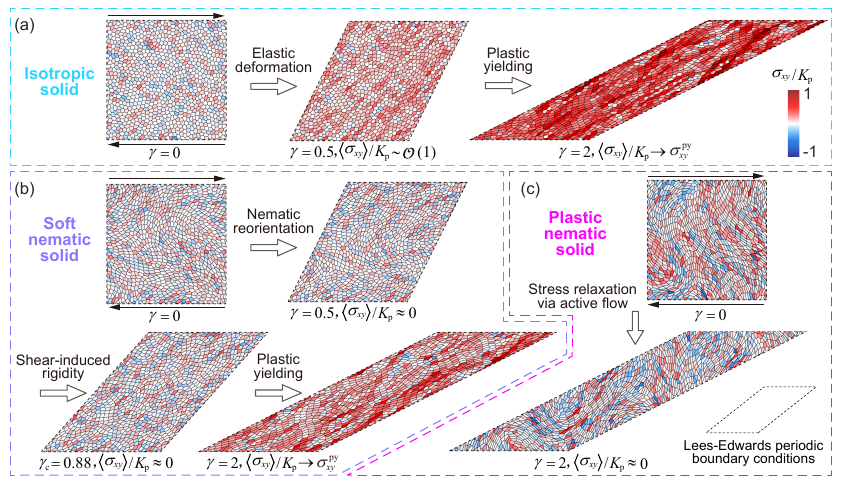}
    \caption{(a--c) Representative snapshots of cell shear stress field under external shear deformation at (a) \(\alpha\beta\!=\!0.6\), (b) \(\alpha\beta\!=\!1.4\), and (c) \(\alpha\beta\!=\!2.2\), which correspond to the snapshots of nematic order field in Figs.~\ref{Figure5}(c--e).
    }
    \label{FigureS4}
\end{figure*}

We perform simple shear deformation by quasi-statically increasing the strain $\gamma(t)$ in the dynamical steady state, using Lees--Edwards periodic boundary conditions. 
In the simulation, the shear strain is incremented by $\Delta\gamma$ at every time interval $\Delta t_\gamma$ as
\begin{equation}
    \gamma(t+\Delta t_\gamma) = \gamma(t) + \Delta\gamma,
\end{equation}
where the initial shear strain is zero and $\Delta\gamma$ is a small incremental strain. 
After updating $\gamma(t)$, each vertex coordinate $(x_i, y_i)$ is first mapped to the sheared position as
\begin{equation}
    \tilde{x}_i = x_i + \Delta\gamma\, y_i,\quad  \tilde{y}_i = y_i,
\end{equation}
and then wrapped back into the simulation box according to
\begin{equation}
\begin{aligned}
        x_i^{\mathrm{new}} &= \tilde{x}_i - L\, \xi_x - \gamma\, L\, \xi_y, \\
        y_i^{\mathrm{new}} &= \tilde{y}_i - L\, \xi_y ,
\end{aligned}
\end{equation}
where $L=\sqrt{N}$ is the initial box length. 
The Lees--Edwards periodic image indices $\xi_y$ and $\xi_x$ are calculated as
\begin{equation}
\begin{aligned}
\xi_x &=
\begin{cases}
0,  & \text{if } \left| (\tilde{x}_i - \gamma\, \tilde{y}_i) / L \right| \le \tfrac12, \\
+1, & \text{if } (\tilde{x}_i - \gamma\, \tilde{y}_i) / L > \tfrac12, \\
-1, & \text{if } (\tilde{x}_i - \gamma\, \tilde{y}_i) / L < -\tfrac12 ,
\end{cases} \\[0.5em]
\xi_y &=
\begin{cases}
0,  & \text{if } \left| \tilde{y}_i / L \right| \le \tfrac12, \\
+1, & \text{if } \tilde{y}_i / L > \tfrac12, \\
-1, & \text{if } \tilde{y}_i / L < -\tfrac12 ,
\end{cases}
\end{aligned}
\end{equation}
which specify the periodic image indices for shear mapping.
We set the time interval for shear increments to $\Delta t_\gamma = 100$, which has been verified to be sufficiently long for the system to relax and thus ensures a quasi-static shear response. The fields of cellular nematic order and shear stress are shown in Fig.~\ref{Figure5} and Fig.~\ref{FigureS4}, respectively. The mean shear stress of the tissue is calculated after each strain increment and relaxation. The data shown in Fig.~\ref{Figure5} and Fig.~\ref{FigureS3}(e) are averaged over three independent simulations.

\section{\label{Appendix_Calculation}Quantitative Measurements}

\subsection{Detection of topological defects}

To detect topological defects in the cellular nematic texture, we first construct a coarse-grained nematic field $\hat{\mathbf{Q}}(\mathbf{R})$ \cite{Lin2023} on a uniform \( n_x \times n_y \) spatial grid by Gaussian-weighted averaging of the full cellular nematic tensor $\mathbf{Q}_J$:
\begin{equation}
\hat{\mathbf{Q}}(\mathbf{R}) = 
\frac{ \displaystyle \sum_{|\mathbf{R} - \mathbf{R}_J| < R_{\text{cut}}} \kern-1.5em w(\mathbf{R} - \mathbf{R}_J) \, \mathbf{Q}_J }{ \displaystyle \sum_{|\mathbf{R} - \mathbf{R}_J| < R_{\text{cut}}} \kern-1.5em w(\mathbf{R} - \mathbf{R}_J) } ,
\end{equation}
where \( \mathbf{R}_J \) is the centroid of cell \( J \). The Gaussian weight function is 
\begin{equation}
w(\mathbf{R} - \mathbf{R}_J) = \frac{1}{\sqrt{2\pi}l_w} \exp\left( -\frac{|\mathbf{R} - \mathbf{R}_J|^2}{2l_w^2} \right),
\end{equation}
with kernel width \(l_w =0.8\) and cutoff radius \( R_{\text{cut}} = 3l_w \). Periodic boundary conditions are applied when computing distances.

We next identify possible defect cores directly from the original cellular nematic field before coarse-graining. A topological defect is associated with a vanishing nematic amplitude, where the director becomes undefined. For each cell, we therefore calculate its nematic order parameter \(q_J= \sqrt{{\rm{Tr(}}{\bf{Q}}_{\!J}^2{\rm{)}}}\), and identify cells with sufficiently small nematic order as candidate defect cores. Since $q_J$ need not reach exactly zero in a discrete numerical system, we identify low-order cells using a small cutoff
\begin{equation}
    q_J < q_{\mathrm{th}},
\end{equation}
with \( q_{\mathrm{th}}=0.2\), corresponding to a nearly isotropic state with very low nematic order in our simulations [cf. Fig.~\ref{Figure1}]. A spatial grid element is marked as a candidate defect region if it contains at least one such low-$q$ cell. Importantly, this screening is performed using the original cell-level $q_J$, rather than the coarse-grained amplitude of $\hat{\mathbf{Q}}$. This prevents a locally small nematic amplitude associated with a possible defect core from being obscured by averaging with surrounding cells of larger nematic order.

For each candidate defect region, we then calculate the winding number around each grid point based on their eight nearest neighbors \cite{Killeen2022}. At each point, the local nematic orientation angle is given by
\begin{equation}
\theta(\mathbf{R}) = \frac{1}{2} \arctan2\left(\hat{Q}_{\!xy}(\mathbf{R}), \hat{Q}_{xx}(\mathbf{R})\right),
\end{equation}
where \(\arctan2\) is the two-argument arctangent function. The total winding number at each grid point is given by
\begin{equation}
k = \frac{1}{2\pi}\sum_{n=1}^{8} \left( \theta_{n+1} - \theta_n + a \right),
\end{equation}
where the continuity correction \( a \) is defined as:
\begin{equation}
    a = 
\begin{cases}
0, & \text{if } |\theta_{n+1} - \theta_n| \leq \pi/2, \\
+\pi, & \text{if } \theta_{n+1} - \theta_n < -\pi/2, \\
-\pi, & \text{if } \theta_{n+1} - \theta_n > \pi/2.
\end{cases}
\end{equation}
Candidate regions satisfying \( |k - 0.5| < 0.05 \) are identified as candidate \( +1/2 \) defects, while those with \( |k + 0.5| < 0.05 \) are marked as candidate \( -1/2 \) defects. 

Finally, at finer coarse-graining resolutions, a single defect core may span several neighboring grid elements and therefore be detected more than once.
To avoid duplication, we apply a filtering step: among pairs of nearby defects of the same type (within 1.5 grid spacings), only the one whose topological charge is closer to the value \( \pm 1/2 \) is retained.

For the coarse-grained nematic fields in 1000-cell and 4000-cell systems shown in Figs.~\ref{Figure4}(a) and \ref{Figure4}(d), respectively, we take \(n_{x,y}=25\). For the 10000-cell system in Fig.~\ref{FigureS2}(c), we take \(n_{x,y}=30\). These choices correspond to grid spacings of approximately $1.3$--$3.3$ in units of $\sqrt{A_0}$ in both the $x$ and $y$ directions. 
In the nematic solid regimes where topological defects are discussed, the cellular nematic order is strong, exhibits long-range correlations, and varies smoothly across the tissue. We have verified that the identified defect patterns and their qualitative statistics are robust to grid resolution and small angular perturbations, provided that the grid spacing remains larger than the single-cell scale and much smaller than the system size.

\begin{figure*}[htb]
    \centering
    \includegraphics[width=0.75\textwidth]{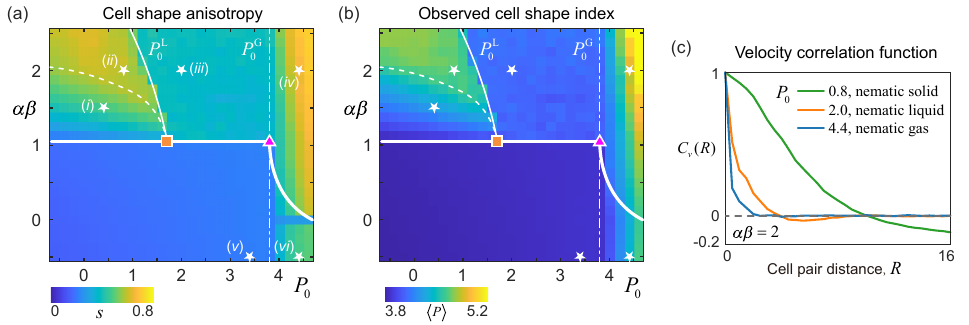}
    \caption{(a,b)~Phase diagrams of (a) cell shape anisotropy \(s\) and (b) observed cell shape index \(\langle P\rangle\!=\!{\left\langle { P_{\!J}}\right\rangle_J}\), depending on \(\alpha\beta\) and \(P_0\). In the isotropic regime ($\alpha\beta < 1$), the \emph{observed} shape index $\langle P\rangle$ indicates the solid-to-fluid transition at the critical \emph{target} shape index $P_0^{\mathrm G}$ \cite{Bi2015}. By contrast, in the active nematic regime ($\alpha\beta > 1$), $\langle P\rangle$ is not a reliable indicator of a fluid-vs-solid state of the tissue: for the nematic liquid, the shape index is \emph{lower} than for the nematic solid. The star labels correspond to the snapshots in Fig.~\ref{Figure1}(c).
    (c)~Velocity correlation function for different tissue states obtained by varying $P_0$ at fixed \(\alpha\beta\!=\!2.0\). All results are for $\alpha=\beta$, $m=-1$ and $N=1000$ cells.
    }
    \label{FigureS5}
\end{figure*}

\subsection{Cell velocity}
The mean cell velocity, which is used to describe the tissue dynamical property, is defined as
\begin{equation}
\langle v \rangle = \left\langle \frac{|\mathbf{R}_J(t_0 + \tau) - \mathbf{R}_J(t_0)|}{\tau} \right\rangle_{J,t_0}
\end{equation}
where \( \mathbf{R}_J(t) \) denotes the centroid position of cell \( J \) at time \( t \), and \(\tau\) is the observation time interval. The average is taken over all cells and 10 independent time points \( t_0 \) in the steady state.

The coarse-grained velocity field shown in Fig.~\ref{Figure1}(c) and Fig.~\ref{Figure4}(a) is similarly obtained by Gaussian-weighted averaging on the uniform \( n_x \times n_y \) spatial grid as
\begin{equation}
\hat{\mathbf{v}}(\mathbf{R}) = \frac{\displaystyle\sum_{|\mathbf{R} - \mathbf{R}_J| < R_{\text{cut}}} \kern-1.5em w(\mathbf{R} - \mathbf{R}_J) \, \mathbf{v}_{\!J}}{\displaystyle \sum_{|\mathbf{R} - \mathbf{R}_J| < R_{\text{cut}}} \kern-1.5em w(\mathbf{R} - \mathbf{R}_J) },
\end{equation}
where \(\mathbf{v}_{\!J} \! =\! \mathrm{d}\mathbf{R}_J / \mathrm{d}t\) is the instantaneous velocity of cell~\( J \).

\subsection{T1 transition rate}
To quantify the dynamics of cell rearrangements, we define the T1 transition rate \cite{Lin2023} following
\begin{equation}
k_{\mathrm{T1}} = \frac{N_{\mathrm{T1}}}{\tau N},
\end{equation}
where \( N_{\mathrm{T1}} \) is the number of T1 transitions that occur during the observation interval \(\tau\), and \( N \) is the total number of cells. The measurement is performed after the system reaches a dynamical steady state.

The T1 transition rate has been quantified in previous studies to characterize the rate of cell rearrangements and tissue fluidization \cite{Lin2023}. In our numerical implementation, a T1 transition is performed only when the edge length falls below the T1 cutoff $\Delta \ell_{\rm T1}$ and the resulting rearrangement lowers the total energy of the system. As a consequence, the newly formed edge tends to further extend rather than revert, and consecutive opposite T1 transitions do not occur. 
Since cell deformation is bounded in vertex models, sustained tissue flow can only occur through cumulative T1 transitions. Accordingly, in our simulations the T1 transition rate closely follows the evolution of the mean cell velocity, supporting its validity as an indicator of tissue flow.

\subsection{Correlation functions}
To characterize the spatial correlation of cellular orientation, we compute the nematic correlation function as
\begin{equation}
C_Q(R) = \frac{\langle \mathbf{Q}_I : \mathbf{Q}_J \rangle_{I,J}}{\langle \mathbf{Q}_J : \mathbf{Q}_J \rangle_J}, \ R - \Delta R < |\mathbf{R}_I - \mathbf{R}_J| \leq R,
\end{equation}
where the average is taken over all cell pairs \( (I, J) \) whose centroid-to-centroid distance falls within the bin \( (R - \Delta R, R] \). We set the bin width as \( \Delta R = 1 \).

Similarly, the velocity correlation function is defined as
\begin{equation}
C_v(R) = \frac{\langle \mathbf{v}_I \cdot \mathbf{v}_J \rangle_{I,J}}{\langle \mathbf{v}_J \cdot \mathbf{v}_J \rangle_J}, \ R - \Delta R < |\mathbf{R}_I - \mathbf{R}_J| \leq R,
\end{equation}
where \( \mathbf{v}_J \) is the velocity of cell \( J \) in the steady state. In Fig.~\ref{Figure6}(c), the mean spatial velocity correlation \( \langle{C}_v \rangle= \langle C_v(R) \rangle_{2 \leq R \leq 4} \) is computed over a neighboring region $R\in[2,4]$, corresponding to distances of approximately two to four effective cell diameters from the reference cell. For states with negligible cell motion (\( \langle v \rangle < 5\times10^{-4} \)), the correlation \( \langle{C}_v \rangle\) is set to 0.

\begin{figure}[htb]
    \centering
    \includegraphics[width=0.48 \textwidth]{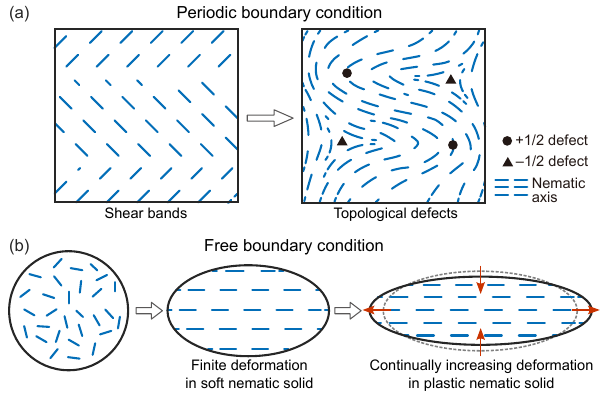}
    \caption{Sketches of the nematic texture and elastic deformation with (a) periodic and (b) free boundary conditions. With periodic boundary conditions, a globally uniform shear displacement field is incompatible with periodicity, so global strain compatibility enforces the formation of opposite shear bands, which gives rise to two pairs of topological defects. With free boundary conditions, the tissue undergoes global shape deformation. A soft nematic solid exhibits finite elongation, while a plastic nematic solid allows continually increasing deformation.}
    \label{FigureS6}
\end{figure}

\nocite{*}

%\bibliography{Reference.bib}

\begin{thebibliography}{121}%
\makeatletter
\providecommand \@ifxundefined [1]{%
 \@ifx{#1\undefined}
}%
\providecommand \@ifnum [1]{%
 \ifnum #1\expandafter \@firstoftwo
 \else \expandafter \@secondoftwo
 \fi
}%
\providecommand \@ifx [1]{%
 \ifx #1\expandafter \@firstoftwo
 \else \expandafter \@secondoftwo
 \fi
}%
\providecommand \natexlab [1]{#1}%
\providecommand \enquote  [1]{``#1''}%
\providecommand \bibnamefont  [1]{#1}%
\providecommand \bibfnamefont [1]{#1}%
\providecommand \citenamefont [1]{#1}%
\providecommand \href@noop [0]{\@secondoftwo}%
\providecommand \href [0]{\begingroup \@sanitize@url \@href}%
\providecommand \@href[1]{\@@startlink{#1}\@@href}%
\providecommand \@@href[1]{\endgroup#1\@@endlink}%
\providecommand \@sanitize@url [0]{\catcode `\\12\catcode `\$12\catcode `\&12\catcode `\#12\catcode `\^12\catcode `\_12\catcode `\%12\relax}%
\providecommand \@@startlink[1]{}%
\providecommand \@@endlink[0]{}%
\providecommand \url  [0]{\begingroup\@sanitize@url \@url }%
\providecommand \@url [1]{\endgroup\@href {#1}{\urlprefix }}%
\providecommand \urlprefix  [0]{URL }%
\providecommand \Eprint [0]{\href }%
\providecommand \doibase [0]{https://doi.org/}%
\providecommand \selectlanguage [0]{\@gobble}%
\providecommand \bibinfo  [0]{\@secondoftwo}%
\providecommand \bibfield  [0]{\@secondoftwo}%
\providecommand \translation [1]{[#1]}%
\providecommand \BibitemOpen [0]{}%
\providecommand \bibitemStop [0]{}%
\providecommand \bibitemNoStop [0]{.\EOS\space}%
\providecommand \EOS [0]{\spacefactor3000\relax}%
\providecommand \BibitemShut  [1]{\csname bibitem#1\endcsname}%
\let\auto@bib@innerbib\@empty
%</preamble>
\bibitem [{\citenamefont {Petridou}\ and\ \citenamefont {Heisenberg}(2019)}]{Petridou2019}%
  \BibitemOpen
  \bibfield  {author} {\bibinfo {author} {\bibfnamefont {N.~I.}\ \bibnamefont {Petridou}}\ and\ \bibinfo {author} {\bibfnamefont {C.-P.}\ \bibnamefont {Heisenberg}},\ }\bibfield  {title} {\bibinfo {title} {Tissue rheology in embryonic organization},\ }\href {https://doi.org/10.15252/embj.2019102497} {\bibfield  {journal} {\bibinfo  {journal} {EMBO J.}\ }\textbf {\bibinfo {volume} {38}},\ \bibinfo {pages} {e102497} (\bibinfo {year} {2019})}\BibitemShut {NoStop}%
\bibitem [{\citenamefont {Lenne}\ and\ \citenamefont {Trivedi}(2022)}]{Lenne2022}%
  \BibitemOpen
  \bibfield  {author} {\bibinfo {author} {\bibfnamefont {P.-F.}\ \bibnamefont {Lenne}}\ and\ \bibinfo {author} {\bibfnamefont {V.}~\bibnamefont {Trivedi}},\ }\bibfield  {title} {\bibinfo {title} {Sculpting tissues by phase transitions},\ }\href {https://doi.org/10.1038/s41467-022-28151-9} {\bibfield  {journal} {\bibinfo  {journal} {Nat. Commun.}\ }\textbf {\bibinfo {volume} {13}},\ \bibinfo {pages} {664} (\bibinfo {year} {2022})}\BibitemShut {NoStop}%
\bibitem [{\citenamefont {Blanchard}\ \emph {et~al.}(2018)\citenamefont {Blanchard}, \citenamefont {{\'E}tienne},\ and\ \citenamefont {Gorfinkiel}}]{Blanchard2018}%
  \BibitemOpen
  \bibfield  {author} {\bibinfo {author} {\bibfnamefont {G.~B.}\ \bibnamefont {Blanchard}}, \bibinfo {author} {\bibfnamefont {J.}~\bibnamefont {{\'E}tienne}},\ and\ \bibinfo {author} {\bibfnamefont {N.}~\bibnamefont {Gorfinkiel}},\ }\bibfield  {title} {\bibinfo {title} {From pulsatile apicomedial contractility to effective epithelial mechanics},\ }\href {https://doi.org/10.1016/j.gde.2018.07.004} {\bibfield  {journal} {\bibinfo  {journal} {Curr. Opin. Genet. Dev.}\ }\textbf {\bibinfo {volume} {51}},\ \bibinfo {pages} {78} (\bibinfo {year} {2018})}\BibitemShut {NoStop}%
\bibitem [{\citenamefont {Mongera}\ \emph {et~al.}(2018)\citenamefont {Mongera}, \citenamefont {Rowghanian}, \citenamefont {Gustafson}, \citenamefont {Shelton}, \citenamefont {Kealhofer}, \citenamefont {Carn}, \citenamefont {Serwane}, \citenamefont {Lucio}, \citenamefont {Giammona},\ and\ \citenamefont {Camp{\`a}s}}]{Mongera2018}%
  \BibitemOpen
  \bibfield  {author} {\bibinfo {author} {\bibfnamefont {A.}~\bibnamefont {Mongera}}, \bibinfo {author} {\bibfnamefont {P.}~\bibnamefont {Rowghanian}}, \bibinfo {author} {\bibfnamefont {H.~J.}\ \bibnamefont {Gustafson}}, \bibinfo {author} {\bibfnamefont {E.}~\bibnamefont {Shelton}}, \bibinfo {author} {\bibfnamefont {D.~A.}\ \bibnamefont {Kealhofer}}, \bibinfo {author} {\bibfnamefont {E.~K.}\ \bibnamefont {Carn}}, \bibinfo {author} {\bibfnamefont {F.}~\bibnamefont {Serwane}}, \bibinfo {author} {\bibfnamefont {A.~A.}\ \bibnamefont {Lucio}}, \bibinfo {author} {\bibfnamefont {J.}~\bibnamefont {Giammona}},\ and\ \bibinfo {author} {\bibfnamefont {O.}~\bibnamefont {Camp{\`a}s}},\ }\bibfield  {title} {\bibinfo {title} {A fluid-to-solid jamming transition underlies vertebrate body axis elongation},\ }\href {https://doi.org/10.1038/s41586-018-0479-2} {\bibfield  {journal} {\bibinfo  {journal} {Nature}\ }\textbf {\bibinfo {volume} {561}},\ \bibinfo {pages} {401} (\bibinfo {year} {2018})}\BibitemShut {NoStop}%
\bibitem [{\citenamefont {Mitchel}\ \emph {et~al.}(2020)\citenamefont {Mitchel}, \citenamefont {Das}, \citenamefont {O'Sullivan}, \citenamefont {Stancil}, \citenamefont {DeCamp}, \citenamefont {Koehler}, \citenamefont {Oca{\~n}a}, \citenamefont {Butler}, \citenamefont {Fredberg}, \citenamefont {Nieto}, \citenamefont {Bi},\ and\ \citenamefont {Park}}]{Mitchel2020}%
  \BibitemOpen
  \bibfield  {author} {\bibinfo {author} {\bibfnamefont {J.~A.}\ \bibnamefont {Mitchel}}, \bibinfo {author} {\bibfnamefont {A.}~\bibnamefont {Das}}, \bibinfo {author} {\bibfnamefont {M.~J.}\ \bibnamefont {O'Sullivan}}, \bibinfo {author} {\bibfnamefont {I.~T.}\ \bibnamefont {Stancil}}, \bibinfo {author} {\bibfnamefont {S.~J.}\ \bibnamefont {DeCamp}}, \bibinfo {author} {\bibfnamefont {S.}~\bibnamefont {Koehler}}, \bibinfo {author} {\bibfnamefont {O.~H.}\ \bibnamefont {Oca{\~n}a}}, \bibinfo {author} {\bibfnamefont {J.~P.}\ \bibnamefont {Butler}}, \bibinfo {author} {\bibfnamefont {J.~J.}\ \bibnamefont {Fredberg}}, \bibinfo {author} {\bibfnamefont {M.~A.}\ \bibnamefont {Nieto}}, \bibinfo {author} {\bibfnamefont {D.}~\bibnamefont {Bi}},\ and\ \bibinfo {author} {\bibfnamefont {J.-A.}\ \bibnamefont {Park}},\ }\bibfield  {title} {\bibinfo {title} {In primary airway epithelial cells, the unjamming transition is distinct from the epithelial-to-mesenchymal transition},\ }\href {https://doi.org/10.1038/s41467-020-18841-7}
  {\bibfield  {journal} {\bibinfo  {journal} {Nat. Commun.}\ }\textbf {\bibinfo {volume} {11}},\ \bibinfo {pages} {5053} (\bibinfo {year} {2020})}\BibitemShut {NoStop}%
\bibitem [{\citenamefont {Hannezo}\ and\ \citenamefont {Heisenberg}(2022)}]{Hannezo2022}%
  \BibitemOpen
  \bibfield  {author} {\bibinfo {author} {\bibfnamefont {E.}~\bibnamefont {Hannezo}}\ and\ \bibinfo {author} {\bibfnamefont {C.-P.}\ \bibnamefont {Heisenberg}},\ }\bibfield  {title} {\bibinfo {title} {Rigidity transitions in development and disease},\ }\href {https://doi.org/10.1016/j.tcb.2021.12.006} {\bibfield  {journal} {\bibinfo  {journal} {Trends Cell Biol.}\ }\textbf {\bibinfo {volume} {32}},\ \bibinfo {pages} {433} (\bibinfo {year} {2022})}\BibitemShut {NoStop}%
\bibitem [{\citenamefont {Villeneuve}\ \emph {et~al.}(2025)\citenamefont {Villeneuve}, \citenamefont {McCreery},\ and\ \citenamefont {Wickstr{\"o}m}}]{Villeneuve2025}%
  \BibitemOpen
  \bibfield  {author} {\bibinfo {author} {\bibfnamefont {C.}~\bibnamefont {Villeneuve}}, \bibinfo {author} {\bibfnamefont {K.~P.}\ \bibnamefont {McCreery}},\ and\ \bibinfo {author} {\bibfnamefont {S.~A.}\ \bibnamefont {Wickstr{\"o}m}},\ }\bibfield  {title} {\bibinfo {title} {Measuring and manipulating mechanical forces during development},\ }\href {https://doi.org/10.1038/s41556-025-01632-x} {\bibfield  {journal} {\bibinfo  {journal} {Nat. Cell Biol.}\ }\textbf {\bibinfo {volume} {27}},\ \bibinfo {pages} {575} (\bibinfo {year} {2025})}\BibitemShut {NoStop}%
\bibitem [{\citenamefont {{Fernandez-Gonzalez}}\ \emph {et~al.}(2009)\citenamefont {{Fernandez-Gonzalez}}, \citenamefont {Simoes}, \citenamefont {R{\"o}per}, \citenamefont {Eaton},\ and\ \citenamefont {Zallen}}]{Fernandez-Gonzalez2009}%
  \BibitemOpen
  \bibfield  {author} {\bibinfo {author} {\bibfnamefont {R.}~\bibnamefont {{Fernandez-Gonzalez}}}, \bibinfo {author} {\bibfnamefont {S.~d.~M.}\ \bibnamefont {Simoes}}, \bibinfo {author} {\bibfnamefont {J.-C.}\ \bibnamefont {R{\"o}per}}, \bibinfo {author} {\bibfnamefont {S.}~\bibnamefont {Eaton}},\ and\ \bibinfo {author} {\bibfnamefont {J.~A.}\ \bibnamefont {Zallen}},\ }\bibfield  {title} {\bibinfo {title} {Myosin {{II dynamics are regulated}} by {{tension}} in {{intercalating cells}}},\ }\href {https://doi.org/10.1016/j.devcel.2009.09.003} {\bibfield  {journal} {\bibinfo  {journal} {Dev. Cell}\ }\textbf {\bibinfo {volume} {17}},\ \bibinfo {pages} {736} (\bibinfo {year} {2009})}\BibitemShut {NoStop}%
\bibitem [{\citenamefont {Shivakumar}\ and\ \citenamefont {Lenne}(2016)}]{Shivakumar2016}%
  \BibitemOpen
  \bibfield  {author} {\bibinfo {author} {\bibfnamefont {P.~C.}\ \bibnamefont {Shivakumar}}\ and\ \bibinfo {author} {\bibfnamefont {P.-F.}\ \bibnamefont {Lenne}},\ }\bibfield  {title} {\bibinfo {title} {Laser {{ablation}} to {{probe}} the {{epithelial mechanics}} in {{Drosophila}}},\ }in\ \href {https://doi.org/10.1007/978-1-4939-6371-3_14} {\emph {\bibinfo {booktitle} {Drosophila: {{Methods}} and {{Protocols}}}}},\ \bibinfo {editor} {edited by\ \bibinfo {editor} {\bibfnamefont {C.}~\bibnamefont {Dahmann}}}\ (\bibinfo  {publisher} {Springer},\ \bibinfo {address} {{New York}},\ \bibinfo {year} {2016})\ pp.\ \bibinfo {pages} {241--251}\BibitemShut {NoStop}%
\bibitem [{\citenamefont {Bambardekar}\ \emph {et~al.}(2015)\citenamefont {Bambardekar}, \citenamefont {Cl{\'e}ment}, \citenamefont {Blanc}, \citenamefont {Chard{\`e}s},\ and\ \citenamefont {Lenne}}]{Bambardekar2015}%
  \BibitemOpen
  \bibfield  {author} {\bibinfo {author} {\bibfnamefont {K.}~\bibnamefont {Bambardekar}}, \bibinfo {author} {\bibfnamefont {R.}~\bibnamefont {Cl{\'e}ment}}, \bibinfo {author} {\bibfnamefont {O.}~\bibnamefont {Blanc}}, \bibinfo {author} {\bibfnamefont {C.}~\bibnamefont {Chard{\`e}s}},\ and\ \bibinfo {author} {\bibfnamefont {P.-F.}\ \bibnamefont {Lenne}},\ }\bibfield  {title} {\bibinfo {title} {Direct laser manipulation reveals the mechanics of cell contacts in vivo},\ }\href {https://doi.org/10.1073/pnas.1418732112} {\bibfield  {journal} {\bibinfo  {journal} {Proc. Natl. Acad. Sci.}\ }\textbf {\bibinfo {volume} {112}},\ \bibinfo {pages} {1416} (\bibinfo {year} {2015})}\BibitemShut {NoStop}%
\bibitem [{\citenamefont {Nishizawa}\ \emph {et~al.}(2023)\citenamefont {Nishizawa}, \citenamefont {Lin}, \citenamefont {Chard{\`e}s}, \citenamefont {Rupprecht},\ and\ \citenamefont {Lenne}}]{Nishizawa2023}%
  \BibitemOpen
  \bibfield  {author} {\bibinfo {author} {\bibfnamefont {K.}~\bibnamefont {Nishizawa}}, \bibinfo {author} {\bibfnamefont {S.-Z.}\ \bibnamefont {Lin}}, \bibinfo {author} {\bibfnamefont {C.}~\bibnamefont {Chard{\`e}s}}, \bibinfo {author} {\bibfnamefont {J.-F.}\ \bibnamefont {Rupprecht}},\ and\ \bibinfo {author} {\bibfnamefont {P.-F.}\ \bibnamefont {Lenne}},\ }\bibfield  {title} {\bibinfo {title} {Two-point optical manipulation reveals mechanosensitive remodeling of cell{\textendash}cell contacts in vivo},\ }\href {https://doi.org/10.1073/pnas.2212389120} {\bibfield  {journal} {\bibinfo  {journal} {Proc. Natl. Acad. Sci.}\ }\textbf {\bibinfo {volume} {120}},\ \bibinfo {pages} {e2212389120} (\bibinfo {year} {2023})}\BibitemShut {NoStop}%
\bibitem [{\citenamefont {Bi}\ \emph {et~al.}(2015)\citenamefont {Bi}, \citenamefont {Lopez}, \citenamefont {Schwarz},\ and\ \citenamefont {Manning}}]{Bi2015}%
  \BibitemOpen
  \bibfield  {author} {\bibinfo {author} {\bibfnamefont {D.}~\bibnamefont {Bi}}, \bibinfo {author} {\bibfnamefont {J.~H.}\ \bibnamefont {Lopez}}, \bibinfo {author} {\bibfnamefont {J.~M.}\ \bibnamefont {Schwarz}},\ and\ \bibinfo {author} {\bibfnamefont {M.~L.}\ \bibnamefont {Manning}},\ }\bibfield  {title} {\bibinfo {title} {A density-independent rigidity transition in biological tissues},\ }\href {https://doi.org/10.1038/nphys3471} {\bibfield  {journal} {\bibinfo  {journal} {Nat. Phys.}\ }\textbf {\bibinfo {volume} {11}},\ \bibinfo {pages} {1074} (\bibinfo {year} {2015})}\BibitemShut {NoStop}%
\bibitem [{\citenamefont {Wang}\ \emph {et~al.}(2020)\citenamefont {Wang}, \citenamefont {Merkel}, \citenamefont {Sutter}, \citenamefont {{Erdemci-Tandogan}}, \citenamefont {Manning},\ and\ \citenamefont {Kasza}}]{Wang2020a}%
  \BibitemOpen
  \bibfield  {author} {\bibinfo {author} {\bibfnamefont {X.}~\bibnamefont {Wang}}, \bibinfo {author} {\bibfnamefont {M.}~\bibnamefont {Merkel}}, \bibinfo {author} {\bibfnamefont {L.~B.}\ \bibnamefont {Sutter}}, \bibinfo {author} {\bibfnamefont {G.}~\bibnamefont {{Erdemci-Tandogan}}}, \bibinfo {author} {\bibfnamefont {M.~L.}\ \bibnamefont {Manning}},\ and\ \bibinfo {author} {\bibfnamefont {K.~E.}\ \bibnamefont {Kasza}},\ }\bibfield  {title} {\bibinfo {title} {Anisotropy links cell shapes to tissue flow during convergent extension},\ }\href {https://doi.org/10.1073/pnas.1916418117} {\bibfield  {journal} {\bibinfo  {journal} {Proc. Natl. Acad. Sci.}\ }\textbf {\bibinfo {volume} {117}},\ \bibinfo {pages} {13541} (\bibinfo {year} {2020})}\BibitemShut {NoStop}%
\bibitem [{\citenamefont {Etournay}\ \emph {et~al.}(2015)\citenamefont {Etournay}, \citenamefont {Popovi{\'c}}, \citenamefont {Merkel}, \citenamefont {Nandi}, \citenamefont {Blasse}, \citenamefont {Aigouy}, \citenamefont {Brandl}, \citenamefont {Myers}, \citenamefont {Salbreux}, \citenamefont {J{\"u}licher},\ and\ \citenamefont {Eaton}}]{Etournay2015}%
  \BibitemOpen
  \bibfield  {author} {\bibinfo {author} {\bibfnamefont {R.}~\bibnamefont {Etournay}}, \bibinfo {author} {\bibfnamefont {M.}~\bibnamefont {Popovi{\'c}}}, \bibinfo {author} {\bibfnamefont {M.}~\bibnamefont {Merkel}}, \bibinfo {author} {\bibfnamefont {A.}~\bibnamefont {Nandi}}, \bibinfo {author} {\bibfnamefont {C.}~\bibnamefont {Blasse}}, \bibinfo {author} {\bibfnamefont {B.}~\bibnamefont {Aigouy}}, \bibinfo {author} {\bibfnamefont {H.}~\bibnamefont {Brandl}}, \bibinfo {author} {\bibfnamefont {G.}~\bibnamefont {Myers}}, \bibinfo {author} {\bibfnamefont {G.}~\bibnamefont {Salbreux}}, \bibinfo {author} {\bibfnamefont {F.}~\bibnamefont {J{\"u}licher}},\ and\ \bibinfo {author} {\bibfnamefont {S.}~\bibnamefont {Eaton}},\ }\bibfield  {title} {\bibinfo {title} {Interplay of cell dynamics and epithelial tension during morphogenesis of the {{Drosophila}} pupal wing},\ }\href {https://doi.org/10.7554/eLife.07090} {\bibfield  {journal} {\bibinfo  {journal} {eLife}\ }\textbf {\bibinfo {volume} {4}},\ \bibinfo {pages}
  {e07090} (\bibinfo {year} {2015})}\BibitemShut {NoStop}%
\bibitem [{\citenamefont {{Piscitello-G{\'o}mez}}\ \emph {et~al.}(2023)\citenamefont {{Piscitello-G{\'o}mez}}, \citenamefont {Gruber}, \citenamefont {Krishna}, \citenamefont {Duclut}, \citenamefont {Modes}, \citenamefont {Popovi{\'c}}, \citenamefont {J{\"u}licher}, \citenamefont {Dye},\ and\ \citenamefont {Eaton}}]{Piscitello-Gomez2023}%
  \BibitemOpen
  \bibfield  {author} {\bibinfo {author} {\bibfnamefont {R.}~\bibnamefont {{Piscitello-G{\'o}mez}}}, \bibinfo {author} {\bibfnamefont {F.~S.}\ \bibnamefont {Gruber}}, \bibinfo {author} {\bibfnamefont {A.}~\bibnamefont {Krishna}}, \bibinfo {author} {\bibfnamefont {C.}~\bibnamefont {Duclut}}, \bibinfo {author} {\bibfnamefont {C.~D.}\ \bibnamefont {Modes}}, \bibinfo {author} {\bibfnamefont {M.}~\bibnamefont {Popovi{\'c}}}, \bibinfo {author} {\bibfnamefont {F.}~\bibnamefont {J{\"u}licher}}, \bibinfo {author} {\bibfnamefont {N.~A.}\ \bibnamefont {Dye}},\ and\ \bibinfo {author} {\bibfnamefont {S.}~\bibnamefont {Eaton}},\ }\bibfield  {title} {\bibinfo {title} {Core {{PCP}} mutations affect short-time mechanical properties but not tissue morphogenesis in the {{Drosophila}} pupal wing},\ }\href {https://doi.org/10.7554/eLife.85581} {\bibfield  {journal} {\bibinfo  {journal} {eLife}\ }\textbf {\bibinfo {volume} {12}},\ \bibinfo {pages} {e85581} (\bibinfo {year} {2023})}\BibitemShut {NoStop}%
\bibitem [{\citenamefont {Rigato}\ \emph {et~al.}(2024)\citenamefont {Rigato}, \citenamefont {Meng}, \citenamefont {Chardes}, \citenamefont {Runions}, \citenamefont {Abouakil}, \citenamefont {Smith},\ and\ \citenamefont {LeGoff}}]{Rigato2024}%
  \BibitemOpen
  \bibfield  {author} {\bibinfo {author} {\bibfnamefont {A.}~\bibnamefont {Rigato}}, \bibinfo {author} {\bibfnamefont {H.}~\bibnamefont {Meng}}, \bibinfo {author} {\bibfnamefont {C.}~\bibnamefont {Chardes}}, \bibinfo {author} {\bibfnamefont {A.}~\bibnamefont {Runions}}, \bibinfo {author} {\bibfnamefont {F.}~\bibnamefont {Abouakil}}, \bibinfo {author} {\bibfnamefont {R.~S.}\ \bibnamefont {Smith}},\ and\ \bibinfo {author} {\bibfnamefont {L.}~\bibnamefont {LeGoff}},\ }\bibfield  {title} {\bibinfo {title} {A mechanical transition from tension to buckling underlies the jigsaw puzzle shape morphogenesis of histoblasts in the {{Drosophila}} epidermis},\ }\href {https://doi.org/10.1371/journal.pbio.3002662} {\bibfield  {journal} {\bibinfo  {journal} {PLOS Biol.}\ }\textbf {\bibinfo {volume} {22}},\ \bibinfo {pages} {e3002662} (\bibinfo {year} {2024})}\BibitemShut {NoStop}%
\bibitem [{\citenamefont {Gierer}\ \emph {et~al.}(1972)\citenamefont {Gierer}, \citenamefont {Berking}, \citenamefont {Bode}, \citenamefont {David}, \citenamefont {Flick}, \citenamefont {Hansmann}, \citenamefont {Schaller},\ and\ \citenamefont {Trenkner}}]{Gierer1972}%
  \BibitemOpen
  \bibfield  {author} {\bibinfo {author} {\bibfnamefont {A.}~\bibnamefont {Gierer}}, \bibinfo {author} {\bibfnamefont {S.}~\bibnamefont {Berking}}, \bibinfo {author} {\bibfnamefont {H.}~\bibnamefont {Bode}}, \bibinfo {author} {\bibfnamefont {C.~N.}\ \bibnamefont {David}}, \bibinfo {author} {\bibfnamefont {K.}~\bibnamefont {Flick}}, \bibinfo {author} {\bibfnamefont {G.}~\bibnamefont {Hansmann}}, \bibinfo {author} {\bibfnamefont {H.}~\bibnamefont {Schaller}},\ and\ \bibinfo {author} {\bibfnamefont {E.}~\bibnamefont {Trenkner}},\ }\bibfield  {title} {\bibinfo {title} {Regeneration of {{Hydra}} from {{reaggregated cells}}},\ }\href {https://doi.org/10.1038/newbio239098a0} {\bibfield  {journal} {\bibinfo  {journal} {Nat. New Biol.}\ }\textbf {\bibinfo {volume} {239}},\ \bibinfo {pages} {98} (\bibinfo {year} {1972})}\BibitemShut {NoStop}%
\bibitem [{\citenamefont {Bode}(2003)}]{Bode2003}%
  \BibitemOpen
  \bibfield  {author} {\bibinfo {author} {\bibfnamefont {H.~R.}\ \bibnamefont {Bode}},\ }\bibfield  {title} {\bibinfo {title} {Head regeneration in {{Hydra}}},\ }\href {https://doi.org/10.1002/dvdy.10225} {\bibfield  {journal} {\bibinfo  {journal} {Dev. Dyn.}\ }\textbf {\bibinfo {volume} {226}},\ \bibinfo {pages} {225} (\bibinfo {year} {2003})}\BibitemShut {NoStop}%
\bibitem [{\citenamefont {Streichan}\ \emph {et~al.}(2018)\citenamefont {Streichan}, \citenamefont {Lefebvre}, \citenamefont {Noll}, \citenamefont {Wieschaus},\ and\ \citenamefont {Shraiman}}]{Streichan2018}%
  \BibitemOpen
  \bibfield  {author} {\bibinfo {author} {\bibfnamefont {S.~J.}\ \bibnamefont {Streichan}}, \bibinfo {author} {\bibfnamefont {M.~F.}\ \bibnamefont {Lefebvre}}, \bibinfo {author} {\bibfnamefont {N.}~\bibnamefont {Noll}}, \bibinfo {author} {\bibfnamefont {E.~F.}\ \bibnamefont {Wieschaus}},\ and\ \bibinfo {author} {\bibfnamefont {B.~I.}\ \bibnamefont {Shraiman}},\ }\bibfield  {title} {\bibinfo {title} {Global morphogenetic flow is accurately predicted by the spatial distribution of myosin motors},\ }\href {https://doi.org/10.7554/eLife.27454} {\bibfield  {journal} {\bibinfo  {journal} {eLife}\ }\textbf {\bibinfo {volume} {7}},\ \bibinfo {pages} {e27454} (\bibinfo {year} {2018})}\BibitemShut {NoStop}%
\bibitem [{\citenamefont {Marchetti}\ \emph {et~al.}(2013)\citenamefont {Marchetti}, \citenamefont {Joanny}, \citenamefont {Ramaswamy}, \citenamefont {Liverpool}, \citenamefont {Prost}, \citenamefont {Rao},\ and\ \citenamefont {Simha}}]{Marchetti2013}%
  \BibitemOpen
  \bibfield  {author} {\bibinfo {author} {\bibfnamefont {M.~C.}\ \bibnamefont {Marchetti}}, \bibinfo {author} {\bibfnamefont {J.~F.}\ \bibnamefont {Joanny}}, \bibinfo {author} {\bibfnamefont {S.}~\bibnamefont {Ramaswamy}}, \bibinfo {author} {\bibfnamefont {T.~B.}\ \bibnamefont {Liverpool}}, \bibinfo {author} {\bibfnamefont {J.}~\bibnamefont {Prost}}, \bibinfo {author} {\bibfnamefont {M.}~\bibnamefont {Rao}},\ and\ \bibinfo {author} {\bibfnamefont {R.~A.}\ \bibnamefont {Simha}},\ }\bibfield  {title} {\bibinfo {title} {Hydrodynamics of soft active matter},\ }\href {https://doi.org/10.1103/RevModPhys.85.1143} {\bibfield  {journal} {\bibinfo  {journal} {Rev. Mod. Phys.}\ }\textbf {\bibinfo {volume} {85}},\ \bibinfo {pages} {1143} (\bibinfo {year} {2013})}\BibitemShut {NoStop}%
\bibitem [{\citenamefont {{L{\'o}pez-Gay}}\ \emph {et~al.}(2020)\citenamefont {{L{\'o}pez-Gay}}, \citenamefont {Nunley}, \citenamefont {Spencer}, \citenamefont {{di Pietro}}, \citenamefont {Guirao}, \citenamefont {Bosveld}, \citenamefont {Markova}, \citenamefont {Gaugue}, \citenamefont {Pelletier}, \citenamefont {Lubensky},\ and\ \citenamefont {Bella{\"i}che}}]{Lopez-Gay2020}%
  \BibitemOpen
  \bibfield  {author} {\bibinfo {author} {\bibfnamefont {J.~M.}\ \bibnamefont {{L{\'o}pez-Gay}}}, \bibinfo {author} {\bibfnamefont {H.}~\bibnamefont {Nunley}}, \bibinfo {author} {\bibfnamefont {M.}~\bibnamefont {Spencer}}, \bibinfo {author} {\bibfnamefont {F.}~\bibnamefont {{di Pietro}}}, \bibinfo {author} {\bibfnamefont {B.}~\bibnamefont {Guirao}}, \bibinfo {author} {\bibfnamefont {F.}~\bibnamefont {Bosveld}}, \bibinfo {author} {\bibfnamefont {O.}~\bibnamefont {Markova}}, \bibinfo {author} {\bibfnamefont {I.}~\bibnamefont {Gaugue}}, \bibinfo {author} {\bibfnamefont {S.}~\bibnamefont {Pelletier}}, \bibinfo {author} {\bibfnamefont {D.~K.}\ \bibnamefont {Lubensky}},\ and\ \bibinfo {author} {\bibfnamefont {Y.}~\bibnamefont {Bella{\"i}che}},\ }\bibfield  {title} {\bibinfo {title} {Apical stress fibers enable a scaling between cell mechanical response and area in epithelial tissue},\ }\href {https://doi.org/10.1126/science.abb2169} {\bibfield  {journal} {\bibinfo  {journal} {Science}\ }\textbf {\bibinfo {volume}
  {370}},\ \bibinfo {pages} {eabb2169} (\bibinfo {year} {2020})}\BibitemShut {NoStop}%
\bibitem [{\citenamefont {{Maroudas-Sacks}}\ \emph {et~al.}(2021)\citenamefont {{Maroudas-Sacks}}, \citenamefont {Garion}, \citenamefont {{Shani-Zerbib}}, \citenamefont {Livshits}, \citenamefont {Braun},\ and\ \citenamefont {Keren}}]{Maroudas-Sacks2021}%
  \BibitemOpen
  \bibfield  {author} {\bibinfo {author} {\bibfnamefont {Y.}~\bibnamefont {{Maroudas-Sacks}}}, \bibinfo {author} {\bibfnamefont {L.}~\bibnamefont {Garion}}, \bibinfo {author} {\bibfnamefont {L.}~\bibnamefont {{Shani-Zerbib}}}, \bibinfo {author} {\bibfnamefont {A.}~\bibnamefont {Livshits}}, \bibinfo {author} {\bibfnamefont {E.}~\bibnamefont {Braun}},\ and\ \bibinfo {author} {\bibfnamefont {K.}~\bibnamefont {Keren}},\ }\bibfield  {title} {\bibinfo {title} {Topological defects in the nematic order of actin fibres as organization centres of {{Hydra}} morphogenesis},\ }\href {https://doi.org/10.1038/s41567-020-01083-1} {\bibfield  {journal} {\bibinfo  {journal} {Nat. Phys.}\ }\textbf {\bibinfo {volume} {17}},\ \bibinfo {pages} {251} (\bibinfo {year} {2021})}\BibitemShut {NoStop}%
\bibitem [{\citenamefont {Lengfeld}\ \emph {et~al.}(2009)\citenamefont {Lengfeld}, \citenamefont {Watanabe}, \citenamefont {Simakov}, \citenamefont {Lindgens}, \citenamefont {Gee}, \citenamefont {Law}, \citenamefont {Schmidt}, \citenamefont {{\"O}zbek}, \citenamefont {Bode},\ and\ \citenamefont {Holstein}}]{Lengfeld2009}%
  \BibitemOpen
  \bibfield  {author} {\bibinfo {author} {\bibfnamefont {T.}~\bibnamefont {Lengfeld}}, \bibinfo {author} {\bibfnamefont {H.}~\bibnamefont {Watanabe}}, \bibinfo {author} {\bibfnamefont {O.}~\bibnamefont {Simakov}}, \bibinfo {author} {\bibfnamefont {D.}~\bibnamefont {Lindgens}}, \bibinfo {author} {\bibfnamefont {L.}~\bibnamefont {Gee}}, \bibinfo {author} {\bibfnamefont {L.}~\bibnamefont {Law}}, \bibinfo {author} {\bibfnamefont {H.~A.}\ \bibnamefont {Schmidt}}, \bibinfo {author} {\bibfnamefont {S.}~\bibnamefont {{\"O}zbek}}, \bibinfo {author} {\bibfnamefont {H.}~\bibnamefont {Bode}},\ and\ \bibinfo {author} {\bibfnamefont {T.~W.}\ \bibnamefont {Holstein}},\ }\bibfield  {title} {\bibinfo {title} {Multiple {{Wnts}} are involved in {{{\emph{Hydra}}}} organizer formation and regeneration},\ }\href {https://doi.org/10.1016/j.ydbio.2009.02.004} {\bibfield  {journal} {\bibinfo  {journal} {Dev. Biol.}\ }\textbf {\bibinfo {volume} {330}},\ \bibinfo {pages} {186} (\bibinfo {year} {2009})}\BibitemShut {NoStop}%
\bibitem [{\citenamefont {Aufschnaiter}\ \emph {et~al.}(2017)\citenamefont {Aufschnaiter}, \citenamefont {{Wedlich-S{\"o}ldner}}, \citenamefont {Zhang},\ and\ \citenamefont {Hobmayer}}]{Aufschnaiter2017}%
  \BibitemOpen
  \bibfield  {author} {\bibinfo {author} {\bibfnamefont {R.}~\bibnamefont {Aufschnaiter}}, \bibinfo {author} {\bibfnamefont {R.}~\bibnamefont {{Wedlich-S{\"o}ldner}}}, \bibinfo {author} {\bibfnamefont {X.}~\bibnamefont {Zhang}},\ and\ \bibinfo {author} {\bibfnamefont {B.}~\bibnamefont {Hobmayer}},\ }\bibfield  {title} {\bibinfo {title} {Apical and basal epitheliomuscular {{F-actin}} dynamics during {{Hydra}} bud evagination},\ }\href {https://doi.org/10.1242/bio.022723} {\bibfield  {journal} {\bibinfo  {journal} {Biol. Open}\ }\textbf {\bibinfo {volume} {6}},\ \bibinfo {pages} {1137} (\bibinfo {year} {2017})}\BibitemShut {NoStop}%
\bibitem [{\citenamefont {{Maroudas-Sacks}}\ \emph {et~al.}(2025)\citenamefont {{Maroudas-Sacks}}, \citenamefont {Suganthan}, \citenamefont {Garion}, \citenamefont {{Ascoli-Abbina}}, \citenamefont {Westfried}, \citenamefont {Dori}, \citenamefont {Pasvinter}, \citenamefont {Popovi{\'c}},\ and\ \citenamefont {Keren}}]{Maroudas-Sacks2025}%
  \BibitemOpen
  \bibfield  {author} {\bibinfo {author} {\bibfnamefont {Y.}~\bibnamefont {{Maroudas-Sacks}}}, \bibinfo {author} {\bibfnamefont {S.}~\bibnamefont {Suganthan}}, \bibinfo {author} {\bibfnamefont {L.}~\bibnamefont {Garion}}, \bibinfo {author} {\bibfnamefont {Y.}~\bibnamefont {{Ascoli-Abbina}}}, \bibinfo {author} {\bibfnamefont {A.}~\bibnamefont {Westfried}}, \bibinfo {author} {\bibfnamefont {N.}~\bibnamefont {Dori}}, \bibinfo {author} {\bibfnamefont {I.}~\bibnamefont {Pasvinter}}, \bibinfo {author} {\bibfnamefont {M.}~\bibnamefont {Popovi{\'c}}},\ and\ \bibinfo {author} {\bibfnamefont {K.}~\bibnamefont {Keren}},\ }\bibfield  {title} {\bibinfo {title} {Mechanical strain focusing at topological defect sites in regenerating {{Hydra}}},\ }\href {https://doi.org/10.1242/dev.204514} {\bibfield  {journal} {\bibinfo  {journal} {Development}\ }\textbf {\bibinfo {volume} {152}},\ \bibinfo {pages} {DEV204514} (\bibinfo {year} {2025})}\BibitemShut {NoStop}%
\bibitem [{\citenamefont {Wang}\ \emph {et~al.}(2023)\citenamefont {Wang}, \citenamefont {Marchetti},\ and\ \citenamefont {Brauns}}]{Wang2023a}%
  \BibitemOpen
  \bibfield  {author} {\bibinfo {author} {\bibfnamefont {Z.}~\bibnamefont {Wang}}, \bibinfo {author} {\bibfnamefont {M.~C.}\ \bibnamefont {Marchetti}},\ and\ \bibinfo {author} {\bibfnamefont {F.}~\bibnamefont {Brauns}},\ }\bibfield  {title} {\bibinfo {title} {Patterning of morphogenetic anisotropy fields},\ }\href {https://doi.org/10.1073/pnas.2220167120} {\bibfield  {journal} {\bibinfo  {journal} {Proc. Natl. Acad. Sci.}\ }\textbf {\bibinfo {volume} {120}},\ \bibinfo {pages} {e2220167120} (\bibinfo {year} {2023})}\BibitemShut {NoStop}%
\bibitem [{\citenamefont {Weevers}\ \emph {et~al.}(2025)\citenamefont {Weevers}, \citenamefont {Falconer}, \citenamefont {Mercker}, \citenamefont {Sadeghi}, \citenamefont {Rozema}, \citenamefont {Ferenc}, \citenamefont {Ma{\^i}tre}, \citenamefont {Ott}, \citenamefont {Oelz}, \citenamefont {{Marciniak-Czochra}},\ and\ \citenamefont {Tsiairis}}]{Weevers2025}%
  \BibitemOpen
  \bibfield  {author} {\bibinfo {author} {\bibfnamefont {S.~L.}\ \bibnamefont {Weevers}}, \bibinfo {author} {\bibfnamefont {A.~D.}\ \bibnamefont {Falconer}}, \bibinfo {author} {\bibfnamefont {M.}~\bibnamefont {Mercker}}, \bibinfo {author} {\bibfnamefont {H.}~\bibnamefont {Sadeghi}}, \bibinfo {author} {\bibfnamefont {D.}~\bibnamefont {Rozema}}, \bibinfo {author} {\bibfnamefont {J.}~\bibnamefont {Ferenc}}, \bibinfo {author} {\bibfnamefont {J.-L.}\ \bibnamefont {Ma{\^i}tre}}, \bibinfo {author} {\bibfnamefont {A.}~\bibnamefont {Ott}}, \bibinfo {author} {\bibfnamefont {D.~B.}\ \bibnamefont {Oelz}}, \bibinfo {author} {\bibfnamefont {A.}~\bibnamefont {{Marciniak-Czochra}}},\ and\ \bibinfo {author} {\bibfnamefont {C.~D.}\ \bibnamefont {Tsiairis}},\ }\bibfield  {title} {\bibinfo {title} {Mechanochemical patterning localizes the organizer of a luminal epithelium},\ }\href {https://doi.org/10.1126/sciadv.adu2286} {\bibfield  {journal} {\bibinfo  {journal} {Sci. Adv.}\ }\textbf {\bibinfo {volume} {11}},\ \bibinfo {pages}
  {eadu2286} (\bibinfo {year} {2025})}\BibitemShut {NoStop}%
\bibitem [{\citenamefont {Ibrahimi}\ and\ \citenamefont {Merkel}(2025)}]{Ibrahimi2025}%
  \BibitemOpen
  \bibfield  {author} {\bibinfo {author} {\bibfnamefont {M.}~\bibnamefont {Ibrahimi}}\ and\ \bibinfo {author} {\bibfnamefont {M.}~\bibnamefont {Merkel}},\ }\bibfield  {title} {\bibinfo {title} {Stabilization of active tissue deformation by a dynamic signaling gradient},\ }\href {https://doi.org/10.1103/yq1s-p2zl} {\bibfield  {journal} {\bibinfo  {journal} {PRX Life}\ }\textbf {\bibinfo {volume} {3}},\ \bibinfo {pages} {043013} (\bibinfo {year} {2025})}\BibitemShut {NoStop}%
\bibitem [{\citenamefont {Rozman}\ \emph {et~al.}(2023)\citenamefont {Rozman}, \citenamefont {Yeomans},\ and\ \citenamefont {Sknepnek}}]{Rozman2023}%
  \BibitemOpen
  \bibfield  {author} {\bibinfo {author} {\bibfnamefont {J.}~\bibnamefont {Rozman}}, \bibinfo {author} {\bibfnamefont {J.~M.}\ \bibnamefont {Yeomans}},\ and\ \bibinfo {author} {\bibfnamefont {R.}~\bibnamefont {Sknepnek}},\ }\bibfield  {title} {\bibinfo {title} {Shape-tension coupling produces nematic order in an epithelium vertex model},\ }\href {https://doi.org/10.1103/PhysRevLett.131.228301} {\bibfield  {journal} {\bibinfo  {journal} {Phys. Rev. Lett.}\ }\textbf {\bibinfo {volume} {131}},\ \bibinfo {pages} {228301} (\bibinfo {year} {2023})}\BibitemShut {NoStop}%
\bibitem [{\citenamefont {Claussen}\ \emph {et~al.}(2024)\citenamefont {Claussen}, \citenamefont {Brauns},\ and\ \citenamefont {Shraiman}}]{Claussen2024}%
  \BibitemOpen
  \bibfield  {author} {\bibinfo {author} {\bibfnamefont {N.~H.}\ \bibnamefont {Claussen}}, \bibinfo {author} {\bibfnamefont {F.}~\bibnamefont {Brauns}},\ and\ \bibinfo {author} {\bibfnamefont {B.~I.}\ \bibnamefont {Shraiman}},\ }\bibfield  {title} {\bibinfo {title} {A geometric-tension-dynamics model of epithelial convergent extension},\ }\href {https://doi.org/10.1073/pnas.2321928121} {\bibfield  {journal} {\bibinfo  {journal} {Proc. Natl. Acad. Sci.}\ }\textbf {\bibinfo {volume} {121}},\ \bibinfo {pages} {e2321928121} (\bibinfo {year} {2024})}\BibitemShut {NoStop}%
\bibitem [{\citenamefont {Mueller}\ \emph {et~al.}(2019)\citenamefont {Mueller}, \citenamefont {Yeomans},\ and\ \citenamefont {Doostmohammadi}}]{Mueller2019}%
  \BibitemOpen
  \bibfield  {author} {\bibinfo {author} {\bibfnamefont {R.}~\bibnamefont {Mueller}}, \bibinfo {author} {\bibfnamefont {J.~M.}\ \bibnamefont {Yeomans}},\ and\ \bibinfo {author} {\bibfnamefont {A.}~\bibnamefont {Doostmohammadi}},\ }\bibfield  {title} {\bibinfo {title} {Emergence of {{active nematic behavior}} in {{monolayers}} of {{isotropic cells}}},\ }\href {https://doi.org/10.1103/PhysRevLett.122.048004} {\bibfield  {journal} {\bibinfo  {journal} {Phys. Rev. Lett.}\ }\textbf {\bibinfo {volume} {122}},\ \bibinfo {pages} {048004} (\bibinfo {year} {2019})}\BibitemShut {NoStop}%
\bibitem [{\citenamefont {Zhang}\ and\ \citenamefont {Yeomans}(2023)}]{Zhang2023}%
  \BibitemOpen
  \bibfield  {author} {\bibinfo {author} {\bibfnamefont {G.}~\bibnamefont {Zhang}}\ and\ \bibinfo {author} {\bibfnamefont {J.~M.}\ \bibnamefont {Yeomans}},\ }\bibfield  {title} {\bibinfo {title} {Active forces in confluent cell monolayers},\ }\href {https://doi.org/10.1103/PhysRevLett.130.038202} {\bibfield  {journal} {\bibinfo  {journal} {Phys. Rev. Lett.}\ }\textbf {\bibinfo {volume} {130}},\ \bibinfo {pages} {038202} (\bibinfo {year} {2023})}\BibitemShut {NoStop}%
\bibitem [{\citenamefont {Chiang}\ \emph {et~al.}(2024)\citenamefont {Chiang}, \citenamefont {Hopkins}, \citenamefont {Loewe}, \citenamefont {Marchetti},\ and\ \citenamefont {Marenduzzo}}]{Chiang2024a}%
  \BibitemOpen
  \bibfield  {author} {\bibinfo {author} {\bibfnamefont {M.}~\bibnamefont {Chiang}}, \bibinfo {author} {\bibfnamefont {A.}~\bibnamefont {Hopkins}}, \bibinfo {author} {\bibfnamefont {B.}~\bibnamefont {Loewe}}, \bibinfo {author} {\bibfnamefont {M.~C.}\ \bibnamefont {Marchetti}},\ and\ \bibinfo {author} {\bibfnamefont {D.}~\bibnamefont {Marenduzzo}},\ }\bibfield  {title} {\bibinfo {title} {Intercellular friction and motility drive orientational order in cell monolayers},\ }\href {https://doi.org/10.1073/pnas.2319310121} {\bibfield  {journal} {\bibinfo  {journal} {Proc. Natl. Acad. Sci.}\ }\textbf {\bibinfo {volume} {121}},\ \bibinfo {pages} {e2319310121} (\bibinfo {year} {2024})}\BibitemShut {NoStop}%
\bibitem [{\citenamefont {Balasubramaniam}\ \emph {et~al.}(2021)\citenamefont {Balasubramaniam}, \citenamefont {Doostmohammadi}, \citenamefont {Saw}, \citenamefont {Narayana}, \citenamefont {Mueller}, \citenamefont {Dang}, \citenamefont {Thomas}, \citenamefont {Gupta}, \citenamefont {Sonam}, \citenamefont {Yap}, \citenamefont {Toyama}, \citenamefont {M{\`e}ge}, \citenamefont {Yeomans},\ and\ \citenamefont {Ladoux}}]{Balasubramaniam2021}%
  \BibitemOpen
  \bibfield  {author} {\bibinfo {author} {\bibfnamefont {L.}~\bibnamefont {Balasubramaniam}}, \bibinfo {author} {\bibfnamefont {A.}~\bibnamefont {Doostmohammadi}}, \bibinfo {author} {\bibfnamefont {T.~B.}\ \bibnamefont {Saw}}, \bibinfo {author} {\bibfnamefont {G.~H. N.~S.}\ \bibnamefont {Narayana}}, \bibinfo {author} {\bibfnamefont {R.}~\bibnamefont {Mueller}}, \bibinfo {author} {\bibfnamefont {T.}~\bibnamefont {Dang}}, \bibinfo {author} {\bibfnamefont {M.}~\bibnamefont {Thomas}}, \bibinfo {author} {\bibfnamefont {S.}~\bibnamefont {Gupta}}, \bibinfo {author} {\bibfnamefont {S.}~\bibnamefont {Sonam}}, \bibinfo {author} {\bibfnamefont {A.~S.}\ \bibnamefont {Yap}}, \bibinfo {author} {\bibfnamefont {Y.}~\bibnamefont {Toyama}}, \bibinfo {author} {\bibfnamefont {R.-M.}\ \bibnamefont {M{\`e}ge}}, \bibinfo {author} {\bibfnamefont {J.~M.}\ \bibnamefont {Yeomans}},\ and\ \bibinfo {author} {\bibfnamefont {B.}~\bibnamefont {Ladoux}},\ }\bibfield  {title} {\bibinfo {title} {Investigating the nature of active forces in
  tissues reveals how contractile cells can form extensile monolayers},\ }\href {https://doi.org/10.1038/s41563-021-00919-2} {\bibfield  {journal} {\bibinfo  {journal} {Nat. Mater.}\ }\textbf {\bibinfo {volume} {20}},\ \bibinfo {pages} {1156} (\bibinfo {year} {2021})}\BibitemShut {NoStop}%
\bibitem [{\citenamefont {{Blanch-Mercader}}\ \emph {et~al.}(2021)\citenamefont {{Blanch-Mercader}}, \citenamefont {Guillamat}, \citenamefont {Roux},\ and\ \citenamefont {Kruse}}]{Blanch-Mercader2021}%
  \BibitemOpen
  \bibfield  {author} {\bibinfo {author} {\bibfnamefont {C.}~\bibnamefont {{Blanch-Mercader}}}, \bibinfo {author} {\bibfnamefont {P.}~\bibnamefont {Guillamat}}, \bibinfo {author} {\bibfnamefont {A.}~\bibnamefont {Roux}},\ and\ \bibinfo {author} {\bibfnamefont {K.}~\bibnamefont {Kruse}},\ }\bibfield  {title} {\bibinfo {title} {Quantifying material properties of cell monolayers by analyzing integer topological defects},\ }\href {https://doi.org/10.1103/PhysRevLett.126.028101} {\bibfield  {journal} {\bibinfo  {journal} {Phys. Rev. Lett.}\ }\textbf {\bibinfo {volume} {126}},\ \bibinfo {pages} {028101} (\bibinfo {year} {2021})}\BibitemShut {NoStop}%
\bibitem [{\citenamefont {Lv}\ \emph {et~al.}(2024)\citenamefont {Lv}, \citenamefont {Li}, \citenamefont {Wang},\ and\ \citenamefont {Li}}]{Lv2024a}%
  \BibitemOpen
  \bibfield  {author} {\bibinfo {author} {\bibfnamefont {C.-L.}\ \bibnamefont {Lv}}, \bibinfo {author} {\bibfnamefont {Z.-Y.}\ \bibnamefont {Li}}, \bibinfo {author} {\bibfnamefont {S.-D.}\ \bibnamefont {Wang}},\ and\ \bibinfo {author} {\bibfnamefont {B.}~\bibnamefont {Li}},\ }\bibfield  {title} {\bibinfo {title} {Morphodynamics of interface between dissimilar cell aggregations},\ }\href {https://doi.org/10.1038/s42005-024-01840-1} {\bibfield  {journal} {\bibinfo  {journal} {Commun. Phys.}\ }\textbf {\bibinfo {volume} {7}},\ \bibinfo {pages} {1} (\bibinfo {year} {2024})}\BibitemShut {NoStop}%
\bibitem [{\citenamefont {Lin}\ \emph {et~al.}(2023)\citenamefont {Lin}, \citenamefont {Merkel},\ and\ \citenamefont {Rupprecht}}]{Lin2023}%
  \BibitemOpen
  \bibfield  {author} {\bibinfo {author} {\bibfnamefont {S.-Z.}\ \bibnamefont {Lin}}, \bibinfo {author} {\bibfnamefont {M.}~\bibnamefont {Merkel}},\ and\ \bibinfo {author} {\bibfnamefont {J.-F.}\ \bibnamefont {Rupprecht}},\ }\bibfield  {title} {\bibinfo {title} {Structure and rheology in vertex models under cell-shape-dependent active stresses},\ }\href {https://doi.org/10.1103/PhysRevLett.130.058202} {\bibfield  {journal} {\bibinfo  {journal} {Phys. Rev. Lett.}\ }\textbf {\bibinfo {volume} {130}},\ \bibinfo {pages} {058202} (\bibinfo {year} {2023})}\BibitemShut {NoStop}%
\bibitem [{\citenamefont {Rozman}\ \emph {et~al.}(2025)\citenamefont {Rozman}, \citenamefont {Chaithanya}, \citenamefont {Yeomans},\ and\ \citenamefont {Sknepnek}}]{Rozman2025}%
  \BibitemOpen
  \bibfield  {author} {\bibinfo {author} {\bibfnamefont {J.}~\bibnamefont {Rozman}}, \bibinfo {author} {\bibfnamefont {K.~V.~S.}\ \bibnamefont {Chaithanya}}, \bibinfo {author} {\bibfnamefont {J.~M.}\ \bibnamefont {Yeomans}},\ and\ \bibinfo {author} {\bibfnamefont {R.}~\bibnamefont {Sknepnek}},\ }\bibfield  {title} {\bibinfo {title} {Vertex model with internal dissipation enables sustained flows},\ }\href {https://doi.org/10.1038/s41467-025-55820-2} {\bibfield  {journal} {\bibinfo  {journal} {Nat. Commun.}\ }\textbf {\bibinfo {volume} {16}},\ \bibinfo {pages} {530} (\bibinfo {year} {2025})}\BibitemShut {NoStop}%
\bibitem [{\citenamefont {Rozman}\ and\ \citenamefont {Yeomans}(2024)}]{Rozman2024a}%
  \BibitemOpen
  \bibfield  {author} {\bibinfo {author} {\bibfnamefont {J.}~\bibnamefont {Rozman}}\ and\ \bibinfo {author} {\bibfnamefont {J.~M.}\ \bibnamefont {Yeomans}},\ }\bibfield  {title} {\bibinfo {title} {Cell sorting in an active nematic vertex model},\ }\href {https://doi.org/10.1103/PhysRevLett.133.248401} {\bibfield  {journal} {\bibinfo  {journal} {Phys. Rev. Lett.}\ }\textbf {\bibinfo {volume} {133}},\ \bibinfo {pages} {248401} (\bibinfo {year} {2024})}\BibitemShut {NoStop}%
\bibitem [{\citenamefont {Bi}\ \emph {et~al.}(2016)\citenamefont {Bi}, \citenamefont {Yang}, \citenamefont {Marchetti},\ and\ \citenamefont {Manning}}]{Bi2016}%
  \BibitemOpen
  \bibfield  {author} {\bibinfo {author} {\bibfnamefont {D.}~\bibnamefont {Bi}}, \bibinfo {author} {\bibfnamefont {X.}~\bibnamefont {Yang}}, \bibinfo {author} {\bibfnamefont {M.~C.}\ \bibnamefont {Marchetti}},\ and\ \bibinfo {author} {\bibfnamefont {M.~L.}\ \bibnamefont {Manning}},\ }\bibfield  {title} {\bibinfo {title} {Motility-driven glass and jamming transitions in biological tissues},\ }\href {https://doi.org/10.1103/PhysRevX.6.021011} {\bibfield  {journal} {\bibinfo  {journal} {Phys. Rev. X}\ }\textbf {\bibinfo {volume} {6}},\ \bibinfo {pages} {021011} (\bibinfo {year} {2016})}\BibitemShut {NoStop}%
\bibitem [{\citenamefont {F.~Staddon}\ \emph {et~al.}(2022)\citenamefont {F.~Staddon}, \citenamefont {P.~Murrell},\ and\ \citenamefont {Banerjee}}]{F.Staddon2022}%
  \BibitemOpen
  \bibfield  {author} {\bibinfo {author} {\bibfnamefont {M.}~\bibnamefont {F.~Staddon}}, \bibinfo {author} {\bibfnamefont {M.}~\bibnamefont {P.~Murrell}},\ and\ \bibinfo {author} {\bibfnamefont {S.}~\bibnamefont {Banerjee}},\ }\bibfield  {title} {\bibinfo {title} {Interplay between substrate rigidity and tissue fluidity regulates cell monolayer spreading},\ }\href {https://doi.org/10.1039/D2SM00757F} {\bibfield  {journal} {\bibinfo  {journal} {Soft Matter}\ }\textbf {\bibinfo {volume} {18}},\ \bibinfo {pages} {7877} (\bibinfo {year} {2022})}\BibitemShut {NoStop}%
\bibitem [{\citenamefont {Ray}\ \emph {et~al.}(2025)\citenamefont {Ray}, \citenamefont {Biswas},\ and\ \citenamefont {Das}}]{Ray2025}%
  \BibitemOpen
  \bibfield  {author} {\bibinfo {author} {\bibfnamefont {S.}~\bibnamefont {Ray}}, \bibinfo {author} {\bibfnamefont {S.}~\bibnamefont {Biswas}},\ and\ \bibinfo {author} {\bibfnamefont {D.}~\bibnamefont {Das}},\ }\bibfield  {title} {\bibinfo {title} {Role of intercellular adhesion in modulating tissue fluidity},\ }\href {https://doi.org/10.7554/eLife.106294.1} {\bibfield  {journal} {\bibinfo  {journal} {eLife}\ }\textbf {\bibinfo {volume} {14}},\ \bibinfo {pages} {RP106294} (\bibinfo {year} {2025})}\BibitemShut {NoStop}%
\bibitem [{\citenamefont {Weliky}\ and\ \citenamefont {Oster}(1990)}]{Weliky1990}%
  \BibitemOpen
  \bibfield  {author} {\bibinfo {author} {\bibfnamefont {M.}~\bibnamefont {Weliky}}\ and\ \bibinfo {author} {\bibfnamefont {G.}~\bibnamefont {Oster}},\ }\bibfield  {title} {\bibinfo {title} {The mechanical basis of cell rearrangement {{I}}. {{Epithelial}} morphogenesis during {{Fundulus}} epiboly},\ }\href {https://doi.org/10.1242/dev.109.2.373} {\bibfield  {journal} {\bibinfo  {journal} {Development}\ }\textbf {\bibinfo {volume} {109}},\ \bibinfo {pages} {373} (\bibinfo {year} {1990})}\BibitemShut {NoStop}%
\bibitem [{\citenamefont {Nagai}\ and\ \citenamefont {Honda}(2001)}]{Nagai2001}%
  \BibitemOpen
  \bibfield  {author} {\bibinfo {author} {\bibfnamefont {T.}~\bibnamefont {Nagai}}\ and\ \bibinfo {author} {\bibfnamefont {H.}~\bibnamefont {Honda}},\ }\bibfield  {title} {\bibinfo {title} {A dynamic cell model for the formation of epithelial tissues},\ }\href {https://doi.org/10.1080/13642810108205772} {\bibfield  {journal} {\bibinfo  {journal} {Philos. Mag. B}\ }\textbf {\bibinfo {volume} {81}},\ \bibinfo {pages} {699} (\bibinfo {year} {2001})}\BibitemShut {NoStop}%
\bibitem [{\citenamefont {Alt}\ \emph {et~al.}(2017)\citenamefont {Alt}, \citenamefont {Ganguly},\ and\ \citenamefont {Salbreux}}]{Alt2017}%
  \BibitemOpen
  \bibfield  {author} {\bibinfo {author} {\bibfnamefont {S.}~\bibnamefont {Alt}}, \bibinfo {author} {\bibfnamefont {P.}~\bibnamefont {Ganguly}},\ and\ \bibinfo {author} {\bibfnamefont {G.}~\bibnamefont {Salbreux}},\ }\bibfield  {title} {\bibinfo {title} {Vertex models: from cell mechanics to tissue morphogenesis},\ }\href {https://doi.org/10.1098/rstb.2015.0520} {\bibfield  {journal} {\bibinfo  {journal} {Philos. Trans. R. Soc. B: Biol. Sci.}\ }\textbf {\bibinfo {volume} {8}},\ \bibinfo {pages} {20150520} (\bibinfo {year} {2017})}\BibitemShut {NoStop}%
\bibitem [{\citenamefont {K{\"a}fer}\ \emph {et~al.}(2007)\citenamefont {K{\"a}fer}, \citenamefont {Hayashi}, \citenamefont {Mar{\'e}e}, \citenamefont {Carthew},\ and\ \citenamefont {Graner}}]{Kafer2007}%
  \BibitemOpen
  \bibfield  {author} {\bibinfo {author} {\bibfnamefont {J.}~\bibnamefont {K{\"a}fer}}, \bibinfo {author} {\bibfnamefont {T.}~\bibnamefont {Hayashi}}, \bibinfo {author} {\bibfnamefont {A.~F.~M.}\ \bibnamefont {Mar{\'e}e}}, \bibinfo {author} {\bibfnamefont {R.~W.}\ \bibnamefont {Carthew}},\ and\ \bibinfo {author} {\bibfnamefont {F.}~\bibnamefont {Graner}},\ }\bibfield  {title} {\bibinfo {title} {Cell adhesion and cortex contractility determine cell patterning in the {{Drosophilaretina}}},\ }\href {https://doi.org/10.1073/pnas.0704235104} {\bibfield  {journal} {\bibinfo  {journal} {Proc. Natl. Acad. Sci.}\ }\textbf {\bibinfo {volume} {104}},\ \bibinfo {pages} {18549} (\bibinfo {year} {2007})}\BibitemShut {NoStop}%
\bibitem [{\citenamefont {Fletcher}\ \emph {et~al.}(2014)\citenamefont {Fletcher}, \citenamefont {Osterfield}, \citenamefont {Baker},\ and\ \citenamefont {Shvartsman}}]{Fletcher2014}%
  \BibitemOpen
  \bibfield  {author} {\bibinfo {author} {\bibfnamefont {A.~G.}\ \bibnamefont {Fletcher}}, \bibinfo {author} {\bibfnamefont {M.}~\bibnamefont {Osterfield}}, \bibinfo {author} {\bibfnamefont {R.~E.}\ \bibnamefont {Baker}},\ and\ \bibinfo {author} {\bibfnamefont {S.~Y.}\ \bibnamefont {Shvartsman}},\ }\bibfield  {title} {\bibinfo {title} {Vertex models of epithelial morphogenesis},\ }\href {https://doi.org/10.1016/j.bpj.2013.11.4498} {\bibfield  {journal} {\bibinfo  {journal} {Biophys. J.}\ }\textbf {\bibinfo {volume} {106}},\ \bibinfo {pages} {2291} (\bibinfo {year} {2014})}\BibitemShut {NoStop}%
\bibitem [{\citenamefont {Osborne}\ \emph {et~al.}(2017)\citenamefont {Osborne}, \citenamefont {Fletcher}, \citenamefont {{Pitt-Francis}}, \citenamefont {Maini},\ and\ \citenamefont {Gavaghan}}]{Osborne2017}%
  \BibitemOpen
  \bibfield  {author} {\bibinfo {author} {\bibfnamefont {J.~M.}\ \bibnamefont {Osborne}}, \bibinfo {author} {\bibfnamefont {A.~G.}\ \bibnamefont {Fletcher}}, \bibinfo {author} {\bibfnamefont {J.~M.}\ \bibnamefont {{Pitt-Francis}}}, \bibinfo {author} {\bibfnamefont {P.~K.}\ \bibnamefont {Maini}},\ and\ \bibinfo {author} {\bibfnamefont {D.~J.}\ \bibnamefont {Gavaghan}},\ }\bibfield  {title} {\bibinfo {title} {Comparing individual-based approaches to modelling the self-organization of multicellular tissues},\ }\href {https://doi.org/10.1371/journal.pcbi.1005387} {\bibfield  {journal} {\bibinfo  {journal} {PLOS Comput. Biol.}\ }\textbf {\bibinfo {volume} {13}},\ \bibinfo {pages} {e1005387} (\bibinfo {year} {2017})}\BibitemShut {NoStop}%
\bibitem [{\citenamefont {Yan}\ and\ \citenamefont {Bi}(2019)}]{Yan2019a}%
  \BibitemOpen
  \bibfield  {author} {\bibinfo {author} {\bibfnamefont {L.}~\bibnamefont {Yan}}\ and\ \bibinfo {author} {\bibfnamefont {D.}~\bibnamefont {Bi}},\ }\bibfield  {title} {\bibinfo {title} {Multicellular rosettes drive fluid-solid transition in epithelial tissues},\ }\href {https://doi.org/10.1103/PhysRevX.9.011029} {\bibfield  {journal} {\bibinfo  {journal} {Phys. Rev. X}\ }\textbf {\bibinfo {volume} {9}},\ \bibinfo {pages} {011029} (\bibinfo {year} {2019})}\BibitemShut {NoStop}%
\bibitem [{\citenamefont {Collinet}\ and\ \citenamefont {Lecuit}(2021)}]{Collinet2021}%
  \BibitemOpen
  \bibfield  {author} {\bibinfo {author} {\bibfnamefont {C.}~\bibnamefont {Collinet}}\ and\ \bibinfo {author} {\bibfnamefont {T.}~\bibnamefont {Lecuit}},\ }\bibfield  {title} {\bibinfo {title} {Programmed and self-organized flow of information during morphogenesis},\ }\href {https://doi.org/10.1038/s41580-020-00318-6} {\bibfield  {journal} {\bibinfo  {journal} {Nat. Rev. Mol. Cell Biol.}\ }\textbf {\bibinfo {volume} {22}},\ \bibinfo {pages} {245} (\bibinfo {year} {2021})}\BibitemShut {NoStop}%
\bibitem [{\citenamefont {Dye}\ \emph {et~al.}(2021)\citenamefont {Dye}, \citenamefont {Popovi{\'c}}, \citenamefont {Iyer}, \citenamefont {Fuhrmann}, \citenamefont {{Piscitello-G{\'o}mez}}, \citenamefont {Eaton},\ and\ \citenamefont {J{\"u}licher}}]{Dye2021}%
  \BibitemOpen
  \bibfield  {author} {\bibinfo {author} {\bibfnamefont {N.~A.}\ \bibnamefont {Dye}}, \bibinfo {author} {\bibfnamefont {M.}~\bibnamefont {Popovi{\'c}}}, \bibinfo {author} {\bibfnamefont {K.~V.}\ \bibnamefont {Iyer}}, \bibinfo {author} {\bibfnamefont {J.~F.}\ \bibnamefont {Fuhrmann}}, \bibinfo {author} {\bibfnamefont {R.}~\bibnamefont {{Piscitello-G{\'o}mez}}}, \bibinfo {author} {\bibfnamefont {S.}~\bibnamefont {Eaton}},\ and\ \bibinfo {author} {\bibfnamefont {F.}~\bibnamefont {J{\"u}licher}},\ }\bibfield  {title} {\bibinfo {title} {Self-organized patterning of cell morphology via mechanosensitive feedback},\ }\href {https://doi.org/10.7554/eLife.57964} {\bibfield  {journal} {\bibinfo  {journal} {eLife}\ }\textbf {\bibinfo {volume} {10}},\ \bibinfo {pages} {e57964} (\bibinfo {year} {2021})}\BibitemShut {NoStop}%
\bibitem [{\citenamefont {Kim}\ and\ \citenamefont {Davidson}(2011)}]{Kim2011}%
  \BibitemOpen
  \bibfield  {author} {\bibinfo {author} {\bibfnamefont {H.~Y.}\ \bibnamefont {Kim}}\ and\ \bibinfo {author} {\bibfnamefont {L.~A.}\ \bibnamefont {Davidson}},\ }\bibfield  {title} {\bibinfo {title} {Punctuated actin contractions during convergent extension and their permissive regulation by the non-canonical {{Wnt-signaling}} pathway},\ }\href {https://doi.org/10.1242/jcs.067579} {\bibfield  {journal} {\bibinfo  {journal} {J. Cell Sci.}\ }\textbf {\bibinfo {volume} {124}},\ \bibinfo {pages} {635} (\bibinfo {year} {2011})}\BibitemShut {NoStop}%
\bibitem [{\citenamefont {Weng}\ \emph {et~al.}(2022)\citenamefont {Weng}, \citenamefont {Huebner},\ and\ \citenamefont {Wallingford}}]{Weng2022}%
  \BibitemOpen
  \bibfield  {author} {\bibinfo {author} {\bibfnamefont {S.}~\bibnamefont {Weng}}, \bibinfo {author} {\bibfnamefont {R.~J.}\ \bibnamefont {Huebner}},\ and\ \bibinfo {author} {\bibfnamefont {J.~B.}\ \bibnamefont {Wallingford}},\ }\bibfield  {title} {\bibinfo {title} {Convergent extension requires adhesion-dependent biomechanical integration of cell crawling and junction contraction},\ }\href {https://doi.org/10.1016/j.celrep.2022.110666} {\bibfield  {journal} {\bibinfo  {journal} {Cell Rep.}\ }\textbf {\bibinfo {volume} {39}},\ \bibinfo {pages} {110666} (\bibinfo {year} {2022})}\BibitemShut {NoStop}%
\bibitem [{\citenamefont {Mao}\ \emph {et~al.}(2022)\citenamefont {Mao}, \citenamefont {Acharya}, \citenamefont {{Rodr{\'i}guez-delaRosa}}, \citenamefont {Marchiano}, \citenamefont {Dehapiot}, \citenamefont {Al~Tanoury}, \citenamefont {Rao}, \citenamefont {{D{\'i}az-Cuadros}}, \citenamefont {Mansur}, \citenamefont {Wagner}, \citenamefont {Chardes}, \citenamefont {Gupta}, \citenamefont {Lenne}, \citenamefont {Habermann}, \citenamefont {Theodoly}, \citenamefont {Pourqui{\'e}},\ and\ \citenamefont {Schnorrer}}]{Mao2022}%
  \BibitemOpen
  \bibfield  {author} {\bibinfo {author} {\bibfnamefont {Q.}~\bibnamefont {Mao}}, \bibinfo {author} {\bibfnamefont {A.}~\bibnamefont {Acharya}}, \bibinfo {author} {\bibfnamefont {A.}~\bibnamefont {{Rodr{\'i}guez-delaRosa}}}, \bibinfo {author} {\bibfnamefont {F.}~\bibnamefont {Marchiano}}, \bibinfo {author} {\bibfnamefont {B.}~\bibnamefont {Dehapiot}}, \bibinfo {author} {\bibfnamefont {Z.}~\bibnamefont {Al~Tanoury}}, \bibinfo {author} {\bibfnamefont {J.}~\bibnamefont {Rao}}, \bibinfo {author} {\bibfnamefont {M.}~\bibnamefont {{D{\'i}az-Cuadros}}}, \bibinfo {author} {\bibfnamefont {A.}~\bibnamefont {Mansur}}, \bibinfo {author} {\bibfnamefont {E.}~\bibnamefont {Wagner}}, \bibinfo {author} {\bibfnamefont {C.}~\bibnamefont {Chardes}}, \bibinfo {author} {\bibfnamefont {V.}~\bibnamefont {Gupta}}, \bibinfo {author} {\bibfnamefont {P.-F.}\ \bibnamefont {Lenne}}, \bibinfo {author} {\bibfnamefont {B.~H.}\ \bibnamefont {Habermann}}, \bibinfo {author} {\bibfnamefont {O.}~\bibnamefont {Theodoly}}, \bibinfo {author}
  {\bibfnamefont {O.}~\bibnamefont {Pourqui{\'e}}},\ and\ \bibinfo {author} {\bibfnamefont {F.}~\bibnamefont {Schnorrer}},\ }\bibfield  {title} {\bibinfo {title} {Tension-driven multi-scale self-organisation in human {{iPSC-derived}} muscle fibers},\ }\href {https://doi.org/10.7554/eLife.76649} {\bibfield  {journal} {\bibinfo  {journal} {eLife}\ }\textbf {\bibinfo {volume} {11}},\ \bibinfo {pages} {e76649} (\bibinfo {year} {2022})}\BibitemShut {NoStop}%
\bibitem [{\citenamefont {Bailles}\ \emph {et~al.}(2025)\citenamefont {Bailles}, \citenamefont {Serafini}, \citenamefont {Andreas}, \citenamefont {Zechner}, \citenamefont {Modes},\ and\ \citenamefont {Tomancak}}]{Bailles2025}%
  \BibitemOpen
  \bibfield  {author} {\bibinfo {author} {\bibfnamefont {A.}~\bibnamefont {Bailles}}, \bibinfo {author} {\bibfnamefont {G.}~\bibnamefont {Serafini}}, \bibinfo {author} {\bibfnamefont {H.}~\bibnamefont {Andreas}}, \bibinfo {author} {\bibfnamefont {C.}~\bibnamefont {Zechner}}, \bibinfo {author} {\bibfnamefont {C.~D.}\ \bibnamefont {Modes}},\ and\ \bibinfo {author} {\bibfnamefont {P.}~\bibnamefont {Tomancak}},\ }\bibfield  {title} {\bibinfo {title} {Anisotropic stretch biases the self-organization of actin fibers in multicellular {{Hydra}} aggregates},\ }\href {https://doi.org/10.1073/pnas.2423437122} {\bibfield  {journal} {\bibinfo  {journal} {Proc. Natl. Acad. Sci.}\ }\textbf {\bibinfo {volume} {122}},\ \bibinfo {pages} {e2423437122} (\bibinfo {year} {2025})}\BibitemShut {NoStop}%
\bibitem [{\citenamefont {Moshe}(2018)}]{Moshe2018}%
  \BibitemOpen
  \bibfield  {author} {\bibinfo {author} {\bibfnamefont {M.}~\bibnamefont {Moshe}},\ }\bibfield  {title} {\bibinfo {title} {Geometric frustration and solid-solid transitions in model 2d tissue},\ }\href {https://doi.org/10.1103/PhysRevLett.120.268105} {\bibfield  {journal} {\bibinfo  {journal} {Phys. Rev. Lett.}\ }\textbf {\bibinfo {volume} {120}},\ \bibinfo {pages} {268105} (\bibinfo {year} {2018})}\BibitemShut {NoStop}%
\bibitem [{\citenamefont {Merkel}\ \emph {et~al.}(2019)\citenamefont {Merkel}, \citenamefont {Baumgarten}, \citenamefont {Tighe},\ and\ \citenamefont {Manning}}]{Merkel2019}%
  \BibitemOpen
  \bibfield  {author} {\bibinfo {author} {\bibfnamefont {M.}~\bibnamefont {Merkel}}, \bibinfo {author} {\bibfnamefont {K.}~\bibnamefont {Baumgarten}}, \bibinfo {author} {\bibfnamefont {B.~P.}\ \bibnamefont {Tighe}},\ and\ \bibinfo {author} {\bibfnamefont {M.~L.}\ \bibnamefont {Manning}},\ }\bibfield  {title} {\bibinfo {title} {A minimal-length approach unifies rigidity in underconstrained materials},\ }\href {https://doi.org/10.1073/pnas.1815436116} {\bibfield  {journal} {\bibinfo  {journal} {Proc. Natl. Acad. Sci.}\ }\textbf {\bibinfo {volume} {116}},\ \bibinfo {pages} {6560} (\bibinfo {year} {2019})}\BibitemShut {NoStop}%
\bibitem [{\citenamefont {Golubovi{\'c}}\ and\ \citenamefont {Lubensky}(1989)}]{Golubovic1989}%
  \BibitemOpen
  \bibfield  {author} {\bibinfo {author} {\bibfnamefont {L.}~\bibnamefont {Golubovi{\'c}}}\ and\ \bibinfo {author} {\bibfnamefont {T.~C.}\ \bibnamefont {Lubensky}},\ }\bibfield  {title} {\bibinfo {title} {Nonlinear elasticity of amorphous solids},\ }\href {https://doi.org/10.1103/PhysRevLett.63.1082} {\bibfield  {journal} {\bibinfo  {journal} {Phys. Rev. Lett.}\ }\textbf {\bibinfo {volume} {63}},\ \bibinfo {pages} {1082} (\bibinfo {year} {1989})}\BibitemShut {NoStop}%
\bibitem [{\citenamefont {Warner}\ \emph {et~al.}(1994)\citenamefont {Warner}, \citenamefont {Bladon},\ and\ \citenamefont {Terentjev}}]{Warner1994}%
  \BibitemOpen
  \bibfield  {author} {\bibinfo {author} {\bibfnamefont {M.}~\bibnamefont {Warner}}, \bibinfo {author} {\bibfnamefont {P.}~\bibnamefont {Bladon}},\ and\ \bibinfo {author} {\bibfnamefont {E.~M.}\ \bibnamefont {Terentjev}},\ }\bibfield  {title} {\bibinfo {title} {``{{Soft}} elasticity'' {\textemdash} deformation without resistance in liquid crystal elastomers},\ }\href {https://doi.org/10.1051/jp2:1994116} {\bibfield  {journal} {\bibinfo  {journal} {J. Phys. II}\ }\textbf {\bibinfo {volume} {4}},\ \bibinfo {pages} {93} (\bibinfo {year} {1994})}\BibitemShut {NoStop}%
\bibitem [{\citenamefont {Koski}\ \emph {et~al.}(2012)\citenamefont {Koski}, \citenamefont {McKiernan}, \citenamefont {Akhenblit},\ and\ \citenamefont {Yarger}}]{Koski2012}%
  \BibitemOpen
  \bibfield  {author} {\bibinfo {author} {\bibfnamefont {K.~J.}\ \bibnamefont {Koski}}, \bibinfo {author} {\bibfnamefont {K.}~\bibnamefont {McKiernan}}, \bibinfo {author} {\bibfnamefont {P.}~\bibnamefont {Akhenblit}},\ and\ \bibinfo {author} {\bibfnamefont {J.~L.}\ \bibnamefont {Yarger}},\ }\bibfield  {title} {\bibinfo {title} {Shear-induced rigidity in spider silk glands},\ }\href {https://doi.org/10.1063/1.4751842} {\bibfield  {journal} {\bibinfo  {journal} {Appl. Phys. Lett.}\ }\textbf {\bibinfo {volume} {101}},\ \bibinfo {pages} {103701} (\bibinfo {year} {2012})}\BibitemShut {NoStop}%
\bibitem [{\citenamefont {Huang}\ \emph {et~al.}(2022)\citenamefont {Huang}, \citenamefont {Cochran}, \citenamefont {Fielding}, \citenamefont {Marchetti},\ and\ \citenamefont {Bi}}]{Huang2022a}%
  \BibitemOpen
  \bibfield  {author} {\bibinfo {author} {\bibfnamefont {J.}~\bibnamefont {Huang}}, \bibinfo {author} {\bibfnamefont {J.~O.}\ \bibnamefont {Cochran}}, \bibinfo {author} {\bibfnamefont {S.~M.}\ \bibnamefont {Fielding}}, \bibinfo {author} {\bibfnamefont {M.~C.}\ \bibnamefont {Marchetti}},\ and\ \bibinfo {author} {\bibfnamefont {D.}~\bibnamefont {Bi}},\ }\bibfield  {title} {\bibinfo {title} {Shear-driven solidification and nonlinear elasticity in epithelial tissues},\ }\href {https://doi.org/10.1103/PhysRevLett.128.178001} {\bibfield  {journal} {\bibinfo  {journal} {Phys. Rev. Lett.}\ }\textbf {\bibinfo {volume} {128}},\ \bibinfo {pages} {178001} (\bibinfo {year} {2022})}\BibitemShut {NoStop}%
\bibitem [{\citenamefont {Fielding}\ \emph {et~al.}(2023)\citenamefont {Fielding}, \citenamefont {Cochran}, \citenamefont {Huang}, \citenamefont {Bi},\ and\ \citenamefont {Marchetti}}]{Fielding2023}%
  \BibitemOpen
  \bibfield  {author} {\bibinfo {author} {\bibfnamefont {S.~M.}\ \bibnamefont {Fielding}}, \bibinfo {author} {\bibfnamefont {J.~O.}\ \bibnamefont {Cochran}}, \bibinfo {author} {\bibfnamefont {J.}~\bibnamefont {Huang}}, \bibinfo {author} {\bibfnamefont {D.}~\bibnamefont {Bi}},\ and\ \bibinfo {author} {\bibfnamefont {M.~C.}\ \bibnamefont {Marchetti}},\ }\bibfield  {title} {\bibinfo {title} {Constitutive model for the rheology of biological tissue},\ }\href {https://doi.org/10.1103/PhysRevE.108.L042602} {\bibfield  {journal} {\bibinfo  {journal} {Phys. Rev. E}\ }\textbf {\bibinfo {volume} {108}},\ \bibinfo {pages} {L042602} (\bibinfo {year} {2023})}\BibitemShut {NoStop}%
\bibitem [{\citenamefont {Brauns}\ \emph {et~al.}(2024)\citenamefont {Brauns}, \citenamefont {Claussen}, \citenamefont {Lefebvre}, \citenamefont {Wieschaus},\ and\ \citenamefont {Shraiman}}]{Brauns2024a}%
  \BibitemOpen
  \bibfield  {author} {\bibinfo {author} {\bibfnamefont {F.}~\bibnamefont {Brauns}}, \bibinfo {author} {\bibfnamefont {N.~H.}\ \bibnamefont {Claussen}}, \bibinfo {author} {\bibfnamefont {M.~F.}\ \bibnamefont {Lefebvre}}, \bibinfo {author} {\bibfnamefont {E.~F.}\ \bibnamefont {Wieschaus}},\ and\ \bibinfo {author} {\bibfnamefont {B.~I.}\ \bibnamefont {Shraiman}},\ }\bibfield  {title} {\bibinfo {title} {The geometric basis of epithelial convergent extension},\ }\href {https://doi.org/10.7554/eLife.95521} {\bibfield  {journal} {\bibinfo  {journal} {eLife}\ }\textbf {\bibinfo {volume} {13}},\ \bibinfo {pages} {RP95521} (\bibinfo {year} {2024})}\BibitemShut {NoStop}%
\bibitem [{\citenamefont {Alert}\ \emph {et~al.}(2020)\citenamefont {Alert}, \citenamefont {Joanny},\ and\ \citenamefont {Casademunt}}]{Alert2020a}%
  \BibitemOpen
  \bibfield  {author} {\bibinfo {author} {\bibfnamefont {R.}~\bibnamefont {Alert}}, \bibinfo {author} {\bibfnamefont {J.-F.}\ \bibnamefont {Joanny}},\ and\ \bibinfo {author} {\bibfnamefont {J.}~\bibnamefont {Casademunt}},\ }\bibfield  {title} {\bibinfo {title} {Universal scaling of active nematic turbulence},\ }\href {https://doi.org/10.1038/s41567-020-0854-4} {\bibfield  {journal} {\bibinfo  {journal} {Nat. Phys.}\ }\textbf {\bibinfo {volume} {16}},\ \bibinfo {pages} {682} (\bibinfo {year} {2020})}\BibitemShut {NoStop}%
\bibitem [{\citenamefont {Grigas}\ \emph {et~al.}(2025)\citenamefont {Grigas}, \citenamefont {Negi}, \citenamefont {Maniou}, \citenamefont {Galea}, \citenamefont {Michaut}, \citenamefont {Mongera},\ and\ \citenamefont {Manning}}]{Grigas2025}%
  \BibitemOpen
  \bibfield  {author} {\bibinfo {author} {\bibfnamefont {A.~T.}\ \bibnamefont {Grigas}}, \bibinfo {author} {\bibfnamefont {R.~S.}\ \bibnamefont {Negi}}, \bibinfo {author} {\bibfnamefont {E.}~\bibnamefont {Maniou}}, \bibinfo {author} {\bibfnamefont {G.~L.}\ \bibnamefont {Galea}}, \bibinfo {author} {\bibfnamefont {A.}~\bibnamefont {Michaut}}, \bibinfo {author} {\bibfnamefont {A.}~\bibnamefont {Mongera}},\ and\ \bibinfo {author} {\bibfnamefont {M.~L.}\ \bibnamefont {Manning}},\ }\bibfield  {title} {\bibinfo {title} {Sparse mesenchymal cell networks as a fluid under tension},\ }\bibfield  {journal} {\bibinfo  {journal} {bioRxiv}\ }\href {https://doi.org/10.64898/2025.12.07.692626} {10.64898/2025.12.07.692626} (\bibinfo {year} {2025})\BibitemShut {NoStop}%
\bibitem [{\citenamefont {Kawaguchi}\ \emph {et~al.}(2017)\citenamefont {Kawaguchi}, \citenamefont {Kageyama},\ and\ \citenamefont {Sano}}]{Kawaguchi2017}%
  \BibitemOpen
  \bibfield  {author} {\bibinfo {author} {\bibfnamefont {K.}~\bibnamefont {Kawaguchi}}, \bibinfo {author} {\bibfnamefont {R.}~\bibnamefont {Kageyama}},\ and\ \bibinfo {author} {\bibfnamefont {M.}~\bibnamefont {Sano}},\ }\bibfield  {title} {\bibinfo {title} {Topological defects control collective dynamics in neural progenitor cell cultures},\ }\href@noop {} {\bibfield  {journal} {\bibinfo  {journal} {Nature}\ }\textbf {\bibinfo {volume} {545}},\ \bibinfo {pages} {327} (\bibinfo {year} {2017})}\BibitemShut {NoStop}%
\bibitem [{\citenamefont {Farhadifar}\ \emph {et~al.}(2007)\citenamefont {Farhadifar}, \citenamefont {R{\"o}per}, \citenamefont {Aigouy}, \citenamefont {Eaton},\ and\ \citenamefont {J{\"u}licher}}]{Farhadifar2007}%
  \BibitemOpen
  \bibfield  {author} {\bibinfo {author} {\bibfnamefont {R.}~\bibnamefont {Farhadifar}}, \bibinfo {author} {\bibfnamefont {J.-C.}\ \bibnamefont {R{\"o}per}}, \bibinfo {author} {\bibfnamefont {B.}~\bibnamefont {Aigouy}}, \bibinfo {author} {\bibfnamefont {S.}~\bibnamefont {Eaton}},\ and\ \bibinfo {author} {\bibfnamefont {F.}~\bibnamefont {J{\"u}licher}},\ }\bibfield  {title} {\bibinfo {title} {The influence of cell mechanics, cell-cell {{Interactions}}, and proliferation on epithelial packing},\ }\href {https://doi.org/10.1016/j.cub.2007.11.049} {\bibfield  {journal} {\bibinfo  {journal} {Curr. Biol.}\ }\textbf {\bibinfo {volume} {17}},\ \bibinfo {pages} {2095} (\bibinfo {year} {2007})}\BibitemShut {NoStop}%
\bibitem [{\citenamefont {Staple}\ \emph {et~al.}(2010)\citenamefont {Staple}, \citenamefont {Farhadifar}, \citenamefont {R{\"o}per}, \citenamefont {Aigouy}, \citenamefont {Eaton},\ and\ \citenamefont {J{\"u}licher}}]{Staple2010}%
  \BibitemOpen
  \bibfield  {author} {\bibinfo {author} {\bibfnamefont {D.~B.}\ \bibnamefont {Staple}}, \bibinfo {author} {\bibfnamefont {R.}~\bibnamefont {Farhadifar}}, \bibinfo {author} {\bibfnamefont {J.~C.}\ \bibnamefont {R{\"o}per}}, \bibinfo {author} {\bibfnamefont {B.}~\bibnamefont {Aigouy}}, \bibinfo {author} {\bibfnamefont {S.}~\bibnamefont {Eaton}},\ and\ \bibinfo {author} {\bibfnamefont {F.}~\bibnamefont {J{\"u}licher}},\ }\bibfield  {title} {\bibinfo {title} {Mechanics and remodelling of cell packings in epithelia},\ }\href {https://doi.org/10.1140/epje/i2010-10677-0} {\bibfield  {journal} {\bibinfo  {journal} {Eur. Phys. J. E}\ }\textbf {\bibinfo {volume} {33}},\ \bibinfo {pages} {117} (\bibinfo {year} {2010})}\BibitemShut {NoStop}%
\bibitem [{\citenamefont {Brodland}(2002)}]{Brodland2002}%
  \BibitemOpen
  \bibfield  {author} {\bibinfo {author} {\bibfnamefont {G.~W.}\ \bibnamefont {Brodland}},\ }\bibfield  {title} {\bibinfo {title} {The differential interfacial tension hypothesis {{(DITH)}}: a comprehensive theory for the self-rearrangement of embryonic cells and tissues},\ }\href {https://doi.org/10.1115/1.1449491} {\bibfield  {journal} {\bibinfo  {journal} {J. Biomech. Eng.}\ }\textbf {\bibinfo {volume} {124}},\ \bibinfo {pages} {188} (\bibinfo {year} {2002})}\BibitemShut {NoStop}%
\bibitem [{\citenamefont {Lecuit}\ and\ \citenamefont {Lenne}(2007)}]{Lecuit2007}%
  \BibitemOpen
  \bibfield  {author} {\bibinfo {author} {\bibfnamefont {T.}~\bibnamefont {Lecuit}}\ and\ \bibinfo {author} {\bibfnamefont {P.-F.}\ \bibnamefont {Lenne}},\ }\bibfield  {title} {\bibinfo {title} {Cell surface mechanics and the control of cell shape, tissue patterns and morphogenesis},\ }\href {https://doi.org/10.1038/nrm2222} {\bibfield  {journal} {\bibinfo  {journal} {Nat. Rev. Mol. Cell Biol.}\ }\textbf {\bibinfo {volume} {8}},\ \bibinfo {pages} {633} (\bibinfo {year} {2007})}\BibitemShut {NoStop}%
\bibitem [{\citenamefont {Yang}\ \emph {et~al.}(2017)\citenamefont {Yang}, \citenamefont {Bi}, \citenamefont {Czajkowski}, \citenamefont {Merkel}, \citenamefont {Manning},\ and\ \citenamefont {Marchetti}}]{Yang2017}%
  \BibitemOpen
  \bibfield  {author} {\bibinfo {author} {\bibfnamefont {X.}~\bibnamefont {Yang}}, \bibinfo {author} {\bibfnamefont {D.}~\bibnamefont {Bi}}, \bibinfo {author} {\bibfnamefont {M.}~\bibnamefont {Czajkowski}}, \bibinfo {author} {\bibfnamefont {M.}~\bibnamefont {Merkel}}, \bibinfo {author} {\bibfnamefont {M.~L.}\ \bibnamefont {Manning}},\ and\ \bibinfo {author} {\bibfnamefont {M.~C.}\ \bibnamefont {Marchetti}},\ }\bibfield  {title} {\bibinfo {title} {Correlating cell shape and cellular stress in motile confluent tissues},\ }\href {https://doi.org/10.1073/pnas.1705921114} {\bibfield  {journal} {\bibinfo  {journal} {Proc. Natl. Acad. Sci.}\ }\textbf {\bibinfo {volume} {114}},\ \bibinfo {pages} {12663} (\bibinfo {year} {2017})}\BibitemShut {NoStop}%
\bibitem [{\citenamefont {{Lawson-Keister}}\ and\ \citenamefont {Manning}(2022)}]{Lawson-Keister2022a}%
  \BibitemOpen
  \bibfield  {author} {\bibinfo {author} {\bibfnamefont {E.}~\bibnamefont {{Lawson-Keister}}}\ and\ \bibinfo {author} {\bibfnamefont {M.~L.}\ \bibnamefont {Manning}},\ }\bibfield  {title} {\bibinfo {title} {Collective chemotaxis in a {{Voronoi}} model for confluent clusters},\ }\href {https://doi.org/10.1016/j.bpj.2022.10.029} {\bibfield  {journal} {\bibinfo  {journal} {Biophys. J.}\ }\textbf {\bibinfo {volume} {121}},\ \bibinfo {pages} {4624} (\bibinfo {year} {2022})}\BibitemShut {NoStop}%
\bibitem [{Note1()}]{Note1}%
  \BibitemOpen
  \bibinfo {note} {While Voronoi-type models offer greater simplicity due to their reduced number of geometric degrees of freedom, they do not correctly capture the (im)balance of forces at vertices. They are therefore not suitable to describe the subtle interplay of active and passive mechanical forces that governs the rich rheological behavior found in our model.}\BibitemShut {Stop}%
\bibitem [{\citenamefont {Tlili}\ \emph {et~al.}(2019)\citenamefont {Tlili}, \citenamefont {Yin}, \citenamefont {Rupprecht}, \citenamefont {{Mendieta-Serrano}}, \citenamefont {Weissbart}, \citenamefont {Verma}, \citenamefont {Teng}, \citenamefont {Toyama}, \citenamefont {Prost},\ and\ \citenamefont {Saunders}}]{Tlili2019a}%
  \BibitemOpen
  \bibfield  {author} {\bibinfo {author} {\bibfnamefont {S.}~\bibnamefont {Tlili}}, \bibinfo {author} {\bibfnamefont {J.}~\bibnamefont {Yin}}, \bibinfo {author} {\bibfnamefont {J.-F.}\ \bibnamefont {Rupprecht}}, \bibinfo {author} {\bibfnamefont {M.~A.}\ \bibnamefont {{Mendieta-Serrano}}}, \bibinfo {author} {\bibfnamefont {G.}~\bibnamefont {Weissbart}}, \bibinfo {author} {\bibfnamefont {N.}~\bibnamefont {Verma}}, \bibinfo {author} {\bibfnamefont {X.}~\bibnamefont {Teng}}, \bibinfo {author} {\bibfnamefont {Y.}~\bibnamefont {Toyama}}, \bibinfo {author} {\bibfnamefont {J.}~\bibnamefont {Prost}},\ and\ \bibinfo {author} {\bibfnamefont {T.~E.}\ \bibnamefont {Saunders}},\ }\bibfield  {title} {\bibinfo {title} {Shaping the zebrafish myotome by intertissue friction and active stress},\ }\href {https://doi.org/10.1073/pnas.1900819116} {\bibfield  {journal} {\bibinfo  {journal} {Proc. Natl. Acad. Sci.}\ }\textbf {\bibinfo {volume} {116}},\ \bibinfo {pages} {25430} (\bibinfo {year} {2019})}\BibitemShut {NoStop}%
\bibitem [{\citenamefont {Lin}\ \emph {et~al.}(2022)\citenamefont {Lin}, \citenamefont {Merkel},\ and\ \citenamefont {Rupprecht}}]{Lin2022}%
  \BibitemOpen
  \bibfield  {author} {\bibinfo {author} {\bibfnamefont {S.-Z.}\ \bibnamefont {Lin}}, \bibinfo {author} {\bibfnamefont {M.}~\bibnamefont {Merkel}},\ and\ \bibinfo {author} {\bibfnamefont {J.-F.}\ \bibnamefont {Rupprecht}},\ }\bibfield  {title} {\bibinfo {title} {Implementation of cellular bulk stresses in vertex models of biological tissues},\ }\href {https://doi.org/10.1140/epje/s10189-021-00154-2} {\bibfield  {journal} {\bibinfo  {journal} {Eur. Phys. J. E}\ }\textbf {\bibinfo {volume} {45}},\ \bibinfo {pages} {4} (\bibinfo {year} {2022})}\BibitemShut {NoStop}%
\bibitem [{\citenamefont {Aditi~Simha}\ and\ \citenamefont {Ramaswamy}(2002)}]{AditiSimha2002}%
  \BibitemOpen
  \bibfield  {author} {\bibinfo {author} {\bibfnamefont {R.}~\bibnamefont {Aditi~Simha}}\ and\ \bibinfo {author} {\bibfnamefont {S.}~\bibnamefont {Ramaswamy}},\ }\bibfield  {title} {\bibinfo {title} {Hydrodynamic fluctuations and instabilities in ordered suspensions of self-propelled particles},\ }\href {https://doi.org/10.1103/PhysRevLett.89.058101} {\bibfield  {journal} {\bibinfo  {journal} {Phys. Rev. Lett.}\ }\textbf {\bibinfo {volume} {89}},\ \bibinfo {pages} {58101} (\bibinfo {year} {2002})}\BibitemShut {NoStop}%
\bibitem [{\citenamefont {Giomi}\ \emph {et~al.}(2013)\citenamefont {Giomi}, \citenamefont {Bowick}, \citenamefont {Ma},\ and\ \citenamefont {Marchetti}}]{Giomi2013}%
  \BibitemOpen
  \bibfield  {author} {\bibinfo {author} {\bibfnamefont {L.}~\bibnamefont {Giomi}}, \bibinfo {author} {\bibfnamefont {M.~J.}\ \bibnamefont {Bowick}}, \bibinfo {author} {\bibfnamefont {X.}~\bibnamefont {Ma}},\ and\ \bibinfo {author} {\bibfnamefont {M.~C.}\ \bibnamefont {Marchetti}},\ }\bibfield  {title} {\bibinfo {title} {Defect annihilation and proliferation in active nematics},\ }\href {https://doi.org/10.1103/PhysRevLett.110.228101} {\bibfield  {journal} {\bibinfo  {journal} {Phys. Rev. Lett.}\ }\textbf {\bibinfo {volume} {110}},\ \bibinfo {pages} {228101} (\bibinfo {year} {2013})}\BibitemShut {NoStop}%
\bibitem [{\citenamefont {Brauns}\ \emph {et~al.}(2026)\citenamefont {Brauns}, \citenamefont {O'Leary}, \citenamefont {Hernandez}, \citenamefont {Bowick},\ and\ \citenamefont {Marchetti}}]{Brauns2026}%
  \BibitemOpen
  \bibfield  {author} {\bibinfo {author} {\bibfnamefont {F.}~\bibnamefont {Brauns}}, \bibinfo {author} {\bibfnamefont {M.}~\bibnamefont {O'Leary}}, \bibinfo {author} {\bibfnamefont {A.}~\bibnamefont {Hernandez}}, \bibinfo {author} {\bibfnamefont {M.~J.}\ \bibnamefont {Bowick}},\ and\ \bibinfo {author} {\bibfnamefont {M.~C.}\ \bibnamefont {Marchetti}},\ }\bibfield  {title} {\bibinfo {title} {Active solids: topological defect self-propulsion without flow},\ }\href {https://doi.org/10.1103/xv94-xpz2} {\bibfield  {journal} {\bibinfo  {journal} {Phys. Rev. Lett.}\ }\textbf {\bibinfo {volume} {136}},\ \bibinfo {pages} {058302} (\bibinfo {year} {2026})}\BibitemShut {NoStop}%
\bibitem [{\citenamefont {Batchelor}(1970)}]{Batchelor1970}%
  \BibitemOpen
  \bibfield  {author} {\bibinfo {author} {\bibfnamefont {G.~K.}\ \bibnamefont {Batchelor}},\ }\bibfield  {title} {\bibinfo {title} {The stress system in a suspension of force-free particles},\ }\href {https://doi.org/10.1017/S0022112070000745} {\bibfield  {journal} {\bibinfo  {journal} {J. Fluid Mech.}\ }\textbf {\bibinfo {volume} {41}},\ \bibinfo {pages} {545} (\bibinfo {year} {1970})}\BibitemShut {NoStop}%
\bibitem [{Note2()}]{Note2}%
  \BibitemOpen
  \bibinfo {note} {Negative $T_e$ would correspond to compressive stress on junctions which would cause them to buckle/wrinkle. This is experimentally observed only in rare exceptions.}\BibitemShut {Stop}%
\bibitem [{\citenamefont {Lecuit}\ \emph {et~al.}(2011)\citenamefont {Lecuit}, \citenamefont {Lenne},\ and\ \citenamefont {Munro}}]{Lecuit2011}%
  \BibitemOpen
  \bibfield  {author} {\bibinfo {author} {\bibfnamefont {T.}~\bibnamefont {Lecuit}}, \bibinfo {author} {\bibfnamefont {P.-F.}\ \bibnamefont {Lenne}},\ and\ \bibinfo {author} {\bibfnamefont {E.}~\bibnamefont {Munro}},\ }\bibfield  {title} {\bibinfo {title} {Force generation, transmission, and integration during cell and tissue morphogenesis},\ }\href {https://doi.org/10.1146/annurev-cellbio-100109-104027} {\bibfield  {journal} {\bibinfo  {journal} {Annu. Rev. Cell Dev. Biol.}\ }\textbf {\bibinfo {volume} {27}},\ \bibinfo {pages} {157} (\bibinfo {year} {2011})}\BibitemShut {NoStop}%
\bibitem [{\citenamefont {Caswell}\ and\ \citenamefont {Zech}(2018)}]{Caswell2018}%
  \BibitemOpen
  \bibfield  {author} {\bibinfo {author} {\bibfnamefont {P.~T.}\ \bibnamefont {Caswell}}\ and\ \bibinfo {author} {\bibfnamefont {T.}~\bibnamefont {Zech}},\ }\bibfield  {title} {\bibinfo {title} {Actin-based cell protrusion in a 3d matrix},\ }\href {https://doi.org/10.1016/j.tcb.2018.06.003} {\bibfield  {journal} {\bibinfo  {journal} {Trends Cell Biol.}\ }\textbf {\bibinfo {volume} {28}},\ \bibinfo {pages} {823} (\bibinfo {year} {2018})}\BibitemShut {NoStop}%
\bibitem [{\citenamefont {Picone}\ \emph {et~al.}(2010)\citenamefont {Picone}, \citenamefont {Ren}, \citenamefont {Ivanovitch}, \citenamefont {Clarke}, \citenamefont {McKendry},\ and\ \citenamefont {Baum}}]{Picone2010}%
  \BibitemOpen
  \bibfield  {author} {\bibinfo {author} {\bibfnamefont {R.}~\bibnamefont {Picone}}, \bibinfo {author} {\bibfnamefont {X.}~\bibnamefont {Ren}}, \bibinfo {author} {\bibfnamefont {K.~D.}\ \bibnamefont {Ivanovitch}}, \bibinfo {author} {\bibfnamefont {J.~D.~W.}\ \bibnamefont {Clarke}}, \bibinfo {author} {\bibfnamefont {R.~A.}\ \bibnamefont {McKendry}},\ and\ \bibinfo {author} {\bibfnamefont {B.}~\bibnamefont {Baum}},\ }\bibfield  {title} {\bibinfo {title} {A polarised population of dynamic microtubules mediates homeostatic length control in animal cells},\ }\href {https://doi.org/10.1371/journal.pbio.1000542} {\bibfield  {journal} {\bibinfo  {journal} {PLOS Biol.}\ }\textbf {\bibinfo {volume} {8}},\ \bibinfo {pages} {e1000542} (\bibinfo {year} {2010})}\BibitemShut {NoStop}%
\bibitem [{\citenamefont {Singh}\ \emph {et~al.}(2018)\citenamefont {Singh}, \citenamefont {Saha}, \citenamefont {Begemann}, \citenamefont {Ricker}, \citenamefont {N{\"u}sse}, \citenamefont {{Thorn-Seshold}}, \citenamefont {Klingauf}, \citenamefont {Galic},\ and\ \citenamefont {Matis}}]{Singh2018}%
  \BibitemOpen
  \bibfield  {author} {\bibinfo {author} {\bibfnamefont {A.}~\bibnamefont {Singh}}, \bibinfo {author} {\bibfnamefont {T.}~\bibnamefont {Saha}}, \bibinfo {author} {\bibfnamefont {I.}~\bibnamefont {Begemann}}, \bibinfo {author} {\bibfnamefont {A.}~\bibnamefont {Ricker}}, \bibinfo {author} {\bibfnamefont {H.}~\bibnamefont {N{\"u}sse}}, \bibinfo {author} {\bibfnamefont {O.}~\bibnamefont {{Thorn-Seshold}}}, \bibinfo {author} {\bibfnamefont {J.}~\bibnamefont {Klingauf}}, \bibinfo {author} {\bibfnamefont {M.}~\bibnamefont {Galic}},\ and\ \bibinfo {author} {\bibfnamefont {M.}~\bibnamefont {Matis}},\ }\bibfield  {title} {\bibinfo {title} {Polarized microtubule dynamics directs cell mechanics and coordinates forces during epithelial morphogenesis},\ }\href {https://doi.org/10.1038/s41556-018-0193-1} {\bibfield  {journal} {\bibinfo  {journal} {Nat. Cell Biol.}\ }\textbf {\bibinfo {volume} {20}},\ \bibinfo {pages} {1126} (\bibinfo {year} {2018})}\BibitemShut {NoStop}%
\bibitem [{\citenamefont {Sadeghipour}\ \emph {et~al.}(2018)\citenamefont {Sadeghipour}, \citenamefont {Garcia}, \citenamefont {Nelson},\ and\ \citenamefont {Pruitt}}]{Sadeghipour2018}%
  \BibitemOpen
  \bibfield  {author} {\bibinfo {author} {\bibfnamefont {E.}~\bibnamefont {Sadeghipour}}, \bibinfo {author} {\bibfnamefont {M.~A.}\ \bibnamefont {Garcia}}, \bibinfo {author} {\bibfnamefont {W.~J.}\ \bibnamefont {Nelson}},\ and\ \bibinfo {author} {\bibfnamefont {B.~L.}\ \bibnamefont {Pruitt}},\ }\bibfield  {title} {\bibinfo {title} {Shear-induced damped oscillations in an epithelium depend on actomyosin contraction and {{E-cadherin}} cell adhesion},\ }\href {https://doi.org/10.7554/eLife.39640} {\bibfield  {journal} {\bibinfo  {journal} {eLife}\ }\textbf {\bibinfo {volume} {7}},\ \bibinfo {pages} {e39640} (\bibinfo {year} {2018})}\BibitemShut {NoStop}%
\bibitem [{\citenamefont {Hertaeg}\ \emph {et~al.}(2024)\citenamefont {Hertaeg}, \citenamefont {Fielding},\ and\ \citenamefont {Bi}}]{Hertaeg2024}%
  \BibitemOpen
  \bibfield  {author} {\bibinfo {author} {\bibfnamefont {M.~J.}\ \bibnamefont {Hertaeg}}, \bibinfo {author} {\bibfnamefont {S.~M.}\ \bibnamefont {Fielding}},\ and\ \bibinfo {author} {\bibfnamefont {D.}~\bibnamefont {Bi}},\ }\bibfield  {title} {\bibinfo {title} {Discontinuous shear thickening in biological tissue rheology},\ }\href {https://doi.org/10.1103/PhysRevX.14.011027} {\bibfield  {journal} {\bibinfo  {journal} {Phys. Rev. X}\ }\textbf {\bibinfo {volume} {14}},\ \bibinfo {pages} {011027} (\bibinfo {year} {2024})}\BibitemShut {NoStop}%
\bibitem [{\citenamefont {Grossman}\ and\ \citenamefont {Joanny}(2025)}]{Grossman2025}%
  \BibitemOpen
  \bibfield  {author} {\bibinfo {author} {\bibfnamefont {D.}~\bibnamefont {Grossman}}\ and\ \bibinfo {author} {\bibfnamefont {J.-F.}\ \bibnamefont {Joanny}},\ }\bibfield  {title} {\bibinfo {title} {Rheology of the vertex model of tissues: simple shear and oscillatory geometries},\ }\href {https://doi.org/10.1103/PhysRevResearch.7.013039} {\bibfield  {journal} {\bibinfo  {journal} {Phys. Rev. Res.}\ }\textbf {\bibinfo {volume} {7}},\ \bibinfo {pages} {013039} (\bibinfo {year} {2025})}\BibitemShut {NoStop}%
\bibitem [{\citenamefont {Nguyen}\ \emph {et~al.}(2025)\citenamefont {Nguyen}, \citenamefont {Huang},\ and\ \citenamefont {Bi}}]{Nguyen2025}%
  \BibitemOpen
  \bibfield  {author} {\bibinfo {author} {\bibfnamefont {A.~Q.}\ \bibnamefont {Nguyen}}, \bibinfo {author} {\bibfnamefont {J.}~\bibnamefont {Huang}},\ and\ \bibinfo {author} {\bibfnamefont {D.}~\bibnamefont {Bi}},\ }\bibfield  {title} {\bibinfo {title} {Origin of yield stress and mechanical plasticity in model biological tissues},\ }\href {https://doi.org/10.1038/s41467-025-58526-7} {\bibfield  {journal} {\bibinfo  {journal} {Nat. Commun.}\ }\textbf {\bibinfo {volume} {16}},\ \bibinfo {pages} {3260} (\bibinfo {year} {2025})}\BibitemShut {NoStop}%
\bibitem [{\citenamefont {Berret}\ \emph {et~al.}(1994)\citenamefont {Berret}, \citenamefont {Roux}, \citenamefont {Porte},\ and\ \citenamefont {Lindner}}]{Berret1994}%
  \BibitemOpen
  \bibfield  {author} {\bibinfo {author} {\bibfnamefont {J.-F.}\ \bibnamefont {Berret}}, \bibinfo {author} {\bibfnamefont {D.~C.}\ \bibnamefont {Roux}}, \bibinfo {author} {\bibfnamefont {G.}~\bibnamefont {Porte}},\ and\ \bibinfo {author} {\bibfnamefont {P.}~\bibnamefont {Lindner}},\ }\bibfield  {title} {\bibinfo {title} {Shear-induced isotropic-to-nematic phase transition in equilibrium polymers},\ }\href {https://doi.org/10.1209/0295-5075/25/7/008} {\bibfield  {journal} {\bibinfo  {journal} {Europhys. Lett.}\ }\textbf {\bibinfo {volume} {25}},\ \bibinfo {pages} {521} (\bibinfo {year} {1994})}\BibitemShut {NoStop}%
\bibitem [{\citenamefont {Finkelmann}\ \emph {et~al.}(1997)\citenamefont {Finkelmann}, \citenamefont {Kundler}, \citenamefont {Terentjev},\ and\ \citenamefont {Warner}}]{Finkelmann1997}%
  \BibitemOpen
  \bibfield  {author} {\bibinfo {author} {\bibfnamefont {H.}~\bibnamefont {Finkelmann}}, \bibinfo {author} {\bibfnamefont {I.}~\bibnamefont {Kundler}}, \bibinfo {author} {\bibfnamefont {E.~M.}\ \bibnamefont {Terentjev}},\ and\ \bibinfo {author} {\bibfnamefont {M.}~\bibnamefont {Warner}},\ }\bibfield  {title} {\bibinfo {title} {Critical stripe-domain instability of nematic elastomers},\ }\href {https://doi.org/10.1051/jp2:1997171} {\bibfield  {journal} {\bibinfo  {journal} {J. Phys. II}\ }\textbf {\bibinfo {volume} {7}},\ \bibinfo {pages} {1059} (\bibinfo {year} {1997})}\BibitemShut {NoStop}%
\bibitem [{\citenamefont {Berezney}\ \emph {et~al.}(2026)\citenamefont {Berezney}, \citenamefont {Ray}, \citenamefont {Kolvin}, \citenamefont {Brauns}, \citenamefont {Chen}, \citenamefont {Bowick}, \citenamefont {Fraden}, \citenamefont {Vitelli},\ and\ \citenamefont {Dogic}}]{Berezney2026}%
  \BibitemOpen
  \bibfield  {author} {\bibinfo {author} {\bibfnamefont {J.}~\bibnamefont {Berezney}}, \bibinfo {author} {\bibfnamefont {S.}~\bibnamefont {Ray}}, \bibinfo {author} {\bibfnamefont {I.}~\bibnamefont {Kolvin}}, \bibinfo {author} {\bibfnamefont {F.}~\bibnamefont {Brauns}}, \bibinfo {author} {\bibfnamefont {S.}~\bibnamefont {Chen}}, \bibinfo {author} {\bibfnamefont {M.}~\bibnamefont {Bowick}}, \bibinfo {author} {\bibfnamefont {S.}~\bibnamefont {Fraden}}, \bibinfo {author} {\bibfnamefont {V.}~\bibnamefont {Vitelli}},\ and\ \bibinfo {author} {\bibfnamefont {Z.}~\bibnamefont {Dogic}},\ }\bibfield  {title} {\bibinfo {title} {Active assembly and non-reciprocal dynamics of elastic membranes},\ }\href {https://doi.org/10.1038/s41567-026-03215-5} {\bibfield  {journal} {\bibinfo  {journal} {Nat. Phys.}\ }\textbf {\bibinfo {volume} {22}},\ \bibinfo {pages} {604} (\bibinfo {year} {2026})}\BibitemShut {NoStop}%
\bibitem [{\citenamefont {Baconnier}\ \emph {et~al.}(2022)\citenamefont {Baconnier}, \citenamefont {Shohat}, \citenamefont {L{\'o}pez}, \citenamefont {Coulais}, \citenamefont {D{\'e}mery}, \citenamefont {D{\"u}ring},\ and\ \citenamefont {Dauchot}}]{Baconnier2022}%
  \BibitemOpen
  \bibfield  {author} {\bibinfo {author} {\bibfnamefont {P.}~\bibnamefont {Baconnier}}, \bibinfo {author} {\bibfnamefont {D.}~\bibnamefont {Shohat}}, \bibinfo {author} {\bibfnamefont {C.~H.}\ \bibnamefont {L{\'o}pez}}, \bibinfo {author} {\bibfnamefont {C.}~\bibnamefont {Coulais}}, \bibinfo {author} {\bibfnamefont {V.}~\bibnamefont {D{\'e}mery}}, \bibinfo {author} {\bibfnamefont {G.}~\bibnamefont {D{\"u}ring}},\ and\ \bibinfo {author} {\bibfnamefont {O.}~\bibnamefont {Dauchot}},\ }\bibfield  {title} {\bibinfo {title} {Selective and collective actuation in active solids},\ }\href {https://doi.org/10.1038/s41567-022-01704-x} {\bibfield  {journal} {\bibinfo  {journal} {Nat. Phys.}\ }\textbf {\bibinfo {volume} {18}},\ \bibinfo {pages} {1234} (\bibinfo {year} {2022})}\BibitemShut {NoStop}%
\bibitem [{\citenamefont {Amiri}\ \emph {et~al.}(2023)\citenamefont {Amiri}, \citenamefont {Duclut}, \citenamefont {J{\"u}licher},\ and\ \citenamefont {Popovi{\'c}}}]{Amiri2023}%
  \BibitemOpen
  \bibfield  {author} {\bibinfo {author} {\bibfnamefont {A.}~\bibnamefont {Amiri}}, \bibinfo {author} {\bibfnamefont {C.}~\bibnamefont {Duclut}}, \bibinfo {author} {\bibfnamefont {F.}~\bibnamefont {J{\"u}licher}},\ and\ \bibinfo {author} {\bibfnamefont {M.}~\bibnamefont {Popovi{\'c}}},\ }\bibfield  {title} {\bibinfo {title} {Random traction yielding transition in epithelial tissues},\ }\href {https://doi.org/10.1103/PhysRevLett.131.188401} {\bibfield  {journal} {\bibinfo  {journal} {Phys. Rev. Lett.}\ }\textbf {\bibinfo {volume} {131}},\ \bibinfo {pages} {188401} (\bibinfo {year} {2023})}\BibitemShut {NoStop}%
\bibitem [{\citenamefont {Zhang}\ \emph {et~al.}(2021)\citenamefont {Zhang}, \citenamefont {Li}, \citenamefont {Nijjer}, \citenamefont {Lu}, \citenamefont {Kothari}, \citenamefont {Alert}, \citenamefont {Cohen},\ and\ \citenamefont {Yan}}]{Zhang2021b}%
  \BibitemOpen
  \bibfield  {author} {\bibinfo {author} {\bibfnamefont {Q.}~\bibnamefont {Zhang}}, \bibinfo {author} {\bibfnamefont {J.}~\bibnamefont {Li}}, \bibinfo {author} {\bibfnamefont {J.}~\bibnamefont {Nijjer}}, \bibinfo {author} {\bibfnamefont {H.}~\bibnamefont {Lu}}, \bibinfo {author} {\bibfnamefont {M.}~\bibnamefont {Kothari}}, \bibinfo {author} {\bibfnamefont {R.}~\bibnamefont {Alert}}, \bibinfo {author} {\bibfnamefont {T.}~\bibnamefont {Cohen}},\ and\ \bibinfo {author} {\bibfnamefont {J.}~\bibnamefont {Yan}},\ }\bibfield  {title} {\bibinfo {title} {Morphogenesis and cell ordering in confined bacterial biofilms},\ }\href {https://doi.org/10.1073/pnas.2107107118} {\bibfield  {journal} {\bibinfo  {journal} {Proc. Natl. Acad. Sci.}\ }\textbf {\bibinfo {volume} {118}},\ \bibinfo {pages} {e2107107118} (\bibinfo {year} {2021})}\BibitemShut {NoStop}%
\bibitem [{\citenamefont {Guillamat}\ \emph {et~al.}(2022)\citenamefont {Guillamat}, \citenamefont {{Blanch-Mercader}}, \citenamefont {Pernollet}, \citenamefont {Kruse},\ and\ \citenamefont {Roux}}]{Guillamat2022}%
  \BibitemOpen
  \bibfield  {author} {\bibinfo {author} {\bibfnamefont {P.}~\bibnamefont {Guillamat}}, \bibinfo {author} {\bibfnamefont {C.}~\bibnamefont {{Blanch-Mercader}}}, \bibinfo {author} {\bibfnamefont {G.}~\bibnamefont {Pernollet}}, \bibinfo {author} {\bibfnamefont {K.}~\bibnamefont {Kruse}},\ and\ \bibinfo {author} {\bibfnamefont {A.}~\bibnamefont {Roux}},\ }\bibfield  {title} {\bibinfo {title} {Integer topological defects organize stresses driving tissue morphogenesis},\ }\href {https://doi.org/10.1038/s41563-022-01194-5} {\bibfield  {journal} {\bibinfo  {journal} {Nat. Mater.}\ }\textbf {\bibinfo {volume} {21}},\ \bibinfo {pages} {588} (\bibinfo {year} {2022})}\BibitemShut {NoStop}%
\bibitem [{\citenamefont {Ravichandran}\ \emph {et~al.}(2025)\citenamefont {Ravichandran}, \citenamefont {Vogg}, \citenamefont {Kruse}, \citenamefont {Pearce},\ and\ \citenamefont {Roux}}]{Ravichandran2025}%
  \BibitemOpen
  \bibfield  {author} {\bibinfo {author} {\bibfnamefont {Y.}~\bibnamefont {Ravichandran}}, \bibinfo {author} {\bibfnamefont {M.}~\bibnamefont {Vogg}}, \bibinfo {author} {\bibfnamefont {K.}~\bibnamefont {Kruse}}, \bibinfo {author} {\bibfnamefont {D.~J.~G.}\ \bibnamefont {Pearce}},\ and\ \bibinfo {author} {\bibfnamefont {A.}~\bibnamefont {Roux}},\ }\bibfield  {title} {\bibinfo {title} {Topology changes of {{Hydra}} define actin orientation defects as organizers of morphogenesis},\ }\href {https://doi.org/10.1126/sciadv.adr9855} {\bibfield  {journal} {\bibinfo  {journal} {Sci. Adv.}\ }\textbf {\bibinfo {volume} {11}},\ \bibinfo {pages} {eadr9855} (\bibinfo {year} {2025})}\BibitemShut {NoStop}%
\bibitem [{\citenamefont {Li}\ \emph {et~al.}(2025)\citenamefont {Li}, \citenamefont {Huebner}, \citenamefont {Williams}, \citenamefont {Sawyer}, \citenamefont {Peifer}, \citenamefont {Wallingford},\ and\ \citenamefont {Thirumalai}}]{Li2025a}%
  \BibitemOpen
  \bibfield  {author} {\bibinfo {author} {\bibfnamefont {X.}~\bibnamefont {Li}}, \bibinfo {author} {\bibfnamefont {R.~J.}\ \bibnamefont {Huebner}}, \bibinfo {author} {\bibfnamefont {M.~L.~K.}\ \bibnamefont {Williams}}, \bibinfo {author} {\bibfnamefont {J.}~\bibnamefont {Sawyer}}, \bibinfo {author} {\bibfnamefont {M.}~\bibnamefont {Peifer}}, \bibinfo {author} {\bibfnamefont {J.~B.}\ \bibnamefont {Wallingford}},\ and\ \bibinfo {author} {\bibfnamefont {D.}~\bibnamefont {Thirumalai}},\ }\bibfield  {title} {\bibinfo {title} {Emergence of cellular nematic order is a conserved feature of gastrulation in animal embryos},\ }\href {https://doi.org/10.1038/s41467-025-61045-0} {\bibfield  {journal} {\bibinfo  {journal} {Nat. Commun.}\ }\textbf {\bibinfo {volume} {16}},\ \bibinfo {pages} {5946} (\bibinfo {year} {2025})}\BibitemShut {NoStop}%
\bibitem [{\citenamefont {Claussen}\ \emph {et~al.}(2026)\citenamefont {Claussen}, \citenamefont {Brauns},\ and\ \citenamefont {Shraiman}}]{Claussen2026}%
  \BibitemOpen
  \bibfield  {author} {\bibinfo {author} {\bibfnamefont {N.~H.}\ \bibnamefont {Claussen}}, \bibinfo {author} {\bibfnamefont {F.}~\bibnamefont {Brauns}},\ and\ \bibinfo {author} {\bibfnamefont {B.~I.}\ \bibnamefont {Shraiman}},\ }\href {https://doi.org/10.48550/arXiv.2601.08968} {\bibinfo {title} {Elasticity without a reference state: continuum mechanics of active tension nets}} (\bibinfo {year} {2026}),\ \Eprint {https://arxiv.org/abs/2601.08968} {arxiv:2601.08968} \BibitemShut {NoStop}%
\bibitem [{\citenamefont {{Gonz{\'a}lez-Albaladejo}}\ and\ \citenamefont {Bonilla}(2024)}]{Gonzalez-Albaladejo2024}%
  \BibitemOpen
  \bibfield  {author} {\bibinfo {author} {\bibfnamefont {R.}~\bibnamefont {{Gonz{\'a}lez-Albaladejo}}}\ and\ \bibinfo {author} {\bibfnamefont {L.~L.}\ \bibnamefont {Bonilla}},\ }\bibfield  {title} {\bibinfo {title} {Power laws of natural swarms as fingerprints of an extended critical region},\ }\href {https://doi.org/10.1103/PhysRevE.109.014611} {\bibfield  {journal} {\bibinfo  {journal} {Phys. Rev. E}\ }\textbf {\bibinfo {volume} {109}},\ \bibinfo {pages} {014611} (\bibinfo {year} {2024})}\BibitemShut {NoStop}%
\bibitem [{\citenamefont {Nelson}(1979)}]{Nelson1979}%
  \BibitemOpen
  \bibfield  {author} {\bibinfo {author} {\bibfnamefont {D.~R.}\ \bibnamefont {Nelson}},\ }\bibfield  {title} {\bibinfo {title} {Dislocation-mediated melting in two dimensions},\ }\href {https://doi.org/10.1103/PhysRevB.19.2457} {\bibfield  {journal} {\bibinfo  {journal} {Phys. Rev. B}\ }\textbf {\bibinfo {volume} {19}},\ \bibinfo {pages} {2457} (\bibinfo {year} {1979})}\BibitemShut {NoStop}%
\bibitem [{\citenamefont {Lin}\ \emph {et~al.}(2025)\citenamefont {Lin}, \citenamefont {Yu},\ and\ \citenamefont {Pathak}}]{Lin2025}%
  \BibitemOpen
  \bibfield  {author} {\bibinfo {author} {\bibfnamefont {W.-J.}\ \bibnamefont {Lin}}, \bibinfo {author} {\bibfnamefont {H.}~\bibnamefont {Yu}},\ and\ \bibinfo {author} {\bibfnamefont {A.}~\bibnamefont {Pathak}},\ }\bibfield  {title} {\bibinfo {title} {Gradients in cell density and shape transitions drive collective cell migration into confining environments},\ }\href {https://doi.org/10.1039/D3SM01240A} {\bibfield  {journal} {\bibinfo  {journal} {Soft Matter}\ }\textbf {\bibinfo {volume} {21}},\ \bibinfo {pages} {719} (\bibinfo {year} {2025})}\BibitemShut {NoStop}%
\bibitem [{\citenamefont {Tahaei}\ \emph {et~al.}(2025)\citenamefont {Tahaei}, \citenamefont {{Piscitello-G{\'o}mez}}, \citenamefont {Suganthan}, \citenamefont {Cwikla}, \citenamefont {Fuhrmann}, \citenamefont {Dye},\ and\ \citenamefont {Popovi{\'c}}}]{Tahaei2025}%
  \BibitemOpen
  \bibfield  {author} {\bibinfo {author} {\bibfnamefont {A.}~\bibnamefont {Tahaei}}, \bibinfo {author} {\bibfnamefont {R.}~\bibnamefont {{Piscitello-G{\'o}mez}}}, \bibinfo {author} {\bibfnamefont {S.}~\bibnamefont {Suganthan}}, \bibinfo {author} {\bibfnamefont {G.}~\bibnamefont {Cwikla}}, \bibinfo {author} {\bibfnamefont {J.~F.}\ \bibnamefont {Fuhrmann}}, \bibinfo {author} {\bibfnamefont {N.~A.}\ \bibnamefont {Dye}},\ and\ \bibinfo {author} {\bibfnamefont {M.}~\bibnamefont {Popovi{\'c}}},\ }\bibfield  {title} {\bibinfo {title} {Cell divisions imprint long lasting elastic strain fields in epithelial tissues},\ }\href {https://doi.org/10.1103/lh3v-v2c9} {\bibfield  {journal} {\bibinfo  {journal} {PRX Life}\ }\textbf {\bibinfo {volume} {3}},\ \bibinfo {pages} {043008} (\bibinfo {year} {2025})}\BibitemShut {NoStop}%
\bibitem [{\citenamefont {Shen}\ \emph {et~al.}(2025)\citenamefont {Shen}, \citenamefont {O'Byrne}, \citenamefont {Schoenit}, \citenamefont {Maitra}, \citenamefont {M{\`e}ge}, \citenamefont {Voituriez},\ and\ \citenamefont {Ladoux}}]{Shen2025}%
  \BibitemOpen
  \bibfield  {author} {\bibinfo {author} {\bibfnamefont {Y.}~\bibnamefont {Shen}}, \bibinfo {author} {\bibfnamefont {J.}~\bibnamefont {O'Byrne}}, \bibinfo {author} {\bibfnamefont {A.}~\bibnamefont {Schoenit}}, \bibinfo {author} {\bibfnamefont {A.}~\bibnamefont {Maitra}}, \bibinfo {author} {\bibfnamefont {R.-M.}\ \bibnamefont {M{\`e}ge}}, \bibinfo {author} {\bibfnamefont {R.}~\bibnamefont {Voituriez}},\ and\ \bibinfo {author} {\bibfnamefont {B.}~\bibnamefont {Ladoux}},\ }\bibfield  {title} {\bibinfo {title} {Flocking and giant fluctuations in epithelial active solids},\ }\href {https://doi.org/10.1073/pnas.2421327122} {\bibfield  {journal} {\bibinfo  {journal} {Proc. Natl. Acad. Sci.}\ }\textbf {\bibinfo {volume} {122}},\ \bibinfo {pages} {e2421327122} (\bibinfo {year} {2025})}\BibitemShut {NoStop}%
\bibitem [{\citenamefont {Michaut}\ \emph {et~al.}(2025)\citenamefont {Michaut}, \citenamefont {Chamolly}, \citenamefont {Villedieu}, \citenamefont {Corson},\ and\ \citenamefont {Gros}}]{Michaut2025}%
  \BibitemOpen
  \bibfield  {author} {\bibinfo {author} {\bibfnamefont {A.}~\bibnamefont {Michaut}}, \bibinfo {author} {\bibfnamefont {A.}~\bibnamefont {Chamolly}}, \bibinfo {author} {\bibfnamefont {A.}~\bibnamefont {Villedieu}}, \bibinfo {author} {\bibfnamefont {F.}~\bibnamefont {Corson}},\ and\ \bibinfo {author} {\bibfnamefont {J.}~\bibnamefont {Gros}},\ }\bibfield  {title} {\bibinfo {title} {A tension-induced morphological transition shapes the avian extra-embryonic territory},\ }\href {https://doi.org/10.1016/j.cub.2025.02.028} {\bibfield  {journal} {\bibinfo  {journal} {Curr. Biol.}\ }\textbf {\bibinfo {volume} {35}},\ \bibinfo {pages} {1681} (\bibinfo {year} {2025})}\BibitemShut {NoStop}%
\bibitem [{\citenamefont {Duclut}\ \emph {et~al.}(2022)\citenamefont {Duclut}, \citenamefont {Paijmans}, \citenamefont {Inamdar}, \citenamefont {Modes},\ and\ \citenamefont {J{\"u}licher}}]{Duclut2022a}%
  \BibitemOpen
  \bibfield  {author} {\bibinfo {author} {\bibfnamefont {C.}~\bibnamefont {Duclut}}, \bibinfo {author} {\bibfnamefont {J.}~\bibnamefont {Paijmans}}, \bibinfo {author} {\bibfnamefont {M.~M.}\ \bibnamefont {Inamdar}}, \bibinfo {author} {\bibfnamefont {C.~D.}\ \bibnamefont {Modes}},\ and\ \bibinfo {author} {\bibfnamefont {F.}~\bibnamefont {J{\"u}licher}},\ }\bibfield  {title} {\bibinfo {title} {Active {{T1}} transitions in cellular networks},\ }\href {https://doi.org/10.1140/epje/s10189-022-00175-5} {\bibfield  {journal} {\bibinfo  {journal} {Eur. Phys. J. E}\ }\textbf {\bibinfo {volume} {45}},\ \bibinfo {pages} {29} (\bibinfo {year} {2022})}\BibitemShut {NoStop}%
\bibitem [{\citenamefont {Sknepnek}\ \emph {et~al.}(2023)\citenamefont {Sknepnek}, \citenamefont {{Djafer-Cherif}}, \citenamefont {Chuai}, \citenamefont {Weijer},\ and\ \citenamefont {Henkes}}]{Sknepnek2023}%
  \BibitemOpen
  \bibfield  {author} {\bibinfo {author} {\bibfnamefont {R.}~\bibnamefont {Sknepnek}}, \bibinfo {author} {\bibfnamefont {I.}~\bibnamefont {{Djafer-Cherif}}}, \bibinfo {author} {\bibfnamefont {M.}~\bibnamefont {Chuai}}, \bibinfo {author} {\bibfnamefont {C.}~\bibnamefont {Weijer}},\ and\ \bibinfo {author} {\bibfnamefont {S.}~\bibnamefont {Henkes}},\ }\bibfield  {title} {\bibinfo {title} {Generating active {{T1}} transitions through mechanochemical feedback},\ }\href {https://doi.org/10.7554/eLife.79862} {\bibfield  {journal} {\bibinfo  {journal} {eLife}\ }\textbf {\bibinfo {volume} {12}},\ \bibinfo {pages} {e79862} (\bibinfo {year} {2023})}\BibitemShut {NoStop}%
\bibitem [{\citenamefont {Lin}\ \emph {et~al.}(2026)\citenamefont {Lin}, \citenamefont {Tlili},\ and\ \citenamefont {Rupprecht}}]{Lin2026}%
  \BibitemOpen
  \bibfield  {author} {\bibinfo {author} {\bibfnamefont {S.-Z.}\ \bibnamefont {Lin}}, \bibinfo {author} {\bibfnamefont {S.}~\bibnamefont {Tlili}},\ and\ \bibinfo {author} {\bibfnamefont {J.-F.}\ \bibnamefont {Rupprecht}},\ }\href {https://doi.org/10.48550/arXiv.2602.13049} {\bibinfo {title} {Viscous vertex model for active epithelial tissues}} (\bibinfo {year} {2026}),\ \Eprint {https://arxiv.org/abs/2602.13049} {arxiv:2602.13049} \BibitemShut {NoStop}%
\bibitem [{\citenamefont {Bera}\ \emph {et~al.}(2026)\citenamefont {Bera}, \citenamefont {Nguyen}, \citenamefont {McCord}, \citenamefont {Bi},\ and\ \citenamefont {Notbohm}}]{Bera2026}%
  \BibitemOpen
  \bibfield  {author} {\bibinfo {author} {\bibfnamefont {P.~K.}\ \bibnamefont {Bera}}, \bibinfo {author} {\bibfnamefont {A.~Q.}\ \bibnamefont {Nguyen}}, \bibinfo {author} {\bibfnamefont {M.}~\bibnamefont {McCord}}, \bibinfo {author} {\bibfnamefont {D.}~\bibnamefont {Bi}},\ and\ \bibinfo {author} {\bibfnamefont {J.}~\bibnamefont {Notbohm}},\ }\href {https://doi.org/10.48550/arXiv.2603.05548} {\bibinfo {title} {Shape-independent fluidization in epithelial cell monolayers}} (\bibinfo {year} {2026}),\ \Eprint {https://arxiv.org/abs/2603.05548} {arxiv:2603.05548} \BibitemShut {NoStop}%
\bibitem [{\citenamefont {Howard}\ \emph {et~al.}(2011)\citenamefont {Howard}, \citenamefont {Grill},\ and\ \citenamefont {Bois}}]{Howard2011}%
  \BibitemOpen
  \bibfield  {author} {\bibinfo {author} {\bibfnamefont {J.}~\bibnamefont {Howard}}, \bibinfo {author} {\bibfnamefont {S.~W.}\ \bibnamefont {Grill}},\ and\ \bibinfo {author} {\bibfnamefont {J.~S.}\ \bibnamefont {Bois}},\ }\bibfield  {title} {\bibinfo {title} {Turing's next steps: the mechanochemical basis of morphogenesis},\ }\href {https://doi.org/10.1038/nrm3120} {\bibfield  {journal} {\bibinfo  {journal} {Nat. Rev. Mol. Cell Biol.}\ }\textbf {\bibinfo {volume} {12}},\ \bibinfo {pages} {392} (\bibinfo {year} {2011})}\BibitemShut {NoStop}%
\bibitem [{\citenamefont {Bailles}\ \emph {et~al.}(2022)\citenamefont {Bailles}, \citenamefont {Gehrels},\ and\ \citenamefont {Lecuit}}]{Bailles2022}%
  \BibitemOpen
  \bibfield  {author} {\bibinfo {author} {\bibfnamefont {A.}~\bibnamefont {Bailles}}, \bibinfo {author} {\bibfnamefont {E.~W.}\ \bibnamefont {Gehrels}},\ and\ \bibinfo {author} {\bibfnamefont {T.}~\bibnamefont {Lecuit}},\ }\bibfield  {title} {\bibinfo {title} {Mechanochemical principles of spatial and temporal patterns in cells and tissues},\ }\href {https://doi.org/10.1146/annurev-cellbio-120420-095337} {\bibfield  {journal} {\bibinfo  {journal} {Annu. Rev. Cell Dev. Biol.}\ }\textbf {\bibinfo {volume} {38}},\ \bibinfo {pages} {321} (\bibinfo {year} {2022})}\BibitemShut {NoStop}%
\bibitem [{\citenamefont {Sussman}(2020)}]{Sussman2020}%
  \BibitemOpen
  \bibfield  {author} {\bibinfo {author} {\bibfnamefont {D.~M.}\ \bibnamefont {Sussman}},\ }\bibfield  {title} {\bibinfo {title} {Interplay of curvature and rigidity in shape-based models of confluent tissue},\ }\href {https://doi.org/10.1103/PhysRevResearch.2.023417} {\bibfield  {journal} {\bibinfo  {journal} {Phys. Rev. Res.}\ }\textbf {\bibinfo {volume} {2}},\ \bibinfo {pages} {023417} (\bibinfo {year} {2020})}\BibitemShut {NoStop}%
\bibitem [{\citenamefont {Luciano}\ \emph {et~al.}(2024)\citenamefont {Luciano}, \citenamefont {Tomba}, \citenamefont {Roux},\ and\ \citenamefont {Gabriele}}]{Luciano2024}%
  \BibitemOpen
  \bibfield  {author} {\bibinfo {author} {\bibfnamefont {M.}~\bibnamefont {Luciano}}, \bibinfo {author} {\bibfnamefont {C.}~\bibnamefont {Tomba}}, \bibinfo {author} {\bibfnamefont {A.}~\bibnamefont {Roux}},\ and\ \bibinfo {author} {\bibfnamefont {S.}~\bibnamefont {Gabriele}},\ }\bibfield  {title} {\bibinfo {title} {How multiscale curvature couples forces to cellular functions},\ }\href {https://doi.org/10.1038/s42254-024-00700-9} {\bibfield  {journal} {\bibinfo  {journal} {Nat. Rev. Phys.}\ }\textbf {\bibinfo {volume} {6}},\ \bibinfo {pages} {246} (\bibinfo {year} {2024})}\BibitemShut {NoStop}%
\bibitem [{\citenamefont {{Torres-S{\'a}nchez}}\ \emph {et~al.}(2022)\citenamefont {{Torres-S{\'a}nchez}}, \citenamefont {Winter},\ and\ \citenamefont {Salbreux}}]{Torres-Sanchez2022}%
  \BibitemOpen
  \bibfield  {author} {\bibinfo {author} {\bibfnamefont {A.}~\bibnamefont {{Torres-S{\'a}nchez}}}, \bibinfo {author} {\bibfnamefont {M.~K.}\ \bibnamefont {Winter}},\ and\ \bibinfo {author} {\bibfnamefont {G.}~\bibnamefont {Salbreux}},\ }\bibfield  {title} {\bibinfo {title} {Interacting active surfaces: {{A}} model for three-dimensional cell aggregates},\ }\href {https://doi.org/10.1371/journal.pcbi.1010762} {\bibfield  {journal} {\bibinfo  {journal} {PLOS Comput. Biol.}\ }\textbf {\bibinfo {volume} {18}},\ \bibinfo {pages} {e1010762} (\bibinfo {year} {2022})}\BibitemShut {NoStop}%
\bibitem [{\citenamefont {Yu}\ \emph {et~al.}(2024)\citenamefont {Yu}, \citenamefont {Li}, \citenamefont {Fang}, \citenamefont {Feng},\ and\ \citenamefont {Li}}]{Yu2024}%
  \BibitemOpen
  \bibfield  {author} {\bibinfo {author} {\bibfnamefont {P.}~\bibnamefont {Yu}}, \bibinfo {author} {\bibfnamefont {Y.}~\bibnamefont {Li}}, \bibinfo {author} {\bibfnamefont {W.}~\bibnamefont {Fang}}, \bibinfo {author} {\bibfnamefont {X.-Q.}\ \bibnamefont {Feng}},\ and\ \bibinfo {author} {\bibfnamefont {B.}~\bibnamefont {Li}},\ }\bibfield  {title} {\bibinfo {title} {Mechanochemical dynamics of collective cells and hierarchical topological defects in multicellular lumens},\ }\href {https://doi.org/10.1126/sciadv.adn0172} {\bibfield  {journal} {\bibinfo  {journal} {Sci. Adv.}\ }\textbf {\bibinfo {volume} {10}},\ \bibinfo {pages} {eadn0172} (\bibinfo {year} {2024})}\BibitemShut {NoStop}%
\bibitem [{\citenamefont {Runser}\ \emph {et~al.}(2024)\citenamefont {Runser}, \citenamefont {Vetter},\ and\ \citenamefont {Iber}}]{Runser2024}%
  \BibitemOpen
  \bibfield  {author} {\bibinfo {author} {\bibfnamefont {S.}~\bibnamefont {Runser}}, \bibinfo {author} {\bibfnamefont {R.}~\bibnamefont {Vetter}},\ and\ \bibinfo {author} {\bibfnamefont {D.}~\bibnamefont {Iber}},\ }\bibfield  {title} {\bibinfo {title} {{{SimuCell3D}}: three-dimensional simulation of tissue mechanics with cell polarization},\ }\href {https://doi.org/10.1038/s43588-024-00620-9} {\bibfield  {journal} {\bibinfo  {journal} {Nat. Comput. Sci.}\ ,\ \bibinfo {pages} {1}} (\bibinfo {year} {2024})}\BibitemShut {NoStop}%
\bibitem [{\citenamefont {Yu}\ \emph {et~al.}(2025)\citenamefont {Yu}, \citenamefont {Zhang},\ and\ \citenamefont {Li}}]{Yu2025}%
  \BibitemOpen
  \bibfield  {author} {\bibinfo {author} {\bibfnamefont {P.}~\bibnamefont {Yu}}, \bibinfo {author} {\bibfnamefont {R.}~\bibnamefont {Zhang}},\ and\ \bibinfo {author} {\bibfnamefont {B.}~\bibnamefont {Li}},\ }\bibfield  {title} {\bibinfo {title} {Cell stiffness-mediated mechanochemical waves in three-dimensional tissues},\ }\href {https://doi.org/10.1103/r4fs-4qqk} {\bibfield  {journal} {\bibinfo  {journal} {Phys. Rev. Lett.}\ }\textbf {\bibinfo {volume} {135}},\ \bibinfo {pages} {108401} (\bibinfo {year} {2025})}\BibitemShut {NoStop}%
\bibitem [{\citenamefont {Salbreux}\ \emph {et~al.}(2012)\citenamefont {Salbreux}, \citenamefont {Charras},\ and\ \citenamefont {Paluch}}]{Salbreux2012}%
  \BibitemOpen
  \bibfield  {author} {\bibinfo {author} {\bibfnamefont {G.}~\bibnamefont {Salbreux}}, \bibinfo {author} {\bibfnamefont {G.}~\bibnamefont {Charras}},\ and\ \bibinfo {author} {\bibfnamefont {E.}~\bibnamefont {Paluch}},\ }\bibfield  {title} {\bibinfo {title} {Actin cortex mechanics and cellular morphogenesis},\ }\href {https://doi.org/10.1016/j.tcb.2012.07.001} {\bibfield  {journal} {\bibinfo  {journal} {Trends Cell Biol.}\ }\textbf {\bibinfo {volume} {22}},\ \bibinfo {pages} {536} (\bibinfo {year} {2012})}\BibitemShut {NoStop}%
\bibitem [{\citenamefont {Mulla}\ \emph {et~al.}(2019)\citenamefont {Mulla}, \citenamefont {MacKintosh},\ and\ \citenamefont {Koenderink}}]{Mulla2019}%
  \BibitemOpen
  \bibfield  {author} {\bibinfo {author} {\bibfnamefont {Y.}~\bibnamefont {Mulla}}, \bibinfo {author} {\bibfnamefont {F.~C.}\ \bibnamefont {MacKintosh}},\ and\ \bibinfo {author} {\bibfnamefont {G.~H.}\ \bibnamefont {Koenderink}},\ }\bibfield  {title} {\bibinfo {title} {Origin of slow stress relaxation in the cytoskeleton},\ }\href {https://doi.org/10.1103/PhysRevLett.122.218102} {\bibfield  {journal} {\bibinfo  {journal} {Phys. Rev. Lett.}\ }\textbf {\bibinfo {volume} {122}},\ \bibinfo {pages} {218102} (\bibinfo {year} {2019})}\BibitemShut {NoStop}%
\bibitem [{\citenamefont {Kim}\ \emph {et~al.}(2021)\citenamefont {Kim}, \citenamefont {Pochitaloff}, \citenamefont {{Stooke-Vaughan}},\ and\ \citenamefont {Camp{\`a}s}}]{Kim2021}%
  \BibitemOpen
  \bibfield  {author} {\bibinfo {author} {\bibfnamefont {S.}~\bibnamefont {Kim}}, \bibinfo {author} {\bibfnamefont {M.}~\bibnamefont {Pochitaloff}}, \bibinfo {author} {\bibfnamefont {G.~A.}\ \bibnamefont {{Stooke-Vaughan}}},\ and\ \bibinfo {author} {\bibfnamefont {O.}~\bibnamefont {Camp{\`a}s}},\ }\bibfield  {title} {\bibinfo {title} {Embryonic tissues as active foams},\ }\href {https://doi.org/10.1038/s41567-021-01215-1} {\bibfield  {journal} {\bibinfo  {journal} {Nat. Phys.}\ }\textbf {\bibinfo {volume} {17}},\ \bibinfo {pages} {859} (\bibinfo {year} {2021})}\BibitemShut {NoStop}%
\bibitem [{\citenamefont {Noll}\ \emph {et~al.}(2017)\citenamefont {Noll}, \citenamefont {Mani}, \citenamefont {Heemskerk}, \citenamefont {Streichan},\ and\ \citenamefont {Shraiman}}]{Noll2017}%
  \BibitemOpen
  \bibfield  {author} {\bibinfo {author} {\bibfnamefont {N.}~\bibnamefont {Noll}}, \bibinfo {author} {\bibfnamefont {M.}~\bibnamefont {Mani}}, \bibinfo {author} {\bibfnamefont {I.}~\bibnamefont {Heemskerk}}, \bibinfo {author} {\bibfnamefont {S.~J.}\ \bibnamefont {Streichan}},\ and\ \bibinfo {author} {\bibfnamefont {B.~I.}\ \bibnamefont {Shraiman}},\ }\bibfield  {title} {\bibinfo {title} {Active tension network model suggests an exotic mechanical state realized in epithelial tissues},\ }\href {https://doi.org/10.1038/nphys4219} {\bibfield  {journal} {\bibinfo  {journal} {Nat. Phys.}\ }\textbf {\bibinfo {volume} {13}},\ \bibinfo {pages} {1221} (\bibinfo {year} {2017})}\BibitemShut {NoStop}%
\bibitem [{\citenamefont {Killeen}\ \emph {et~al.}(2022)\citenamefont {Killeen}, \citenamefont {Bertrand},\ and\ \citenamefont {Lee}}]{Killeen2022}%
  \BibitemOpen
  \bibfield  {author} {\bibinfo {author} {\bibfnamefont {A.}~\bibnamefont {Killeen}}, \bibinfo {author} {\bibfnamefont {T.}~\bibnamefont {Bertrand}},\ and\ \bibinfo {author} {\bibfnamefont {C.~F.}\ \bibnamefont {Lee}},\ }\bibfield  {title} {\bibinfo {title} {Polar fluctuations lead to extensile nematic behavior in confluent tissues},\ }\href {https://doi.org/10.1103/PhysRevLett.128.078001} {\bibfield  {journal} {\bibinfo  {journal} {Phys. Rev. Lett.}\ }\textbf {\bibinfo {volume} {128}},\ \bibinfo {pages} {078001} (\bibinfo {year} {2022})}\BibitemShut {NoStop}%
\end{thebibliography}
%\bibliographystyle{apsrev4-2.bst}


\end{document}